\documentclass{article}

\usepackage{graphicx}

\usepackage{epsf}

\usepackage{longtable}

\usepackage{epubtk}

\newcommand{\pd}[2]{\frac{\partial {#1}}{\partial {#2}}}
\newcommand{\s}{{\rm s}}
\newcommand{\n}{{\rm n}}

\newcommand{\X}{{\rm x}}
\newcommand{\Y}{{\rm y}}
\newcommand{\Z}{{\rm z}}

\newcommand{\A}{{\cal A}}
\newcommand{\B}{{\cal B}}
\newcommand{\C}{{\cal C}}
\newcommand{\E}{{\cal E}}

\begin{document}

\title{Relativistic fluid dynamics: physics for many different scales}

\author{%
\epubtkAuthorData{N.~Andersson}{%
School of Mathematics \\ 
University of Southampton \\ 
Southampton SO17 1BJ, United Kingdom}{%
na@maths.soton.ac.uk}{%
} 
\and
\epubtkAuthorData{G.~L.~Comer}{%
Department of Physics \& Center for Fluids at All Scales \\ 
Saint Louis University \\ 
St.~Louis, MO, 63156-0907, USA}{%
comergl@slu.edu}{%
}
}

\epubtkKeywords{fluid dynamics,relativistic hydrodynamics,relativistic%
astrophysics,variational methods,classical field theory}

\date{\today}

\maketitle

\begin{abstract}
The relativistic fluid is a highly successful model used to describe the 
dynamics of many-particle, relativistic systems. \ It takes as input 
basic physics from microscopic scales and yields as output predictions of 
bulk, macroscopic motion. \ By inverting the process, an understanding of bulk 
features can lead to insight into physics on the microscopic scale. \ 
Relativistic fluids have been used to model systems as ``small'' as heavy 
ions in collisions, and as large as the universe itself, with ``intermediate'' 
sized objects like neutron stars being considered along the way. \ The purpose 
of this review is to discuss the mathematical and theoretical physics 
underpinnings of the relativistic (multiple) fluid model. \ We focus on 
the variational principle approach championed by Brandon Carter and his 
collaborators, in which a crucial element is to distinguish the momenta that 
are conjugate to the particle number density currents. \ This approach 
differs from the ``standard'' text-book derivation of the equations of motion 
from the divergence of the stress-energy tensor, in that one explicitly 
obtains the relativistic Euler equation as an ``integrability'' condition on 
the relativistic vorticity. \ We discuss the conservation laws and the 
equations of motion in detail, and provide a number of (in our opinion) 
interesting and relevant applications of the general theory. 
\end{abstract}

\newpage

\section{Introduction}
\label{sec:intro}

\subsection{Setting the stage}

If one does a search on the topic of relativistic fluids on any of the major 
physics article databases one is overwhelmed by the number of ``hits''. \ This 
is a manifestation of the importance that the fluid model has long had for 
modern physics and engineering. \ For relativistic physics, in particular, the 
fluid model is essential. \ After all, many-particle astrophysical and 
cosmological systems are the best sources of detectable effects associated 
with General Relativity. \ Two obvious examples, the expansion of the universe 
and oscillations of neutron stars, indicate the vast range of scales 
on which relativistic fluid models are relevant. \ A particularly exciting 
context for general relativistic fluids today is their use in the modelling of 
sources of gravitational waves. \ This includes the compact binary inspiral 
problem, either of two neutron stars or a neutron star and a black hole, the 
collapse of stellar cores during supernovae, or various neutron star 
instabilities. \ One should also not forget the use of special relativistic 
fluids in modelling collisions of heavy nuclei, astrophysical jets, and 
gamma-ray burst sources. 

This review provides an introduction to the modelling of fluids in General 
Relativity. \ Our main target audience is graduate students with a need for an 
understanding of relativistic fluid dynamics. \ Hence, we have made an effort 
to keep the presentation pedagogical. \ The article will (hopefully) also be 
useful to researchers who work in areas outside of General Relativity and 
gravitation per se (eg.~a nuclear physicist who develops neutron star 
equations of state), but who require a working knowledge of relativistic fluid 
dynamics.

Throughout most of the article we will assume that General Relativity is 
the proper description of gravity. \ Although not too severe, this is a 
restriction since the problem of fluids in other theories of gravity has 
interesting aspects. \ As we hope that the article will be used by students 
and researchers who are not necessarily experts in General Relativity and 
techniques of differential geometry, we have included some discussion of the 
mathematical tools required to build models of relativistic objects. \ Even 
though our summary is not a proper introduction to General Relativity we have 
tried to define the tools that are required for the discussion that 
follows. \ Hopefully our description is sufficiently self-contained to 
provide a less experienced reader with a working understanding of (at least 
some of) the mathematics involved. \ In particular, the reader will find 
an extended discussion of the covariant and Lie derivatives. \ This is natural 
since many important properties of fluids, both relativistic and 
non-relativistic, can be established and understood by the use of parallel 
transport and Lie-dragging. \ But it is vital to appreciate the 
distinctions between the two. 

Ideally, the reader should have some familiarity with standard fluid 
dynamics, eg.~at the level of the discussion in Landau \& Lifshitz 
\cite{landau59:_fluid_mech}, basic thermodynamics \cite{reichl98:_book}, 
and the mathematics of action principles and how they are used to 
generate equations of motion \cite{lanczos49:_var_mechs}. \ Having stated 
this, it is clear that we are facing a facing a real challenge. \ We are 
trying to introduce a topic on which numerous books have been written 
(eg.~\cite{tolman34:_book,landau59:_fluid_mech,lichnerowica67:_book,%
anile90:_relfl_book,wilson03:_book}), 
and which requires an understanding of much of theoretical physics. \ 
Yet, one can argue that an article of this kind is timely. \ In 
particular, there has recently been exciting developments for 
multi-constituent systems, such as superfluid/superconducting neutron 
star cores.\epubtkFootnote{In this article we use ``superfluid'' to refer to 
any system which has the ability to flow without friction. \ In this sense, 
superfluids and superconductors are viewed in the same way. \ When we wish to 
distinguish charge carrying fluids, we will call them superconductors.} \ 
Much of this work has been guided by the geometric approach to fluid dynamics 
championed by Carter 
\cite{carter83:_in_random_walk,carter89:_covar_theor_conduc,carter92:_brane}. 
\ This provides a powerful framework that makes extensions to multi-fluid 
situations quite intuitive. \ A typical example of a phenomenon that 
arises naturally is the so-called entrainment effect, which plays a 
crucial role in a superfluid neutron star core. \ Given the potential for 
future applications of this formalism, we have opted to base much of our 
description on the work of Carter and his colleagues. 

Even though the subject of relativistic fluids is far from new, a number 
of issues remain to be resolved. \ The most obvious shortcoming of the 
present theory concerns dissipative effects. \ As we will discuss, 
dissipative effects are (at least in principle) easy to incorporate in 
Newtonian theory but the extension to General Relativity is problematic 
(see, for instance, Hiscock and Lindblom \cite{hiscock85:_rel_diss_fluids}). 
\ Following early work by Eckart \cite{eckart40:_rel_diss_fluid}, a 
significant effort was made by Israel and Stewart 
\cite{Israel79:_kintheo1,Israel79:_kintheo2} and Carter 
\cite{carter83:_in_random_walk,carter89:_covar_theor_conduc}. \ 
Incorporation of dissipation is still an active enterprise, and of key 
importance for future gravitational-wave asteroseismology which requires 
detailed estimates of the role of viscosity in suppressing possible 
instabilities. 

\subsection{A brief history of fluids}

The two fluids air and water are essential to human survival. \ This obvious 
fact implies a basic human need to divine their innermost secrets. \ Homo 
Sapiens have always needed to anticipate air and water behaviour under a 
myriad of circumstances, such as those that concern water supply, weather, and 
travel. \ The essential importance of fluids for survival, and that they can 
be exploited to enhance survival, implies that the study of fluids probably 
reaches as far back into antiquity as the human race itself. \ 
Unfortunately, our historical record of this ever-ongoing study is not so 
great that we can reach very far accurately. 

A wonderful account (now in Dover print) is ``A History and Philosophy of 
Fluid Mechanics,'' by G.~A.~Tokaty \cite{tokaty}. \ He points out that while 
early cultures may not have had universities, government sponsored labs, or 
privately funded centres pursuing fluids research (nor the Living Reviews 
archive on which to communicate results!), there was certainly some 
collective understanding. After all,  there is a clear connection between the 
viability of early civilizations and their access to water. \ For example, we 
have the societies associated with the Yellow and Yangtze rivers in China, the 
Ganges in India, the Volga in Russia, the Thames in England, and the Seine in 
France, to name just a few. \ We should also not forget the Babylonians and 
their amazing technological (irrigation) achievements in the land between the 
Tigris and Euphrates, and  the Egyptians, whose intimacy with the flooding 
of the Nile is well documented. \ In North America, we have the so-called 
Mississippians, who left behind their mound-building accomplishments. \ 
For example, the Cahokians (in Collinsville, Illinois) constructed Monk's 
Mound, the largest precolumbian earthen structure in existence that is (see 
\url{http://en.wikipedia.org/wiki/Monk's\_Mound}) ``... over 100 feet tall, 
1000 feet long, and 800 feet wide (larger at its base than the Great Pyramid 
of Giza).'' 

In terms of ocean and sea travel, we know that the maritime ability of the 
Mediterranean people was a main mechanism for ensuring cultural and economic 
growth and societal stability. \ The finely-tuned skills of the Polynesians 
in the South Pacific allowed them to travel great distances, perhaps reaching 
as far as South America itself, and certainly making it to the ``most remote 
spot on the Earth,'' Easter Island. \ Apparently, they were remarkably 
adept at reading the smallest of signs --- water color, views of weather on 
the horizon, subtleties of wind patterns, floating objects, birds, etc --- as 
indication of nearby land masses. \ Finally, the harsh climate of the North 
Atlantic was overcome by the highly accomplished Nordic sailors, whose skills 
allowed them to reach several sites in North America. \ Perhaps it would be 
appropriate to think of these early explorers as adept geophysical fluid 
dynamicists/oceanographers?

\begin{figure}[t]
\centering
\includegraphics[width=5cm,clip]{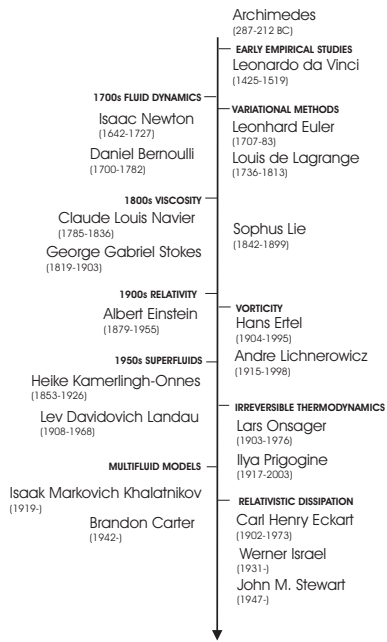}
\caption{A ``timeline'' focussed on the topics covered in this
	review, including chemists, engineers, mathematicians,
	philosophers, and physicists who have contributed to the
	development of non-relativistic fluids, their relativistic
	counterparts, multi-fluid versions of both, and exotic
	phenomena such as superfluidity.}
\label{time}
\end{figure}

Many great scientists are associated with the study of fluids. \ Lost are the 
names of those individuals who, almost 400,000 years ago, carved 
``aerodynamically correct'' \cite{gadelhak} wooden spears. \ Also lost are 
those who developed boomerangs and fin-stabilized arrows. \ Among those not 
lost is Archimedes, the Greek Mathematician (287-212 BC), who provided a 
mathematical expression for the buoyant force on bodies. \ Earlier, Thales of 
Miletus (624-546 BC) asked the simple question: What {\em is} air and 
water? \ His question is profound since it represents a clear departure 
from the main, myth-based modes of inquiry at that time. \ Tokaty ranks Hero 
of Alexandria as one of the great, early contributors. \ Hero (c.~10-70) was a 
Greek scientist and engineer, who left behind many writings and drawings that, 
from today's perspective, indicate a good grasp of basic fluid mechanics. \ 
To make complete account of individual contributions to our present 
understanding of fluid dynamics is, of course, impossible. \ Yet it is 
useful to list some of the contributors to the field. \ We provide a highly 
subjective ``timeline'' in Fig.~\ref{time}. \ Our list is to a large extent 
focussed on the topics covered in this review, and includes chemists, 
engineers, mathematicians, philosophers, and physicists. \ It recognizes those 
who have contributed to the development of non-relativistic fluids, their 
relativistic counterparts, multi-fluid versions of both, and exotic phenomena 
such as superfluidity. \ We have provided this list with the hope that the 
reader can use these names as key words in a modern, web-based literature 
search whenever more information is required. 

Tokaty \cite{tokaty} discusses the human propensity for destruction when it 
comes to our water resources. \ Depletion and pollution are the main 
offenders. \ He refers to a ``Battle of the Fluids'' as a struggle between 
their destruction and protection. \ His context for this discussion was the 
Cold War. \ He rightly points out the failure to protect our water and air 
resource by the two predominant participants --- the USA and USSR. \ In an 
ironic twist, modern study of the relativistic properties of fluids has also 
a ``Battle of the Fluids.'' \ A self-gravitating mass can become absolutely 
unstable and collapse to a black hole, the ultimate destruction of any form of 
matter.

\subsection{Notation and conventions}

Throughout the article we assume the ``MTW'' \cite{mtw73} conventions. \ 
We also generally assume geometrized units $c=G=1$, unless specifically noted 
otherwise. \ A coordinate basis will always be used, with spacetime 
indices denoted by lowercase Greek letters that range over $\{0,1,2,3\}$ 
(time being the zeroth coordinate), and purely spatial indices denoted by 
lowercase Latin letters that range over $\{1,2,3\}$. \ Unless otherwise 
noted, we assume the Einstein summation convention.

\newpage

\section{Physics in a Curved Spacetime}
\label{sec:gr}

There is an extensive literature on special and general relativity, and 
the spacetime-based view (i.e.~three space and one time dimensions that 
form a manifold) of the laws of physics. \ For the student at any 
level interested in developing a working understanding we recommend 
Taylor and Wheeler \cite{taylorwheeler92:_book} for an introduction, 
followed by Hartle's excellent recent text \cite{hartle03:_book} designed for 
students at the undergraduate level. \ For the more advanced students, we 
suggest two of the classics, ``MTW'' \cite{mtw73} and Weinberg 
\cite{weinberg72:_book}, or the more contemporary book by Wald 
\cite{wald84:_book}. \ Finally, let us not forget the Living Reviews 
archive as a premier online source of up-to-date information!

In terms of the experimental and/or observational support for special and 
general relativity, we recommend two articles by Will that were written 
for the 2005 World Year of Physics celebration \cite{will05:_sr,will05:_gr}. 
\ They summarize a variety of tests that have been designed to expose 
breakdowns in both theories. \ (We also recommend Will's popular book 
{\em Was Einstein Right?} \cite{will86:_rightbook} and his technical 
exposition {\em Theory and Experiment in Gravitational Physics} 
\cite{will93:_theorybook}.) \ To date, Einstein's theoretical edifice is 
still standing! 

For special relativity, this is not surprising, given its long list of 
successes: explanation of the Michaelson-Morley result, the prediction 
and subsequent discovery of anti-matter, and the standard model of 
particle physics, to name a few. \ Will \cite{will05:_sr} offers the 
observation that genetic mutations via cosmic rays require special 
relativity, since otherwise muons would decay before making it to the 
surface of the Earth. \ On a more somber note, we may consider the Trinity 
site in New Mexico, and the tragedies of Hiroshima and Nagasaki, as 
reminders of $E = m c^2$. 

In support of general relativity, there are E\"otv\"os-type experiments 
testing the equivalence of inertial and gravitational mass, detection of 
gravitational red-shifts of photons, the passing of the solar system 
tests, confirmation of energy loss via gravitational radiation in the 
Hulse-Taylor binary pulsar, and the expansion of the universe. \ 
Incredibly, general relativity even finds  a practical application in the 
GPS system: if general relativity is neglected an error of about 15 meters 
results when trying to resolve the location of an object 
\cite{will05:_sr}. \ Definitely enough to make driving dangerous!

The evidence is thus overwhelming that general relativity, or at least 
something that passes the same tests, is the proper description of 
gravity. \ Given this, we assume the Einstein Equivalence Principle, i.e.~that 
\cite{will05:_sr,will05:_gr,will93:_theorybook}: 
\begin{itemize}
\item test bodies fall with the same acceleration independently of their 
internal structure or composition;
\item the outcome of any local non-gravitational experiment is 
independent of the velocity of the freely-falling reference frame in 
which it is performed;
\item the outcome of any local non-gravitational experiment is 
independent of where and when in the universe it is performed. 
\end{itemize}
If the Equivalence Principle holds, then gravitation must be described by a 
metric-based theory \cite{will05:_sr}. \ This means that 
\begin{enumerate}
\item spacetime is endowed with a symmetric metric, 
\item the trajectories of freely falling bodies are geodesics of that 
metric, and
\item in local freely falling reference frames, the non-gravitational 
laws of physics are those of special relativity.
\end{enumerate}
For our present purposes this is very good news. \ The availability of a 
metric means that we can develop the theory without requiring much of the 
differential geometry edifice that would be needed in a more general case. \ 
We will develop the description of relativistic fluids with this in mind. \ 
Readers that find our approach too ``pedestrian'' may want to consult
the recent article by Gourgoulhon \cite{gour06}, which serves as a useful 
complement to our description.

\subsection{The Metric and Spacetime Curvature}

Our strategy is to provide a ``working understanding'' of the 
mathematical objects that enter the Einstein equations of general 
relativity. \ We assume that the metric is the fundamental ``field'' of 
gravity. \ For four-dimensional spacetime it determines the distance 
between two spacetime points along a given curve, which can generally be 
written as a one parameter function with, say, components 
$x^\mu(\lambda)$. \ As we will see, once a notion of parallel transport 
is established, the metric also encodes  information about the 
curvature of its spacetime, which is taken to be of the pseudo-Riemannian 
type, meaning that the signature of the metric is $-+++$ 
(cf.~Eq.~(\ref{sig}) below). 

In a coordinate basis, which we will assume throughout this review, the metric 
is denoted by $g_{\mu \nu} = g_{\nu \mu}$. \ The symmetry implies that 
there are in general ten independent components (modulo the freedom to set 
arbitrarily four components that is inherited from coordinate transformations; 
cf.~Eqs.~(\ref{ct1}) and (\ref{ct2}) below). \ The spacetime version of the 
Pythagorean theorem takes the form 
\begin{equation}
     {\rm d} s^2 = g_{\mu \nu} {\rm d} x^\mu {\rm d} x^\nu \ ,
\end{equation}
and in a local set of Minkowski coordinates $\{t,x,y,z\}$ (i.e. in a local 
inertial frame) it looks like 
\begin{equation}
     {\rm d} s^2 = - \left({\rm d} t\right)^2 + \left({\rm d} x\right)^2 + 
                   \left({\rm d} y\right)^2 + \left({\rm d} z\right)^2 \ . 
                   \label{sig} 
\end{equation}
This illustrates the $-+++$ signature. \ The inverse metric $g^{\mu \nu}$ is 
such that 
\begin{equation}
     g^{\mu \rho} g_{\rho \nu} = \delta^\mu{}_\nu \ , 
\end{equation}
where $\delta^\mu{}_\nu$ is the unit tensor. \ The metric is also used to 
raise and lower spacetime indices, i.e.~if we let $V^\mu$ denote a 
contravariant vector, then its associated covariant vector (also known as a 
covector or one-form) $V_\mu$ is obtained as 
\begin{equation}
    V_\mu = g_{\mu \nu} V^\nu \Leftrightarrow V^\mu = g^{\mu \nu} V_\nu 
            \ . 
\end{equation} 

We can now consider three different classes of curves: timelike, null, and 
spacelike. \ A vector is said to be timelike if $g_{\mu \nu} V^\mu V^\nu < 
0$, null if $g_{\mu \nu} V^\mu V^\nu = 0$, and spacelike if $g_{\mu \nu} 
V^\mu V^\nu > 0$. \ We can naturally define timelike, null, and spacelike 
curves in terms of the congruence of tangent vectors that they generate. \ A 
particularly useful timelike curve for fluids is one that is parameterized by 
the so-called proper time, i.e.~$x^\mu(\tau)$ where 
\begin{equation}
    {\rm d} \tau^2 = - {\rm d} s^2 \ . 
\end{equation}
The tangent $u^\mu$ to such a curve has unit magnitude; specifically, 
\begin{equation} 
    u^\mu \equiv \frac{{\rm d} x^\mu}{{\rm d} \tau} \ , \label{uvec} 
\end{equation} 
and thus 
\begin{equation} 
    g_{\mu \nu} u^\mu u^\nu = g_{\mu \nu} \frac{{\rm d} x^\mu}{{\rm d} \tau} 
    \frac{{\rm d} x^\nu}{{\rm d} \tau} = \frac{{\rm d} s^2}{{\rm d} \tau^2} = 
    - 1 \ . 
\end{equation}

Under a coordinate transformation $x^\mu \to \overline{x}^\mu$, 
contravariant vectors transform as 
\begin{equation} 
     \overline{V}^\mu = \frac{\partial \overline{x}^\mu}{\partial x^\nu} 
                        V^\nu \label{ct1} 
\end{equation} 
and covariant vectors as 
\begin{equation}
     \overline{V}_\mu = \frac{\partial x^\nu}{\partial \overline{x}^\mu} 
                        V_\nu \ . \label{ct2} 
\end{equation} 
Tensors with a greater rank (i.e.~a greater number of indices), transform  
similarly by acting linearly on each index using the above two rules. 

When integrating, as we need to when we discuss conservation laws for fluids, 
we must be careful to have an appropriate measure that ensures the coordinate 
invariance of the integration. \ In the context of three-dimensional 
Euclidean space the measure is referred to as the Jacobian. \ For spacetime, 
we use the so-called volume form $\epsilon_{\mu \nu \rho \tau}$. \ It is 
completely antisymmetric, and for four-dimensional spacetime, it has only one 
independent component, which is 
\begin{equation} 
     \epsilon_{0 1 2 3} = \sqrt{- g} \qquad \mbox{and} \qquad  
     \epsilon^{0 1 2 3} = \frac{1}{\sqrt{- g}} \ , 
\end{equation}
where $g$ is the determinant of the metric. \ The minus sign is required 
under the square root because of the metric signature. \ By contrast, for 
three-dimensional Euclidean space (i.e.~when considering the fluid equations 
in the Newtonian limit) we have 
\begin{equation}
     \epsilon_{1 2 3} = \sqrt{g} \qquad \mbox{and} \qquad  
     \epsilon^{1 2 3} = \frac{1}{\sqrt{g}} \ , 
\end{equation}
where $g$ is the determinant of the three-dimensional space metric. \ A 
general identity that is extremely useful is \cite{wald84:_book} 
\begin{equation} 
     \epsilon^{\mu_1 ... \mu_j \mu_{j+1} ... \mu_n} 
     \epsilon_{\mu_1 ... \mu_j \nu_{j+1} ... \nu_n} = \left(- 1\right)^s 
     \left(n - j\right)! j! \delta^{[\mu_{j + 1}}{}_{\nu_{j+1}}^{~.~.~.~} 
     \delta^{\mu_n}{}^{]}_{\nu_n} \label{epsmup} 
\end{equation} 
where $s$ is the number of minus signs in the metric (eg.~$s = 1$ here). 
\ As an application consider three-dimensional, Euclidean space. \ Then 
$s = 0$ and $n = 3$. \ Using lower-case Latin indices we see (for $j = 1$ 
in Eq.~(\ref{epsmup})) 
\begin{equation} 
     \epsilon^{m i j} \epsilon_{m k l} = \delta^i{}_k \delta^j{}_l - 
     \delta^j{}_k \delta^i{}_l \ . 
\end{equation}
The general identity will be used in our discussion of the variational  
principle, while the three-dimensional version is often needed in discussions 
of Newtonian vorticity. 

\subsection{Parallel Transport and the Covariant Derivative}

In order to have a generally covariant prescription for fluids, i.e.~written 
in terms of spacetime tensors, we must have a notion of derivative 
$\nabla_\mu$ that is itself covariant. \ For example, when $\nabla_\mu$ acts 
on a vector $V^\nu$ a rank-two tensor of mixed indices must result: 
\begin{equation}
    \overline{\nabla}_\mu \overline{V}^\nu = \frac{\partial x^\sigma} 
    {\partial \overline{x}^\mu} \frac{\partial \overline{x}^\nu} 
    {\partial x^\rho} \nabla_\sigma V^\rho \ . 
\end{equation} 
The ordinary partial derivative does not work because under a general 
coordinate transformation 
\begin{equation} 
    \frac{\partial \overline{V}^\nu}{\partial \overline{x}^\mu} = 
    \frac{\partial x^\sigma}{\partial \overline{x}^\mu} 
    \frac{\partial \overline{x}^\nu}{\partial x^\rho} 
    \frac{\partial V^\rho}{\partial x^\sigma} + 
    \frac{\partial x^\sigma}{\partial \overline{x}^\mu} 
    \frac{\partial^2 \overline{x}^\nu}{\partial x^\sigma 
    \partial x^\rho} V^\rho \ . 
\end{equation} 
The second term spoils the general covariance, since it vanishes only for the 
restricted set of rectilinear transformations 
\begin{equation} 
    \overline{x}^\mu = a^\mu_\nu x^\nu + b^\mu \ , 
\end{equation} 
where $a^\mu_\nu$ and $b^\mu$ are constants. 

For both physical and mathematical reasons, one expects a covariant derivative 
to be defined in terms of a limit. \ This is, however, a bit problematic. \ In 
three-dimensional Euclidean space limits can be defined uniquely in that 
vectors can be moved around without their lengths and directions changing, for 
instance, via the use of Cartesian coordinates, the 
$\{{\bf i},{\bf j},{\bf k}\}$ set of basis vectors, and the usual dot product. 
\ Given these limits those corresponding to more general curvilinear 
coordinates can be established. \ The same is not true of curved spaces and/or 
spacetimes because they do {\em not} have an a priori notion of parallel 
transport. 

\begin{figure}[t]
\centering
\includegraphics[width=10cm,clip]{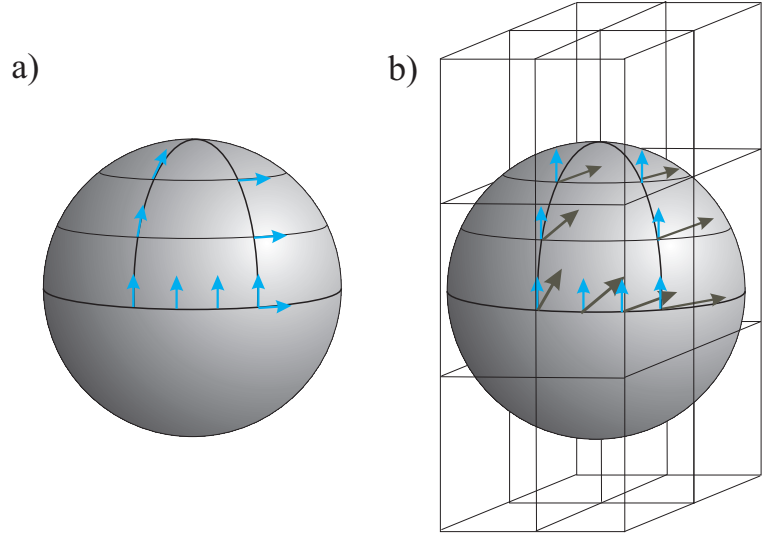} 
\caption{A schematic illustration of two possible versions of parallel 
transport. \ In the first case (a) a vector is transported along great circles 
on the sphere locally maintaining the same angle with the path. \ If the 
contour is closed, the final orientation of the vector will differ from the 
original one. \ In case (b) the sphere is considered to be embedded in a 
three-dimensional Euclidean space, and the vector on the sphere results from 
projection. \ In this case, the vector returns to the original orientation for 
a closed contour.} 
\label{sphere} 
\end{figure}

Consider the classic example of a vector on the surface of a sphere 
(illustrated in Fig.~\ref{sphere}). \ Take that vector and move it along 
some great circle from the equator to the North pole in such a way as to 
always keep the vector pointing along the circle. \ Pick a different great 
circle, and without allowing the vector to rotate, by forcing it to maintain 
the same angle with the locally straight portion of the great circle that it 
happens to be on, move it back to the equator. \ Finally, move the vector in 
a similar way along the equator until it gets back to its starting point. \ 
The vector's spatial orientation will be different from its original 
direction, and the difference is directly related to the particular path 
that the vector followed. 

On the other hand, we could consider that the sphere is embedded in a 
three-dimensional Euclidean space, and that the two-dimensional vector on 
the sphere results from projection of a three-dimensional vector. \ Move 
the projection so that its higher-dimensional counterpart always maintains 
the same orientation with respect to its original direction in the embedding 
space. \ When the projection returns to its starting place it will have 
exactly the same orientation as it started with, see Fig.~\ref{sphere}. \ 
It is thus clear that a derivative operation that depends on comparing a 
vector at one point to that of a nearby point is not unique, because it 
depends on the choice of parallel transport. 

Pauli \cite{pauli81:_rel_book} notes that Levi-Civita 
\cite{levicivita03:_partrans} is the first to have formulated the concept 
of parallel ``displacement,'' with Weyl \cite{weyl52:_rel_book} 
generalizing it to manifolds that do not have a metric. \ The point of 
view expounded in the books of Weyl and Pauli is that parallel transport 
is best defined as a mapping of the ``totality of all vectors'' that 
``originate'' at one point of a manifold with the totality at another 
point. \ (In modern texts, one will find discussions based on fiber 
bundles.) \ Pauli points out that we cannot simply require equality of 
vector components as the mapping. 

Let us examine the parallel transport of the force-free, point particle 
velocity in Euclidean three-dimensional space as a means for motivating 
the form of the mapping. \ As the velocity is constant, we know that the 
curve traced out by the particle will be a straight line. \ In fact, we can 
turn this around and say that the velocity parallel transports itself 
because the path traced out is a geodesic (i.e.~the straightest possible 
curve allowed by Euclidean space). \ In our analysis we will borrow 
liberally from the excellent discussion of Lovelock and Rund 
\cite{lovelock89:_tensor_book}. \ Their text is comprehensive yet 
readable for one who is not well-versed with differential geometry. \ 
Finally, we note that this analysis will be of use later when we obtain 
the Newtonian limit of the general relativistic equations, but in an 
arbitrary coordinate basis. 

We are all well aware that the points on the curve traced out by the 
particle are describable, in Cartesian coordinates, by three functions 
$x^i(t)$ where $t$ is the standard Newtonian time. \ Likewise, we know 
the tangent vector at each point of the curve is given by the velocity 
components $v^i(t) = {\rm d} x^i/{\rm d} t$, and that the force-free condition 
is equivalent to 
\begin{equation}
    a^i(t) = \frac{{\rm d} v^i}{{\rm d} t} = 0 \quad \Rightarrow \quad 
             v^i(t) = {\rm const} \ . 
\end{equation} 
Hence, the velocity components $v^i(0)$ at the point $x^i(0)$ are equal 
to those at any other point along the curve, say $v^i(T)$ at $x^i(T)$, 
and so we could simply take $v^i(0) = v^i(T)$ as the mapping. \ But as 
Pauli warns, we only need to reconsider this example using spherical 
coordinates to see that the velocity components 
$\{\dot{r},\dot{\theta},\dot{\phi}\}$ must necessarily change as they 
undergo parallel transport along a straight-line path (assuming the 
particle does not pass through the origin). \ The question is what should be 
used in place of component equality? \ The answer follows once we find a 
curvilinear coordinate version of ${\rm d}v^i/{\rm d}t = 0$. 

What we need is a new ``time'' derivative $\overline{\rm D}/{\rm d} t$, 
that yields a generally covariant statement 
\begin{equation}
    \frac{\overline{\rm D} \overline{v}^i}{{\rm d} t} = 0 \ , 
\end{equation} 
where the $\overline{v}^i(t) = {\rm d} \overline{x}^i/{\rm d} t$ are  
the velocity components in a curvilinear system of coordinates. \ Let us 
consider an infinitesimal displacement of the velocity from $x^i(t)$ to 
$x^i(t + \delta t)$. \ In Cartesian coordinates we know that 
\begin{equation}
    \delta v^i = \frac{\partial v^i}{\partial x^j} \delta x^j = 0 \ . 
\end{equation} 
Consider now a coordinate transformation to the curvilinear coordinate 
system $\overline{x}^i$, the inverse being $x^i = x^i(\overline{x}^j)$. \ 
Given that 
\begin{equation}
    v^i = \frac{\partial x^i}{\partial \overline{x}^j} \overline{v}^j 
\end{equation}
we can write 
\begin{equation} 
    \frac{{\rm d} v^i}{{\rm d} t} = \left(\frac{\partial x^i}{\partial 
         \overline{x}^j} \frac{\partial \overline{v}^j}{\partial 
         \overline{x}^k} + \frac{\partial^2 x^i}{\partial \overline{x}^k 
         \partial \overline{x}^j} \overline{v}^j\right) \overline{v}^k \ . 
\end{equation}

The metric $\overline{g}_{i j}$ for our curvilinear coordinate system is 
obtainable from 
\begin{equation}
    \overline{g}_{i j} = \frac{\partial x^k}{\partial \overline{x}^i} 
                         \frac{\partial x^l}{\partial \overline{x}^j} 
                         g_{k l} \ , \label{mettrans} 
\end{equation} 
where 
\begin{equation} 
    g_{i j} = \left\{\begin{array}{cc} 
              1 & i = j \\ 
              0 & i \neq j 
              \end{array}\right. \ . \label{flatmet} 
\end{equation}
Differentiating Eq.~(\ref{mettrans}) with respect to $\overline{x}$, and 
permuting indices, we can show that 
\begin{equation}
    \frac{\partial^2 x^h}{\partial \overline{x}^i \partial \overline{x}^j} 
    \frac{\partial x^l}{\partial \overline{x}^k} g_{h l}= \frac{1}{2}  
    \left(\overline{g}_{i k,j} + \overline{g}_{j k,i} - 
    \overline{g}_{i j,k}\right) \equiv \overline{g}_{i l} 
    \overline{\left\{\scriptstyle{l \atop j~k}\right\}} \ , 
\end{equation}
where 
\begin{equation}
     g_{i j , k} \equiv \frac{\partial g_{i j}}{\partial x^k} \ . 
\end{equation}

With some tensor algebra we find 
\begin{equation} 
    \frac{{\rm d} v^i}{{\rm d} t} = \frac{\partial x^i}{\partial  
         \overline{x}^j} \frac{\overline{\rm D} \overline{v}^i}{{\rm d} t} 
         \ , 
\end{equation} 
where 
\begin{equation} 
    \frac{\overline{\rm D} \overline{v}^i}{{\rm d} t} = \overline{v}^j 
         \left(\frac{\partial \overline{v}^i}{\partial \overline{x}^j} + 
         \overline{\left\{\scriptstyle{i \atop k~j}\right\}} 
         \overline{v}^k\right) \ . \label{geodesic} 
\end{equation} 
One can verify that the operator $\overline{\rm D}/{\rm d} t$ is 
covariant with respect to general transformations of curvilinear 
coordinates. 

We now identify our generally covariant derivative (dropping the 
overline) as 
\begin{equation} 
    \nabla_j v^i = \frac{\partial v^i}{\partial x^j} + 
                   \left\{\scriptstyle{i \atop k~j}\right\} v^k \equiv 
                   v^i{}_{; j} \ . \label{covdevcon} 
\end{equation} 
Similarly, the covariant derivative of a covector is 
\begin{equation} 
    \nabla_j v_i = \frac{\partial v_i}{\partial x^j} - 
                   \left\{\scriptstyle{k \atop i~j}\right\} v_k \equiv 
                   v_{i ; j} \ . \label{covdevcov} 
\end{equation} 
One extends the covariant derivative to higher rank tensors by adding to 
the partial derivative each term that results by acting linearly on each 
index with $\left\{\scriptstyle{i \atop k~j}\right\}$ using the two rules 
given above. 

By relying on our understanding of the force-free point particle, we have 
built a notion of parallel transport that is consistent with our 
intuition based on equality of components in Cartesian coordinates. \ We 
can now expand this intuition and see how the vector components in a 
curvilinear coordinate system must change under an infinitesimal, 
parallel displacement from $x^i(t)$ to $x^i(t + \delta t)$. \ Setting 
Eq.~(\ref{geodesic}) to zero, and noting that $v^i \delta t = \delta 
x^i$, implies 
\begin{equation} 
    \delta v^i \equiv \frac{\partial v^i}{\partial x^j} \delta x^j = - 
    \left\{\scriptstyle{i \atop k~j}\right\} v^k \delta x^j \ . 
\end{equation} 
In general relativity we assume that under an infinitesimal parallel 
transport from a spacetime point $x^\mu(\lambda)$ on a given curve to a 
nearby point $x^\mu(\lambda + \delta \lambda)$ on the same curve, the 
components of a vector $V^\mu$ will change in an analogous way, namely 
\begin{equation}
    \delta V^\mu_\parallel \equiv \frac{\partial V^\mu}
    {\partial x^\nu} \delta x^\nu = - \Gamma^\mu_{\rho \nu} V^\rho 
    \delta x^\nu \ , \label{partransgr} 
\end{equation} 
where 
\begin{equation} 
    \delta x^\mu \equiv \frac{{\rm d} x^\nu}{{\rm d} \lambda} \delta 
    \lambda \ . 
\end{equation} 
Weyl \cite{weyl52:_rel_book} refers to the symbol $\Gamma^\mu_{\rho \nu}$ 
as the ``components of the affine relationship,'' but we will use the 
modern terminology and call it the connection. \ In the language of Weyl 
and Pauli, this is the mapping that we were looking for. 

For Euclidean space, we can verify that the metric satisfies 
\begin{equation}
    \nabla_i g_{j k} = 0 
\end{equation} 
for a general, curvilinear coordinate system. \ The metric is thus said 
to be ``compatible'' with the covariant derivative. \ We impose metric 
compatibility in general relativity. \ This results in the so-called 
Christoffel symbol for the connection, defined as 
\begin{equation}
     \Gamma^\rho_{\mu \nu} = \frac{1}{2} g^{\rho \sigma} \left(
                             g_{\mu \sigma , \nu} + g_{\nu \sigma , \mu} 
                             - g_{\mu \nu , \sigma}\right) \ . 
\end{equation} 
The rules for the covariant derivative of a contravariant vector and a 
covector are the same as in Eqs.~(\ref{covdevcon}) and (\ref{covdevcov}), 
except that all indices are for full spacetime. 

\subsection{The Lie Derivative and Spacetime Symmetries}

Another important tool for measuring changes in tensors from point to point 
in spacetime is the Lie derivative. \ It requires a vector field, but no 
connection, and is a more natural definition in the sense that it does not 
even require a metric. \ It yields a tensor of the same type and rank as 
the tensor on which the derivative operated (unlike the covariant 
derivative, which increases the rank by one). \ The Lie derivative is as 
important for Newtonian, non-relativistic fluids as for relativistic ones 
(a fact which needs to be continually emphasized as it has not yet 
permeated the fluid literature for chemists, engineers, and physicists). \ 
For instance, the classic papers on the Chandrasekhar-Friedman-Schutz 
instability \cite{friedman78:_lagran,friedman78:_secul_instab} in rotating 
stars are great illustrations of the use of the Lie derivative in Newtonian 
physics. \ We recommend the book by Schutz \cite{schutz80:_geomet} for a 
complete discussion and derivation of the Lie derivative and its role in 
Newtonian fluid dynamics (see also the recent series of papers by Carter 
and Chamel \cite{Carter03:_newtI,Carter03:_newtII,Carter04:_newtIII}). \ We 
will adapt here the coordinate-based discussion of Schouten 
\cite{schouten89:_tenanal}, as it may be more readily understood by those 
not well-versed in differential geometry. 

In a first course on classical mechanics, when students encounter rotations, 
they are introduced to the idea of active and passive transformations. \ An 
\underline{active} transformation would be to fix the origin and 
axis-orientations of a given coordinate system with respect to some external 
observer, and then move an object from one point to another point of the 
same coordinate system. \ A \underline{passive} transformation would be to 
place an object so that it remains fixed with respect to some external 
observer, and then induce a rotation of the object with respect to a given 
coordinate system, by rotating the coordinate system itself with respect to 
the external observer. \ We will derive the Lie derivative of a vector by 
first performing an active transformation and then following this with a 
passive transformation and finding how the final vector differs from its 
original form. \ In the language of differential geometry, we will first 
``push-forward'' the vector, and then subject it to a ``pull-back''.  

In the active, push-forward sense we imagine that there are two 
spacetime points connected by a smooth curve $x^\mu(\lambda)$. \ Let the 
first point be at $\lambda = 0$, and the second, nearby point at $\lambda 
= \epsilon$, i.e.~$x^\mu(\epsilon)$; that is, 
\begin{equation} 
     x^\mu_\epsilon \equiv x^\mu(\epsilon) \approx x^\mu_0 + 
                           \epsilon~\xi^\mu \ , \label{push} 
\end{equation} 
where $x^\mu_0 \equiv x^\mu(0)$ and 
\begin{equation} 
    \xi^\mu = \left.\frac{{\rm d} x^\mu}{{\rm d} \lambda} 
              \right|_{\lambda = 0} 
\end{equation} 
is the tangent to the curve at $\lambda = 0$. \ In the passive, 
pull-back sense we imagine that the coordinate system itself is 
changed to $\overline{x}{}^\mu =\overline{x}{}^\mu(x^\nu)$, but in the 
very special form 
\begin{equation}
    \overline{x}{}^\mu = x^\mu - \epsilon~\xi^\mu \ . \label{pull} 
\end{equation} 
It is in this last step that the Lie derivative differs from the 
covariant derivative. \ In fact, if we insert Eq.~(\ref{push}) into 
(\ref{pull}) we find the result $\overline{x}{}^\mu_\epsilon = x^\mu_0$. 
\ This is called ``Lie-dragging'' of the coordinate frame, meaning that 
the coordinates at $\lambda = 0$ are carried along so that at $\lambda = 
\epsilon$, and in the new coordinate system, the coordinate labels take 
the same numerical values. 

\begin{figure}[t]
\centering 
\includegraphics[height=7cm,clip]{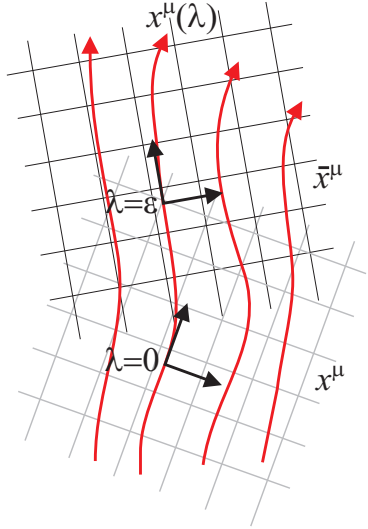} 
\caption{A schematic illustration of the Lie derivative. \ The coordinate 
system is dragged along with the flow, and one can imagine an observer 
``taking derivatives'' as he/she moves with the flow (see the discussion in 
the text).} 
\label{transport} 
\end{figure} 

As an interesting aside it is worth noting that Arnold 
\cite{arnold95:_mathbook}, only a little whimsically, refers to 
this construction as the ``fisherman's derivative''. \ He imagines a fisherman 
sitting in a boat on a river, ``taking derivatives'' as the boat moves with 
the current. \ Playing on this imagery, the covariant derivative is cast with 
the high-test Zebco parallel transport fishing pole, the Lie derivative with 
the Shimano, Lie-dragging ultra-light. \ Let us now see how Lie-dragging reels 
in vectors. 

For some given vector field that takes values $V^\mu(\lambda)$, say, 
along the curve, we write 
\begin{equation}
    V^\mu_0 = V^\mu(0) 
\end{equation}
for the value of $V^\mu$ at $\lambda = 0$ and 
\begin{equation} 
    V^\mu_\epsilon = V^\mu(\epsilon) 
\end{equation} 
for the value at $\lambda = \epsilon$. \ Because the two points $x^\mu_0$ 
and $x^\mu_\epsilon$ are infinitesimally close $\epsilon << 1$, and we 
can thus write 
\begin{equation} 
    V^\mu_\epsilon \approx V^\mu_0 + \epsilon~\xi^\nu 
    \left.\frac{\partial V^\mu}{\partial x^\nu}\right|_{\lambda = 0} \ , 
    \label{nearbyvec} 
\end{equation} 
for the value of $V^\mu$ at the nearby point and in the {\em same} 
coordinate system. \ However, in the new coordinate system at the nearby 
point, we find 
\begin{equation} 
    \overline{V}{}^\mu_\epsilon = \left.\left(\frac{\partial 
    \overline{x}{}^\mu}{\partial x^\nu} V^\nu\right)\right|_{\lambda = 
    \epsilon} \approx V^\mu_\epsilon - \epsilon~V^\nu_0 
    \left.\frac{\partial \xi^\mu}{\partial x^\nu}\right|_{\lambda = 0} \ . 
\end{equation} 
The Lie derivative now is defined to be 
\begin{eqnarray} 
     {\cal L}_\xi V^\mu &=& \lim_{\epsilon \to 0} 
     \frac{\overline{V}{}^\mu_\epsilon - V^\mu}{\epsilon} \cr 
     && \cr 
     &=& \xi^\nu \frac{\partial V^\mu}{\partial x^\nu} - 
     V^\nu \frac{\partial \xi^\mu}{\partial x^\nu} \cr 
     && \cr 
     &=& \xi^\nu \nabla_\nu V^\mu - V^\nu \nabla_\nu \xi^\mu 
     \ , \label{liedev} 
\end{eqnarray} 
where we have dropped the ``$0$'' subscript and the last equality follows 
easily by noting $\Gamma^\rho_{\mu \nu} = \Gamma^\rho_{\nu \mu}$. 

We introduced in Eq.~(\ref{partransgr}) the effect of parallel transport 
on vector components. \ By contrast, the Lie-dragging of a vector causes 
its components to change as 
\begin{equation}
    \delta V^\mu_{\cal L} = {\cal L}_\xi V^\mu~\epsilon \ . 
\end{equation} 
We see that if ${\cal L}_\xi V^\mu = 0$, then the components of the 
vector do not change as the vector is Lie-dragged. \ Suppose now that 
$V^\mu$ represents a vector field and that there exists a corresponding 
congruence of curves with tangent given by $\xi^\mu$. \ If the components 
of the vector field do not change under Lie-dragging we can show that 
this implies a symmetry, meaning that a coordinate system can be found such 
that the vector components will not depend on one of the coordinates. \ 
This is a potentially very powerful statement. 

Let $\xi^\mu$ represent the tangent to the curves drawn out by, say, the 
$\mu = a$ coordinate. \ Then we can write $x^a(\lambda) = \lambda$ which 
means 
\begin{equation} 
    \xi^\mu = \delta^\mu{}_a \ . 
\end{equation} 
If the Lie derivative of $V^\mu$ with respect to $\xi^\nu$ vanishes we find 
\begin{equation}
    \xi^\nu \frac{\partial V^\mu}{\partial x^\nu} = V^\nu \frac{\partial 
    \xi^\mu}{\partial x^\nu} = 0 \ . 
\end{equation} 
Using this in Eq.~(\ref{nearbyvec}) implies $V^\mu_\epsilon = V^\mu_0$, 
that is to say, the vector field $V^\mu(x^\nu)$ does not depend on the 
$x^a$ coordinate. \ Generally speaking, every $\xi^\mu$ that exists that 
causes the Lie derivative of a vector to vanish represents a symmetry. 
 
\subsection{Spacetime Curvature}

The main message of the previous two subsections is that one must have 
an a priori idea of how vectors and higher rank tensors are moved from 
point to point in spacetime. \ Another manifestation of the complexity 
associated with carrying tensors about in spacetime is that the covariant 
derivative does not commute. \ For a vector we find 
\begin{equation}
     \nabla_\rho \nabla_\sigma V^\mu - \nabla_\sigma \nabla_\rho V^\mu = 
     R^\mu{}_{\nu \rho \sigma} V^\nu \ , \label{covcom} 
\end{equation}
where $R^\mu{}_{\nu \rho \sigma}$ is the Riemann tensor. \ It is obtained from 
\begin{equation} 
     R^\mu{}_{\nu \rho \sigma} = \Gamma^\mu_{\nu \sigma , \rho} - 
     \Gamma^\mu_{\nu \rho , \sigma} + \Gamma^\mu_{\tau \rho} 
     \Gamma^\tau_{\nu \sigma} - \Gamma^\mu_{\tau \sigma} 
     \Gamma^\tau_{\nu \rho} \ . 
\end{equation} 
Closely associated are the Ricci tensor $R_{\nu \sigma} = R_{\sigma \nu}$  
and scalar $R$ that are defined by the contractions 
\begin{equation} 
     R_{\nu \sigma} = R^\mu{}_{\nu \mu \sigma} \quad , \quad 
     R = g^{\nu \sigma} R_{\nu \sigma} \ . 
\end{equation} 
We will later also have need for the Einstein tensor, which is given by 
\begin{equation} 
     G_{\mu \nu} = R_{\mu \nu} - \frac{1}{2} R g_{\mu \nu} \ . 
\end{equation} 
It is such that $\nabla_\mu G^\mu{}_\nu$ vanishes identically (this is known 
as the Bianchi Identity). 

A more intuitive understanding of the Riemann tensor is obtained by seeing 
how its presence leads to a path-dependence in the changes that a vector 
experiences as it moves from point to point in spacetime. \ Such a situation 
is known as a ``non-integrability'' condition, because the result depends on 
the whole path and not just the initial and final points. \ That is, it is 
unlike a total derivative which can be integrated and thus depends on only the 
lower and upper limits of the integration. \ Geometrically we say that the 
spacetime is curved, which is why the Riemann tensor is also known as the 
curvature tensor. 

To illustrate the meaning of the curvature tensor, let us suppose that we are 
given a surface that can be parameterized by the two parameters $\lambda$ and 
$\eta$. \ Points that live on this surface will have coordinate labels 
$x^\mu(\lambda,\eta)$. \ We want to consider an infinitesimally small 
``parallelogram'' whose four corners (moving counterclockwise with the first 
corner at the lower left) are given by $x^\mu(\lambda,\eta)$, 
$x^\mu(\lambda,\eta + \delta \eta)$, $x^\mu(\lambda + \delta \lambda,\eta + 
\delta \eta)$, and $x^\mu(\lambda + \delta \lambda,\eta)$. \ Generally 
speaking, any ``movement'' towards the right of the parallelogram is effected 
by varying $\eta$, and that towards the top results by varying $\lambda$. \ 
The plan is to take a vector $V^\mu(\lambda,\eta)$ at the lower-left 
corner $x^\mu(\lambda,\eta)$, parallel transport it along a $\lambda = 
constant$ curve to the lower-right corner at $x^\mu(\lambda,\eta + \delta 
\eta)$ where it will have the components 
$V^\mu(\lambda,\eta + \delta \eta)$, and end up by parallel transporting 
$V^\mu$ at $x^\mu(\lambda,\eta + \delta \eta)$ along an $\eta = constant$
curve to the upper-right corner at 
$x^\mu(\lambda + \delta \lambda,\eta + \delta \eta)$. \ We will call this 
path I and denote the final component values of the vector as 
$V^\mu_{\rm I}$. \ We repeat the same process except that the path will 
go from the lower-left to the upper-left and then on to the upper-right 
corner. \ We will call this path II and denote the final component values 
as $V^\mu_{\rm II}$. 

Recalling Eq.~(\ref{partransgr}) as the definition of parallel transport, 
we first of all have  
\begin{equation} 
    V^\mu(\lambda,\eta + \delta \eta) \approx V^\mu(\lambda,\eta) + 
    \delta_\eta V^\mu_\parallel (\lambda,\eta) = 
    V^\mu(\lambda,\eta) - \Gamma^\mu_{\nu \rho} V^\nu \delta_\eta x^\rho 
\end{equation} 
and 
\begin{equation} 
    V^\mu(\lambda + \delta \lambda,\eta) \approx V^\mu(\lambda,\eta) + 
    \delta_\lambda V^\mu_\parallel (\lambda,\eta) = 
    V^\mu(\lambda,\eta) - \Gamma^\mu_{\nu \rho} V^\nu 
    \delta_\lambda x^\rho \ , 
\end{equation} 
where 
\begin{equation} 
    \delta_\eta x^\mu \approx x^\mu(\lambda,\eta + \delta \eta) - 
                        x^\mu(\lambda,\eta) 
                  \quad , \quad 
    \delta_\lambda x^\mu \approx x^\mu(\lambda + \delta \lambda,\eta) - 
                   x^\mu(\lambda,\eta) \ . 
\end{equation} 
Next, we need 
\begin{eqnarray} 
    V^\mu_{\rm I} &\approx& V^\mu(\lambda,\eta + \delta \eta) + 
    \delta_\lambda V^\mu_\parallel(\lambda,\eta + \delta \eta) 
\end{eqnarray} 
and 
\begin{eqnarray} 
    V^\mu_{\rm II} &\approx& V^\mu(\lambda + \delta \lambda,\eta) + 
    \delta_\eta V^\mu_\parallel(\lambda + \delta 
    \lambda,\eta) \ . 
\end{eqnarray}
Working things out, we find that the difference between the two paths is 
\begin{equation} 
    \Delta V^\mu \equiv V^\mu_{\rm I} - V^\mu_{\rm II} = 
    R^\mu{}_{\nu \rho \sigma} V^\nu \delta_\lambda x^\sigma \delta_\eta 
    x^\rho \ , 
\end{equation} 
which follows because $\delta_\lambda \delta_\eta x^\mu = \delta_\eta 
\delta_\lambda x^\mu$, i.e.~we have closed the parallelogram. 

\newpage

\section{The Stress-Energy-Momentum Tensor and the Einstein Equations} 

Any discussion of relativistic physics must include the 
stress-energy-momentum tensor $T_{\mu \nu}$. \ It is as important for 
general relativity as $G_{\mu \nu}$ in that it enters the 
Einstein equations in as direct a way as possible, i.e. 
\begin{equation}
    G_{\mu \nu} = 8 \pi T_{\mu \nu} \ . 
\end{equation}
Misner, Thorne, and Wheeler \cite{mtw73} refer to $T_{\mu \nu}$ as ``...a 
machine that contains a knowledge of the energy density, momentum 
density, and stress as measured by any and all observers at that event.'' 

Without an a priori, physically-based specification for $T_{\mu \nu}$, 
solutions to the Einstein equations are devoid of physical content, a 
point which has been emphasized, for instance, by Geroch and Horowitz 
(in \cite{hawk1979:_gr}). \ Unfortunately, the following algorithm for 
producing ``solutions'' has been much abused: (i) specify the form 
of the metric, typically by imposing some type of symmetry, or 
symmetries, (ii) work out the components of $G_{\mu \nu}$ based on this 
metric, (iii) define the energy density to be $G_{0 0}$ and the pressure 
to be $G_{1 1}$, say, and thereby ``solve'' those two equations, and (iv) 
based on the ``solutions'' for the energy density and pressure solve the 
remaining Einstein equations. \ The problem is that this algorithm is little 
more than a mathematical game. \ It is only by sheer luck that it will 
generate a physically viable solution for a non-vacuum spacetime. \ As such, 
the strategy is antithetical to the {\em raison d'\^etre} of 
gravitational-wave astrophysics, which is to use gravitational-wave data as 
a probe of all the wonderful microphysics, say, in the cores of neutron 
stars. \ Much effort is currently going into taking given microphysics and 
combining it with the Einstein equations to model gravitational-wave emission 
from realistic neutron stars. \ To achieve this aim, we need an appreciation 
of the stress-energy tensor and how it is obtained from microphysics.  

Those who are familiar with Newtonian fluids will be aware of the roles 
that total internal energy, particle flux, and the stress tensor play in 
the fluid equations. \ In special relativity we learn that in order to 
have spacetime covariant theories (e.g.~well-behaved with respect to the 
Lorentz transformations) energy and momentum must be combined into a 
spacetime vector, whose zeroth component is the energy and the spatial 
components give the momentum. \ The fluid stress must also be 
incorporated into a spacetime object, hence the necessity for 
$T_{\mu \nu}$. \ Because the Einstein tensor's covariant divergence 
vanishes identically, we must have also $\nabla_\mu T^\mu{}_\nu = 0$ 
(which we will later see happens automatically once the fluid field equations 
are satisfied). 

\begin{figure}[h] 
\centering
\includegraphics[height=6cm,clip]{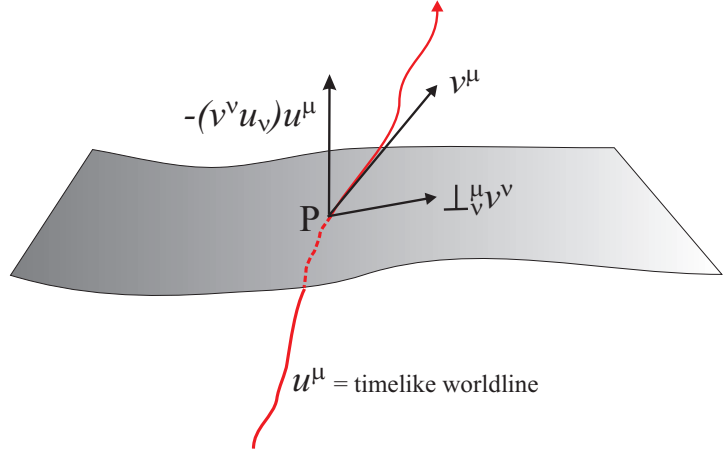} 
\caption{The projections at point P of a vector $V^\mu$ onto the  
worldline defined by $U^\mu$ and into the perpendicular hypersurface 
(obtained from the action of $\perp^\mu_\nu$).} 
\label{projection}
\end{figure}

To understand what the various components of $T_{\mu \nu}$ mean 
physically we will write them in terms of projections into the timelike 
and spacelike directions associated with a given observer. \ In order to 
project a tensor index along the observer's timelike direction we 
contract that index with the observer's unit four-velocity $U^\mu$. \ A 
projection of an index into spacelike directions perpendicular to the 
timelike direction defined by $U^\mu$ (see \cite{smarr78:_kinematic} for the 
idea from a ``3+1'' point of view, or \cite{carter92:_brane} from the 
``brane'' point of view) is realized via the operator $\perp^\mu_\nu$, 
defined as 
\begin{equation}
    \perp^\mu_\nu = \delta^\mu{}_\nu + U^\mu U_\nu 
                    \quad , \quad 
    U^\mu U_\mu = - 1 \ . 
\end{equation}
Any tensor index that has been ``hit'' with the projection operator will 
be perpendicular to the timelike direction associated with $U^\mu$ in the 
sense that $\perp^\mu_\nu U^\nu = 0$. \ Fig.~\ref{projection} is a local 
view of both projections of a vector $V^\mu$ for an observer with unit 
four-velocity $U^\mu$. \ More general tensors are projected by acting 
with $U^\mu$ or $\perp^\mu_\nu$ on each index separately
(i.e.~multi-linearly). 

The energy density $\rho$ as perceived by the observer is (see Eckart 
\cite{eckart40:_rel_diss_fluid} for one of the earliest discussions)
\begin{equation}
     \rho = T_{\mu \nu} U^\mu U^\nu \ , 
\end{equation} 
\begin{equation}
    {\cal P}_\mu = - \perp^\rho_\mu U^\nu T_{\rho \nu} 
\end{equation} 
is the spatial momentum density, and  the spatial stress ${\cal S}_{\mu \nu}$ 
is 
\begin{equation}
    {\cal S}_{\mu \nu} = \perp^\sigma_\mu \perp^\rho_\nu T_{\sigma \rho} 
                         \ . 
\end{equation} 
The manifestly spatial component $S_{i j}$ is understood to be the 
$i^{\rm th}$-component of the force across a unit area that is 
perpendicular to the $j^{\rm th}$-direction. \ With respect to the 
observer, the stress-energy-momentum tensor can be written in full 
generality as the decomposition 
\begin{equation}
     T_{\mu \nu} = \rho U_\mu U_\nu + 2 U_{(\mu} {\cal P}_{\nu)} + 
                   {\cal S}_{\mu \nu} \ , 
\end{equation} 
where $2 U_{(\mu} {\cal P}_{\nu)} \equiv U_\mu {\cal P}_\nu + U_\nu 
{\cal P}_\mu$. \ Because $U^\mu {\cal P}_\mu = 0$, we see that the trace 
$T = T^\mu{}_\mu$ is  
\begin{equation} 
     T = {\cal S} - \rho \ , 
\end{equation} 
where ${\cal S} = {\cal S}^\mu{}_\mu$. \ We should point out that use of 
$\rho$ for the energy density is not universal. \ Many authors prefer to use 
the symbol $\varepsilon$ and reserve $\rho$ for the mass-density. \ We will 
later (in Sec.~\ref{shelve}) use the above decomposition as motivation for 
the simplest perfect fluid model. 

\newpage

\section{Why are fluids useful models?}
\label{sec:fluidsuseful}

The Merriam-Webster online dictionary (\url{http://www.m-w.com/}) defines a 
fluid as ``$\dots$a substance (as a liquid or gas) tending to flow or 
conform to the outline of its container'' when taken as a noun and 
``$\dots$having particles that easily move and change their relative 
position without a separation of the mass and that easily yield to 
pressure: capable of flowing'' when taken as an adjective. \ The best 
model of physics is the Standard Model which is ultimately the 
description of the ``substance'' that will make up our fluids. \ The 
substance of the Standard Model consists of remarkably few classes of 
elementary particles: leptons, quarks, and so-called ``force'' carriers 
(gauge-vector bosons). \ Each elementary particle is quantum 
mechanical, but the Einstein equations require explicit trajectories. \ 
Moreover, cosmology and neutron stars are basically many particle systems 
and, even forgetting about quantum mechanics, it is not practical to 
track each and every ``particle'' that makes them up. \ Regardless of 
whether these are elementary (leptons, quarks, etc) or collections of 
elementary particles (eg.~stars in galaxies and galaxies in cosmology). \ 
The fluid model is such that the inherent quantum mechanical behaviour, 
and the existence of many particles are averaged over in such a way that 
it can be implemented consistently in the Einstein equations. 

\begin{figure}[h]
\centering
\includegraphics[height=6cm,clip]{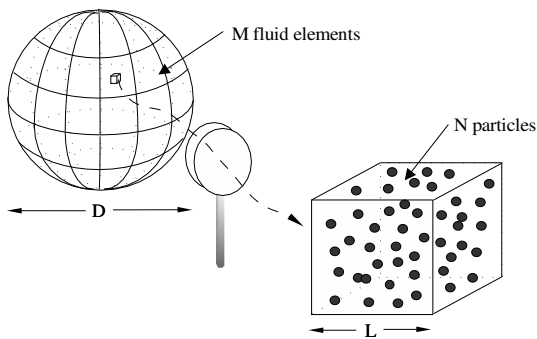} 
\caption{An object with a characteristic size $D$ is modelled as a fluid 
that contains $M$ fluid elements. \ From inside the object we magnify a 
generic fluid element of characteristic size $L$. \ In order for the 
fluid model to work we require $M >> N >> 1$ and $D >> L$.} 
\label{fluidparticle} 
\end{figure}

Central to the model is the notion of ``fluid particle,'' also known as a 
fluid element. \ It is an imaginary, local ``box'' that is infinitesimal 
with respect to the system {\em en masse} and yet large enough to contain 
a large number of particles (eg.~an Avogadro's number of particles). \ 
This is illustrated in Fig.~\ref{fluidparticle}. \ In order for the fluid 
model to work we require $M >> N >> 1$ and $D >> L$. \ Strictly speaking, 
our model has $L$ infinitesimal, $M \to \infty$, but with the total 
number of particles remaining finite. \ The explicit trajectories that 
enter the Einstein equations are those of the fluid elements, {\em not} 
the much smaller (generally fundamental) particles that are ``confined,'' 
on average, to the elements. \ Hence, when we speak later of the fluid 
velocity, we mean the velocity of fluid elements. 

In this sense, the use of the phrase ``fluid particle'' is very apt. \ 
For instance, each fluid element will trace out a timelike trajectory in 
spacetime. \ This is illustrated in Fig.~\ref{pullback} for a number of 
fluid elements. \ An object like a neutron star is a collection of 
worldlines that fill out continuously a portion of spacetime. \ In fact, 
we will see later that the relativistic Euler equation is little more 
than an ``integrability'' condition that guarantees that this filling (or 
fibration) of spacetime can be performed. \ The dual picture to this is 
to consider the family of three-dimensional hypersurfaces that are 
pierced by the worldlines at given instants of time, as illustrated in 
Fig.~\ref{pullback}. \ The integrability condition in this case will 
guarantee that the family of hypersurfaces continuously fill out a 
portion of spacetime. \ In this view, a fluid is a so-called three-brane 
(see \cite{carter92:_brane} for a general discussion of branes). \ In 
fact the method used in Sec.~\ref{sec:pullback} to derive the relativistic 
fluid equations is based on thinking of a fluid as living in a 
three-dimensional ``matter'' space (i.e.~the left-hand-side of 
Fig.~\ref{pullback}). 

Once one understands how to build a fluid model using the matter space, 
it is straight-forward to extend the technique to single fluids with 
several constituents, as in Sec.~\ref{pbonemc}, and multiple fluid 
systems, as in Sec.~\ref{2fluids}. \ An example of the former would be 
a fluid with one species of particles at a non-zero temperature, 
i.e.~non-zero entropy, that does not allow for heat conduction relative 
to the particles. \ (Of course, entropy does flow through spacetime.) \ 
The latter example can be obtained by relaxing the constraint of no heat 
conduction. \ In this case the particles and the entropy are both 
considered to be fluids that are dynamically independent, meaning that 
the entropy will have a four-velocity that is generally different from 
that of the particles. \ There is thus an associated collection of fluid 
elements for the particles and another for the entropy. \ At each point 
of spacetime that the system occupies there will be two fluid elements, 
in other words, there are two matter spaces (cf.~Sec.~\ref{2fluids}). \ 
Perhaps the most important consequence of this is that there can be a 
relative flow of the entropy with respect to the particles (which in general 
leads to the so-called entrainment effect). \ The canonical example of such 
a two fluid model is superfluid ${\rm He}^4$ \cite{putterman74:_sfhydro}. 

\newpage

\section{A Primer on Thermodynamics and Equations of State} \label{thermo} 
\label{sec:thermo} 
 
Fluids consist of many fluid elements, and each fluid element consists of 
many particles. \ The state of matter in a given fluid element is 
determined thermodynamically \cite{reichl98:_book}, meaning that only a 
few parameters are tracked as the fluid element evolves. \ Generally, not 
all the thermodynamic variables are independent, being connected through 
the so-called equation of state. \ The number of independent variables 
can also be reduced if the system has an overall additivity property. \ As 
this is a very instructive example, we will now illustrate such additivity 
in detail. 

\subsection{Fundamental, or Euler, Relation} 

Consider the standard form of the combined first and second laws for a 
simple, single-species system:  
\begin{equation} 
    {\rm d} E = T {\rm d} S - p {\rm d} V + \mu {\rm d} N \ . 
\end{equation} 
This follows because there is an equation of state, meaning that $E = 
E(S,V,N)$ where 
\begin{equation} 
    T = \left.\frac{\partial E}{\partial S}\right|_{V,N} \quad , \quad 
    p = - \left.\frac{\partial E}{\partial V}\right|_{S,N} \quad , \quad 
    \mu = \left.\frac{\partial E}{\partial N}\right|_{S,V} \ . 
\end{equation} 
The total energy $E$, entropy $S$, volume $V$, and particle number $N$ 
are said to be extensive if when $S$, $V$, and $N$ are doubled, say, then 
$E$ will also double. \ Conversely, the temperature $T$, pressure $p$, 
and chemical potential $\mu$ are called intensive if they do not change 
their values when $V$, $N$, and $S$ are doubled. \ This is the additivity 
property and we will now show that it implies an Euler relation (also 
known as the ``fundamental relation'' \cite{reichl98:_book}) among the 
thermodynamic variables. 
 
Let a tilde represent the change in thermodynamic variables when $S$, $V$, 
and $N$ are all increased by the same amount $\lambda$, i.e. 
\begin{equation}
    \tilde{S} = \lambda S \ , \ \tilde{V} = \lambda V \ , \ \tilde{N} = 
                \lambda N \ . 
\end{equation} 
Taking $E$ to be extensive then means 
\begin{equation} 
    \tilde{E}(\tilde{S},\tilde{V},\tilde{N}) = \lambda E(S,V,N) \ . 
\end{equation} 
Of course we have for the intensive variables 
\begin{equation} 
     \tilde{T} = T \ , \ \tilde{p} = p \ , \ \tilde{\mu} = \mu \ . 
\end{equation} 
Now, 
\begin{eqnarray} 
    {\rm d} \tilde{E} &=& \lambda {\rm d} E + E {\rm d} \lambda \cr 
                 &=& \tilde{T} {\rm d} \tilde{S} - \tilde{p} {\rm d} 
                     \tilde{V} + \tilde{\mu} {\rm d} \tilde{N} \cr 
                 &=& \lambda \left(T {\rm d} S - p {\rm d} V + \mu {\rm d} 
                     N\right) + \left(T S - p V + \mu N\right) {\rm d} 
                     \lambda \ , 
\end{eqnarray} 
and therefore we find the Euler relation 
\begin{equation} 
    E = T S - p V + \mu N \ . 
\end{equation} 
If we let $\rho = E / V$ denote the total energy density, $s = S / V$ 
the total entropy density, and $n = N / V$ the total particle number 
density, then 
\begin{equation} 
    p + \rho = T s + \mu n \ . \label{funrel} 
\end{equation} 

The nicest feature of an extensive system is that the number of 
parameters required for a complete specification of the thermodynamic 
state can be reduced by one, and in such a way that only intensive 
thermodynamic variables remain. \ To see this, let $\lambda = 1/V$, in 
which case 
\begin{equation} 
    \tilde{S} = s \ , \ \tilde{V} = 1 \ , \ \tilde{N} = n \ . 
\end{equation} 
The re-scaled energy becomes just the total energy density, 
i.e.~$\tilde{E} = E / V = \rho$, and moreover $\rho = \rho(s,n)$ since 
\begin{equation} 
     \rho = \tilde{E}(\tilde{S},\tilde{V},\tilde{N}) =  
            \tilde{E}(S/V,1,N/V) = \tilde{E}(s,n) \ . 
\end{equation} 
The first law thus becomes 
\begin{equation}
     {\rm d} \tilde{E} = \tilde{T} {\rm d} \tilde{S} - \tilde{p} {\rm d} 
                         \tilde{V} + \tilde{\mu} {\rm d} \tilde{N} = T {\rm d} 
                         s + \mu {\rm d} n \ , 
\end{equation} 
or 
\begin{equation} 
     {\rm d} \rho = T {\rm d} s + \mu {\rm d} n \ . \label{1stlaw} 
\end{equation} 
This implies 
\begin{equation} 
    T = \left.\frac{\partial \rho}{\partial s}\right|_n \ , \ 
    \mu = \left.\frac{\partial \rho}{\partial n}\right|_s \ . 
\end{equation} 
The Euler relation Eq.~(\ref{funrel}) then yields the pressure as 
\begin{equation} 
     p = - \rho  + s \left.\frac{\partial \rho}{\partial s}\right|_n 
         + n \left.\frac{\partial \rho}{\partial n}\right|_s \ . 
         \label{eulerrel} 
\end{equation} 
 
We can think of a given relation  $\rho(s,n)$ as the equation of state, to 
be determined in the flat, tangent space at each point of the manifold, or, 
physically, a small enough region across which the changes in the 
gravitational field are negligible, but also large enough to contain 
a large number of particles. \ For example, for a neutron star 
Glendenning \cite{glendenning97:_compact_stars} has reasoned that the 
relative change in the metric over the size of a nucleon with respect to 
the change over the entire star is about $10^{- 19}$, and thus one must 
consider many internucleon spacings before a substantial change in the 
metric occurs. \ In other words, it is sufficient to determine the properties 
of matter in special relativity, neglecting effects due to spacetime 
curvature. \ The equation of state is the major link between the 
microphysics that governs the local fluid behaviour and global 
quantities (such as the mass and radius of a star). 

In what follows we will use a thermodynamic formulation that satisfies the 
fundamental scaling relation, meaning that the local thermodynamic state 
(modulo entrainment) is a function of the variables $N/V$, $S/V$, etcetera. 
\ This is in contrast to the fluid formulation of ``MTW'' \cite{mtw73}. \ 
In their approach one fixes from the outset the total number of particles 
$N$, meaning that one simply sets ${\rm d} N = 0$ in the first law of 
thermodynamics. \ Thus without imposing any scaling relation, one can write 
\begin{equation} 
    {\rm d} \rho = {\rm d} \left(E/V\right) = T {\rm d} s + \frac{1}{n} 
                   \left(p + \rho - T \s\right) {\rm d} n \ . 
\end{equation} 
This is consistent with our starting point for fluids, because we assume 
that the extensive variables associated with a fluid element do not 
change as the fluid element moves through spacetime. \ However, we feel 
that the use of scaling is necessary in that the fully conservative, or 
non-dissipative, fluid formalism presented below can be adapted to 
non-conservative, or dissipative, situations where ${\rm d} N = 0$ cannot be 
imposed. 

\subsection{From Microscopic Models to Fluid Equation of State} 

Let us now briefly discuss how an equation of state is constructed. \ For 
simplicity, we focus on a one-parameter system, with that parameter being 
the particle number density. \ The equation of state will then be of the form 
$\rho = \rho(n)$. \ In many-body physics (such as studied in condensed 
matter, nuclear, and particle physics) one can in principle construct the 
quantum mechanical particle number density $n_{{\rm QM}}$, 
stress-energy-momentum tensor $T^{{\rm QM}}_{\mu \nu}$, and associated 
conserved particle number density current $n^\mu_{{\rm QM}}$ (starting with 
some fundamental Lagrangian, say; 
cf.~\cite{Walecka:1995mi,glendenning97:_compact_stars,Weber:1999qn}). \ But 
unlike in quantum field theory in curved spacetime \cite{Birrell:1982ix}, one 
assumes that the matter exists in an infinite Minkowski spacetime (cf.~the 
discussion following Eq.~(\ref{eulerrel})). \ If the reader likes, the 
application of $T^{{\rm QM}}_{\mu \nu}$ at a spacetime point means that 
$T^{{\rm QM}}_{\mu \nu}$ has been determined with respect to a flat tangent 
space at that point.

Once $T^{{\rm QM}}_{\mu \nu}$ is obtained, and after (quantum mechanical and 
statistical) expectation values with respect to the system's (quantum and 
statistical) states are taken, one defines the energy density as 
\begin{equation} 
    \rho = u^\mu u^\nu \langle T^{{\rm QM}}_{\mu \nu} \rangle \ , 
\end{equation} 
where 
\begin{equation} 
    u^\mu \equiv \frac{1}{n} \langle n^\mu_{{\rm QM}} \rangle 
          \quad , \quad 
    n = \langle n_{{\rm QM}} \rangle \ . 
\end{equation} 
At sufficiently small temperatures, $\rho$ will just be a function of the 
number density of particles $n$ at the spacetime point in question, 
i.e.~$\rho = \rho(n)$. \ Similarly, the pressure is obtained as 
\begin{equation} 
     p = \frac{1}{3} \left(\langle T^{{\rm QM} \mu}{}_\mu \rangle + 
         \rho\right) 
\end{equation} 
and it will also be a function of $n$. 

One must be very careful to distinguish $T^{{\rm QM}}_{\mu \nu}$ from 
$T_{\mu \nu}$. \ The former describes the states of elementary 
particles with respect to a fluid element, whereas the latter describes 
the states of fluid elements with respect to the system. \ Comer and 
Joynt \cite{comer03:_rel_ent} have shown how this line of reasoning 
applies to the two-fluid case. 

\newpage

\section{An Overview of the Perfect Fluid} 
\label{sec:perfect_fluid} 

There are many different ways of constructing general relativistic fluid 
equations. \ Our purpose here is not to review all possible methods, but 
rather to focus on a couple: (i) an ``off-the-shelve'' consistency analysis 
for the simplest fluid a la Eckart \cite{eckart40:_rel_diss_fluid}, to 
establish some key ideas, and then (ii) a more powerful method based on 
an action principle that varies fluid element world lines. \ The ideas behind 
this variational approach can be traced back to Taub  
\cite{taub54:_gr_variat_princ} (see also \cite{schutz_var}). \ Our description 
of the method relies heavily on the work of Brandon Carter, his students, and 
collaborators \cite{carter89:_covar_theor_conduc,comer93:_hamil_multi_con,%
comer94:_hamil_sf,carter95:_kalb_ramond,carter98:_relat_supercond_superfl,%
langlois98:_differ_rotat_superfl_ns,prix00:_these,prix04:_multi_fluid}. \ We 
prefer this approach as it utilizes as much as possible the tools of the trade 
of relativistic fields, i.e.~no special tricks or devices will be required 
(unlike even in the case of our ``off-the-shelve'' approach). \ One's footing 
is then always made sure by well-grounded, action-based derivations. \ As 
Carter has always made clear: when there are multiple fluids, of both the 
charged and uncharged variety, it is essential to distinguish the fluid 
momenta from the velocities. \ In particular, in order to make the geometrical 
and physical content of the equations transparent. \ A well-posed action is, 
of course, perfect for systematically constructing the momenta. 

\subsection{Rates-of-Change and Eulerian Versus Lagrangian Observers} 

The key geometric difference between generally covariant Newtonian fluids and 
their general relativistic counterparts is that the former have an a priori 
notion of time \cite{Carter03:_newtI,Carter03:_newtII,Carter04:_newtIII}. \ 
Newtonian fluids also have an a priori notion of space (which can be seen 
explicitly in the Newtonian covariant derivative introduced earlier; cf.~the 
discussion in \cite{Carter03:_newtI}). \ Such a structure has clear advantages 
for evolution problems, where one needs to be unambiguous about the 
rate-of-change of the system. \ But once a problem requires, say, 
electromagnetism, then the a priori Newtonian time is at odds with the full 
spacetime covariance of electromagnetic fields. \ Fortunately, for spacetime 
covariant theories there is the so-called ``3+1'' formalism (see, for 
instance, \cite{smarr78:_kinematic}) that allows one to define 
``rates-of-change'' in an unambiguous manner, by introducing a family 
of spacelike hypersurfaces (the ``3'') given as the level surfaces of 
a spacetime scalar (the ``1''). 

Something that Newtonian and relativistic fluids have in common is that there 
are preferred frames for measuring changes---those that are attached to the 
fluid elements. \ In the parlance of hydrodynamics, one refers to Lagrangian 
and Eulerian frames, or observers. \ A Newtonian Eulerian observer is one who 
sits at a fixed point in space, and watches fluid elements pass by, all the 
while taking measurements of their densities, velocities, etc.~at the given 
location. \ In contrast, a Lagrangian observer rides along with a particular 
fluid element and records changes of that element as it moves through space 
and time. \ A relativistic Lagrangian observer is the same, but the 
relativistic Eulerian observer is more complicated to define. \ Smarr and York 
\cite{smarr78:_kinematic} define such an observer as one who would follow
along a worldline that remains everywhere orthogonal to the family of
spacelike hypersurfaces.
 
The existence of a preferred frame for a one fluid system can be used to great 
advantage. \ In the very next sub-section we will use an ``off-the-shelve'' 
analysis that exploits the fact of one preferred frame to derive the standard 
perfect fluid equations. \ Later, we will use Eulerian and Lagrangian 
variations to build an action principle for the single and multiple fluid 
systems. \ These same variations can also be used as the foundation for a 
linearized perturbation analysis of neutron stars \cite{kkbs}. \ As we will 
see, the use of Lagrangian variations is absolutely essential for establishing 
instabilities in rotating fluids 
\cite{friedman78:_lagran,friedman78:_secul_instab}. \ Finally, we 
point out that multiple fluid systems can have as many notions of 
Lagrangian observer as there are fluids in the system. 

\subsection{The Single, Perfect Fluid Problem: ``Off-the-shelve'' 
Consistency Analysis} \label{shelve} 

We earlier took the components of a general stress-energy-momentum tensor 
and projected them onto the axes of a coordinate system carried by an observer 
moving with four-velocity $U^\mu$. \ As mentioned above, the simplest fluid is 
one for which there is only one four-velocity $u^\mu$. \ Hence, there is a 
preferred frame defined by $u^\mu$ and if we want the observer to sit in this 
frame we can simply take $U^\mu = u^\mu$. \ With respect to the fluid there 
will be no momentum flux, i.e.~${\cal P}_\mu = 0$. \ Since we use a fully 
spacetime covariant formulation, i.e.~there are only spacetime indices, 
the resulting stress-energy-momentum tensor will transform properly under 
general coordinate transformations, and hence can be used for any 
observer. 

The spatial stress is a two-index, symmetric tensor, and the only objects 
that can be used to carry the indices are the four-velocity $u^\mu$ and 
the metric $g_{\mu \nu}$. \ Furthermore, because the spatial stress must 
also be symmetric, the only possibility is a linear combination of 
$g_{\mu \nu}$ and $u^\mu u^\nu$. \ Given that $u^\mu {\cal S}_{\mu \nu} = 
0$, we find
\begin{equation} 
     {\cal S}_{\mu \nu} = \frac{1}{3} {\cal S} (g_{\mu \nu} + 
                          u_\mu u_\nu) \ . 
\end{equation} 
If we identify the pressure as $p = {\cal S} / 3$ 
\cite{eckart40:_rel_diss_fluid}, then 
\begin{equation}
     T_{\mu \nu} = \left(\rho + p\right) u_\mu u_\nu + p g_{\mu \nu} \ , 
\end{equation} 
which is the well-established result for a perfect fluid. 

Given a relation $p = p(\rho)$, there are four independent fluid variables. \ 
Because of this the equations of motion are often understood to be 
$\nabla_\mu T^\mu{}_\nu = 0$, which follows immediately from the Einstein 
equations and the fact that $\nabla_\mu G^\mu{}_\nu = 0$. \ To simplify 
matters, we take as equation of state a relation of the form $\rho = \rho(n)$ 
where $n$ is the particle number density. \ The chemical potential is then 
given by 
\begin{equation} 
     {\rm d} \rho = \frac{\partial \rho}{\partial n} {\rm d} n \equiv \mu 
                    {\rm d} n \ , 
\end{equation} 
and we see from the Euler relation Eq.~(\ref{funrel}) that 
\begin{equation} 
     \mu n = p + \rho \ . \label{funrel1} 
\end{equation} 

Let us now get rid of the free index of $\nabla_\mu T^\mu{}_\nu = 0$ in 
two ways: first, by contracting it with $u^\nu$ and second, by 
projecting it with $\perp^\nu_\rho$ (letting $U^\mu = u^\mu$). \ 
Recalling the fact that $u^\mu u_\mu = - 1$ we have the identity 
\begin{equation} 
    \nabla_\mu \left(u^\nu u_\nu\right) = 0 \quad \Rightarrow \quad 
    u_\nu \nabla_\mu u^\nu = 0 \ . 
\end{equation} 
Contracting with $u^\nu$ and using this identity gives 
\begin{equation} 
    u^\mu \nabla_\mu \rho + (\rho + p) \nabla_\mu u^\mu = 0 \ . 
\end{equation} 
The definition of the chemical potential $\mu$ and the Euler relation allow 
us to rewrite this as
\begin{equation}
    \mu u^\mu \nabla_\mu n + \mu n \nabla_\mu u^\mu \quad \Rightarrow 
    \quad \nabla_\mu n^\mu = 0 \ , 
\end{equation} 
where $n^\mu \equiv n u^\mu$. \ Projection of the free index using 
$\perp^\rho_\mu$ leads to 
\begin{equation} 
    D_\mu p = - (\rho + p) a_\mu \ , 
\end{equation} 
where $D_\mu \equiv \perp^\rho_\mu \nabla_\rho$ is a purely spatial 
derivative and $a_\mu \equiv u^\nu \nabla_\nu u_\mu$ is the acceleration. 
\ This is reminiscent of the familiar Euler equation for Newtonian 
fluids. 

However, we should not be too quick to think that this is the only way to 
understand $\nabla_\mu T^\mu{}_\nu = 0$. \ There is an alternative 
form that makes the perfect fluid have much in common with vacuum 
electromagnetism. \ If we define 
\begin{equation} 
     \mu_\mu = \mu u_\mu \ , 
\end{equation} 
and note that $u_\mu {\rm d} u^\mu = 0$ (because $u^\mu u_\mu = - 1$), then 
\begin{equation} 
     {\rm d} \rho = - \mu_\mu {\rm d} n^\mu \ . \label{frstlw1} 
\end{equation} 
The stress-energy-momentum tensor can now be written in the form 
\begin{equation} 
    T^\mu{}_\nu = p \delta^\mu{}_\nu + n^\mu \mu_\nu \ . 
\end{equation} 
We have here our first encounter with the momentum $\mu_\mu$ that is  
conjugate to the particle number density current $n^\mu$. \ Its 
importance will become clearer as this review develops, particularly 
when we discuss the two-fluid case. \ If we now project onto the free 
index of $\nabla_\mu T^\mu{}_\nu = 0$ using $\perp^\mu_\nu$, as before, 
we find 
\begin{equation} 
     f_\nu + \left(\nabla_\mu n^\mu\right) \mu_\nu = 0 \ , \label{diveq} 
\end{equation} 
where the force density $f_\nu$ is 
\begin{equation} 
     f_\nu = 2 n^\mu \omega_{\mu \nu} \ , \label{fdens} 
\end{equation} 
and the vorticity $\omega_{\mu \nu}$ is defined as 
\begin{equation} 
     \omega_{\mu \nu} \equiv \nabla_{[ \mu} \mu_{\nu ]} = \frac{1}{2} 
     \left(\nabla_\mu \mu_\nu - \nabla_\nu \mu_\mu\right) \ . 
\end{equation} 
Contracting Eq.~(\ref{diveq}) with $n^\nu$ implies (since $\omega_{\mu \nu} 
= - \omega_{\nu \mu}$) that 
\begin{equation} 
     \nabla_\mu n^\mu = 0 
\label{cons1} 
\end{equation} 
and as a consequence 
\begin{equation} 
     f_\nu = 2 n^\mu \omega_{\mu \nu} = 0 \ . \label{euler1f} 
\end{equation} 
The vorticity two-form $\omega_{\mu \nu}$ has emerged quite naturally as 
an essential ingredient of the fluid dynamics 
\cite{lichnerowica67:_book,carter89:_covar_theor_conduc,%
bekenstein87:_helicity,katz84:_vorticity}. \ Those who are familiar with 
Newtonian fluids should be inspired by this, as the vorticity is often used 
to establish theorems on fluid behaviour (for instance the Kelvin-Helmholtz 
Theorem \cite{landau59:_fluid_mech}) and is at the heart of turbulence 
modelling \cite{pullin98:_vortex_turb}. 

\begin{figure}[h]
\centering
\includegraphics[width=8cm,clip]{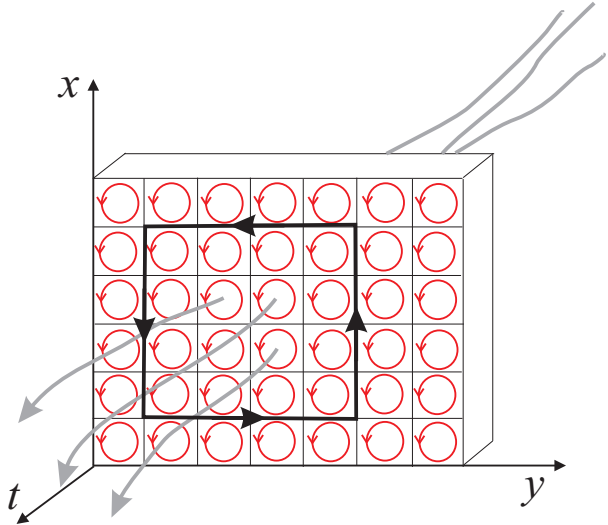}
\caption{A local, geometrical view of the Euler equation as an 
integrability condition of the vorticity for a single-constituent perfect 
fluid.}
\label{vorticity}
\end{figure}

To demonstrate the role of $\omega_{\mu \nu}$ as the vorticity, consider a 
small region of the fluid where the time direction $t^\mu$, in local 
Cartesian coordinates, is adjusted to be the same as that of the fluid 
four-velocity so that $u^\mu = t^\mu = (1,0,0,0)$. \ Eq.~(\ref{euler1f}) and 
the antisymmetry then imply that $\omega_{\mu \nu}$ can only have purely 
spatial components. \ Because the rank of $\omega_{\mu \nu}$ is two, there 
are two ``nulling'' vectors, meaning their contraction with either index of 
$\omega_{\mu \nu}$ yields zero (a condition which is true also of vacuum 
electromagnetism). \ We have arranged already that $t^\mu$ be one such 
vector. \ By a suitable rotation of the coordinate system the other one can 
be taken as $z^\mu = (0,0,0,1)$, thus implying that the only non-zero
component of $\omega_{\mu \nu}$ is $\omega_{x y}$. \ ``MTW'' \cite{mtw73} 
points out that such a two-form can be pictured geometrically as a collection 
of oriented worldtubes, whose walls lie in the $x = constant$ and $y = 
constant$ planes. \ Any contraction of a vector with a two-form that does not 
yield zero implies that the vector pierces the walls of the worldtubes. \ But 
when the contraction {\em is} zero, as in Eq.~(\ref{euler1f}), the vector 
{\em does not} pierce the walls. \ This is illustrated in 
Fig.~\ref{vorticity}, where the red circles indicate the orientation of 
each world-tube. \ The individual fluid element four-velocities lie in the 
centers of the world-tubes. \ Finally, consider the dashed, closed contour 
in Fig.~\ref{vorticity}. \ If that contour is attached to fluid-element 
worldlines, then the number of worldtubes contained within the contour will 
not change because the worldlines cannot pierce the walls of the worldtubes. \ 
This is essentially the Kelvin-Helmholtz Theorem on conservation of
vorticity. \ From this we learn that the Euler equation is an integrability 
condition which ensures that the vorticity two-surfaces mesh together to fill 
out spacetime. 

As we have just seen, the form Eq.~(\ref{euler1f}) of the equations of motion 
can be used to discuss the conservation of vorticity in an elegant way. \ 
It can also be used as the basis for a derivation of other known theorems 
in fluid mechanics. \ To illustrate this, let us derive a generalized form of 
Bernoulli's theorem. \ Let us assume that the flow is invariant with respect
to transport by some vector field $k^\rho$. \ That is, we have 
\begin{equation} 
     \mathcal{L}_k \mu_\rho = 0 \ , \qquad \longrightarrow 
     \qquad (\nabla_\rho \mu_\sigma - \nabla_\sigma \mu_\rho) k^\sigma 
     = \nabla_\rho ( k^\sigma \mu_\sigma) \ . 
\end{equation}
Here one may consider two particular situations. \ If $k^\sigma$ is taken to 
be the four-velocity, then the scalar $k^\sigma \mu_\sigma$ represents the 
``energy per particle''. \ If instead $k^\sigma$ represents an axial generator 
of rotation, then the scalar will correspond to an angular momentum. \ For the 
purposes of the present discussion we can leave $k^\sigma$ unspecified, but it 
is still useful to keep these possibilities in mind. \ Now contract the 
equation of motion Eq.~(\ref{euler1f}) with $k^\sigma$ and assume that the 
conservation law Eq.~(\ref{cons1}) holds. \ Then it is easy to show that we 
have
\begin{equation}
    n^\rho\nabla_\rho \left(\mu_\sigma k^\sigma\right) = \nabla_\rho 
    \left(n^\rho \mu_\sigma k^\sigma\right) = 0 \ . 
\end{equation} 
In other words, we have shown that $n^\rho \mu_\sigma k^\sigma$ is a conserved 
quantity. 


Given that we have just inferred the equations of motion from the 
identity that $\nabla_\mu T^\mu{}_\nu = 0$, we now emphatically state 
that while the equations are correct the reasoning is severely limited. \ In 
fact, from a field theory point of view it is completely wrong! \ The proper 
way to think about the identity is that the equations of motion are satisfied 
first, which then guarantees that $\nabla_\mu T^\mu{}_\nu = 0$. \ There is no 
clearer way to understand this than to study the multi-fluid case: Then the 
vanishing of the covariant divergence represents only four equations, whereas 
the multi-fluid problem clearly requires more information (as there are more 
velocities that must be determined). \ We have reached the end of the road as 
far as the ``off-the-shelf'' strategy is concerned, and now move on to an 
action-based derivation of the fluid equations of motion. 

\newpage

\section{Setting the Context: The Point Particle} \label{ppart} 
\label{sec:point}

The simplest physics problem, i.e.~the point particle, has always served 
as a guide to deep principles that are used in much harder problems. \ We 
have used it already to motivate parallel transport as the foundation for the 
covariant derivative. \ Let us call upon the point particle again to set the 
context for the action-based derivation of the fluid field equations. \ We 
will simplify the discussion by considering only motion in one dimension. \ 
We assure the reader that we have good reasons, and ask for patience while 
we remind him/her of what may be very basic facts. 

Early on we learn that an action appropriate for the point particle is 
\begin{equation} 
    I = \int^{t_f}_{t_i} {\rm d} t~T = \int^{t_f}_{t_i} {\rm d} t 
    \left(\frac{1}{2} m \dot{x}^2\right) \ , 
\end{equation} 
where $m$ is the mass and $T$ the kinetic energy. \ A first-order variation 
of the action with respect to $x(t)$ yields 
\begin{equation} 
   \delta I = - \int^{t_f}_{t_i} {\rm d} t \left(m \ddot{x}\right) 
              \delta x + \left.\left(m \dot{x} \delta x\right) 
              \right|^{t_f}_{t_i} \ . 
\end{equation} 
If this is all the physics to be incorporated, i.e.~if there are no forces 
acting on the particle, then we impose d'Alembert's principle of 
least action \cite{lanczos49:_var_mechs}, which states that those 
trajectories $x(t)$ that make the action stationary, i.e.~$\delta I = 0$, 
are those that yield the true motion. \ We see from the above that functions 
$x(t)$ that satisfy the boundary conditions 
\begin{equation} 
    \delta x(t_i) = 0 = \delta x(t_f) \ , 
\end{equation} 
and the equation of motion 
\begin{equation} 
    m \ddot{x} = 0 \ , 
\end{equation} 
will indeed make $\delta I = 0$. \ This same reasoning applies in the 
substantially more difficult fluid actions that will be considered later. 

But, of course, forces need to be included. \ First on the list are the 
so-called conservative forces, describable by a potential $V(x)$, which are 
placed into the action according to: 
\begin{equation} 
    I = \int^{t_f}_{t_i} {\rm d} t L(x,\dot{x}) = \int^{t_f}_{t_i} {\rm d} t 
        \left(\frac{1}{2} m \dot{x}^2 - V(x)\right) \ , 
\end{equation} 
where $L = T - V$ is the so-called Lagrangian. \ The variation now leads to
\begin{equation} 
    \delta I = - \int^{t_f}_{t_i} {\rm d} t \left(m \ddot{x} + 
              \frac{\partial V}{\partial x}\right) \delta x 
              + \left.\left(m \dot{x} \delta x\right) 
              \right|^{t_f}_{t_i} \ . 
\end{equation} 
Assuming no externally applied forces, d'Alembert's principle yields the 
equation of motion 
\begin{equation} 
    m \ddot{x} + \frac{\partial V}{\partial x} = 0 \ . 
\end{equation} 
An alternative way to write this is to introduce the momentum $p$ (not to 
be confused with the fluid pressure introduced earlier) defined as 
\begin{equation} 
     p = \frac{\partial L}{\partial \dot{x}} = m \dot{x} \ , 
\end{equation} 
in which case 
\begin{equation} 
     \dot{p} + \frac{\partial V}{\partial x} = 0 \ . 
\end{equation} 

In the most honest applications, one has the obligation to incorporate 
dissipative, i.e.~non-conservative, forces. \ Unfortunately, dissipative 
forces $F_d$ cannot be put into action principles. \ Fortunately, 
Newton's second law is of great guidance, since it states 
\begin{equation} 
    m \ddot{x} + \frac{\partial V}{\partial x} = F_d \ , \label{newteq} 
\end{equation} 
when conservative and dissipative forces act. \ A crucial observation of 
Eq.~(\ref{newteq}) is that the ``kinetic'' ($m \ddot{x} = \dot{p}$) 
and conservative ($\partial V/\partial x$) forces, which enter the 
left-hand side, do follow from the action, i.e. 
\begin{equation} 
    \frac{\delta I}{\delta x} = - \left(m \ddot{x} + 
    \frac{\partial V}{\partial x}\right) \ . 
\end{equation} 
When there are {\em no} dissipative forces acting, the action principle 
gives us the appropriate equation of motion. \ When there {\em are} 
dissipative forces, the action {\em defines} for us the kinetic and 
conservative force terms that are to be balanced by the dissipative 
terms. \ It also defines for us the momentum. 

We should  emphasize that this way of using the action to define the 
kinetic and conservative pieces of the equation of motion, as well as the 
momentum, can also be used in a context where the system experiences 
an externally applied force $F_{ext}$. \ The force can be conservative or 
dissipative, and will enter the equation of motion in the same way as 
$F_d$ did above. \ That is 
\begin{equation} 
    - \frac{\delta I}{\delta x} = F_d + F_{ext} \ . 
\end{equation}
Like a dissipative force, the main effect of the external force can be to 
siphon kinetic energy from the system. \ Of course, whether a force is 
considered to be external or not depends on the {\em a priori} definition 
of the system. \ To summarize: The variational argument leads to equations 
of motion of the form
\begin{equation} 
    - \frac{\delta I}{\delta x} = m \ddot{x} + 
    \frac{\partial V}{\partial x} = \dot{p} + 
    \frac{\partial V}{\partial x} 
    \left\{\begin{array}{cc} 
         = 0 & {\rm Conservative} \\ 
         \neq 0 & {\rm Dissipative~and/or~External~Forces} 
    \end{array}\right. \label{ppmodel} 
\end{equation} 

\newpage

\section{The ``Pull-back'' Formalism for a Single Fluid} 
\label{sec:pullback} 
 
In this section the equations of motion and the stress-energy-momentum 
tensor for a one-component, general relativistic fluid are obtained from 
an action principle. \ Specifically a so-called ``pull-back'' approach 
(see, for instance, \cite{comer93:_hamil_multi_con,comer94:_hamil_sf,%
comer02:_zero_freq_subspace}) is used to construct a Lagrangian displacement 
of the number density four-current $n^\mu$, whose magnitude $n$ is the 
particle number density. \ This will form the basis for the variations of 
the fundamental fluid variables in the action principle. 

\begin{figure}[h] 
\centering 
\includegraphics[height=6cm,clip]{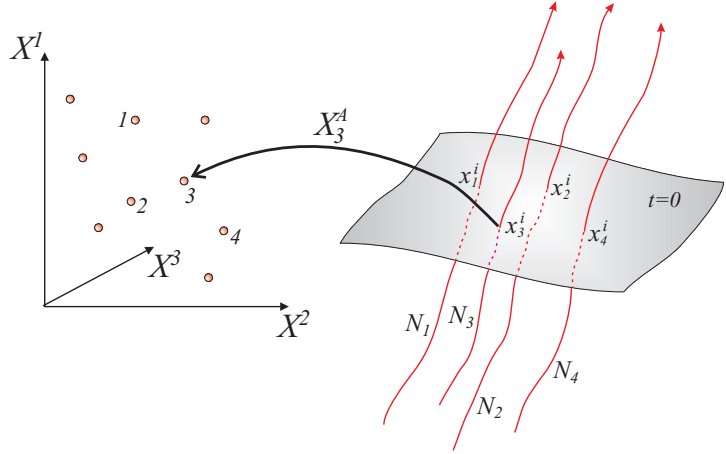} 
\caption{The pull-back from ``fluid-particle'' worldlines in spacetime, 
on the right, to ``fluid-particle'' points in the three-dimensional 
matter space labelled by the coordinates $\{X^1,X^2,X^3\}$. \ Here, the 
pull-back of the ``$I^{\rm th}$'' ($I = 1,2,...,n$) worldline into the 
abstract space (for the $t = 0$ configuration of ``fluid-particle'' 
points) is $X^A_I = X^A(0,x^i_I)$ where $x^i_I$ is the spatial position 
of the intersection of the worldline with the $t = 0$ time slice.} 
\label{pullback} 
\end{figure}

As there is only one species of particle considered here, $n^{\mu}$ is 
conserved, meaning that once a number of particles $N$ is assigned to a 
particular fluid element, then that number is the same at each point of the 
fluid element's worldline. \ This would correspond to attaching a given  
number of particles (i.e.~$N_1$, $N_2$, etc.) to each of the worldlines 
in Fig.~\ref{pullback}. \ Mathematically, one can write this as a standard 
particle-flux conservation equation: 
\begin{equation} 
    \nabla_{\mu} n^{\mu} = 0 \ . \label{consv} 
\end{equation} 
For reasons that will become clear shortly, it is useful to rewrite this 
conservation law in what may (if taken out context) seem a rather obscure
way. \ We introduce the dual $n_{\nu \lambda \tau}$ to $n^{\mu}$, i.e. 
\begin{equation}
     n_{\nu \lambda \tau} = \epsilon_{\nu \lambda \tau \mu} n^{\mu} 
                            \quad , \quad  
     n^{\mu} = \frac{1}{3!} \epsilon^{\mu \nu \lambda \tau} 
               n_{\nu \lambda \tau} \ , \label{ndual} 
\end{equation} 
such that 
\begin{equation} 
     n^2 = \frac{1}{3!} n_{\nu \lambda \tau} \n^{\nu \lambda \tau} \ , 
\end{equation} 
In Fig.~\ref{vorticity} we have seen that a two-form gives worldtubes. \ A 
three-form is the next higher-ranked object and it can be thought of, in an 
analogous way, as leading to boxes \cite{mtw73}. \ The key step here is to 
realize that the conservation rule is equivalent to having the three-form 
$n_{\nu \lambda \tau}$ be closed. \ In differential geometry this means 
that 
\begin{equation} 
    \nabla_{[\mu} n_{\nu \lambda \tau]} = 0 \ . 
\end{equation} 
This can be shown to be equal to Eq.~(\ref{consv}) by contracting with 
$\epsilon^{\mu \nu \lambda \tau}$. 

The main reason for introducing the dual is that it is straightforward to 
construct a particle number density three-form that is automatically 
closed, since the conservation of the particle number density current 
should not---speaking from a strict field theory point of view---be a 
part of the equations of motion, but rather should be automatically 
satisfied when evaluated on a solution of the ``true'' equations.  

This can be made to happen by introducing a three-dimensional ``matter'' 
space---the left-hand part of Fig.~\ref{pullback}---which can be 
labelled by coordinates $X^A = X^A(x^\mu)$, where $A,B,C~{\rm etc} = 
1,2,3$. \ For each time slice in spacetime, there will be a corresponding 
configuration in the matter space; that is, as time goes forward, the 
weaving of worldlines in spacetime will be matched by the fluid particle 
positions in the matter space. \ In this sense we are ``pulling-back'' 
from spacetime to the matter space (cf.~the previous discussion of the 
Lie derivative). \ The three-form can be pulled-back to its 
three-dimensional matter space by using the mappings $X^A$, as indicated 
in Fig.~\ref{pullback}. \ This allows us to construct a three-form that 
is automatically closed on spacetime. \ Hence, we let 
\begin{equation} 
     n_{\nu \lambda \tau} = N_{A B C} \left(\nabla_\nu X^A\right) 
                            \left(\nabla_\lambda X^B\right) \nabla_\tau 
                            X^C \ , 
\end{equation} 
where $N_{A B C}$ is completely antisymmetric in its indices and is a 
function only of the $X^A$. \ The $X^A$ will also be comoving in the 
sense that 
\begin{equation}
     n^\mu \nabla_\mu X^A = \frac{1}{3!} \epsilon^{\mu \nu \lambda \tau} 
                            N_{B C D} \left(\nabla_\nu X^B\right) 
                            \left(\nabla_\lambda X^C\right) 
                            \left(\nabla_\tau X^D\right) \nabla_\mu X^A 
                          = 0 \ . \label{comove} 
\end{equation} 
The time part of the spacetime dependence of the $X^A$ is thus somewhat ad 
hoc; that is, if we would take the flow of time to be the proper time of the 
worldlines, then the $X^A$ would not change, and hence there would be no 
``motion'' of the fluid particles in the matter space. 

Because the matter space indices are three-dimensional and the closure 
condition involves four spacetime indices, and also the $X^A$ are 
scalars on spacetime (and thus two covariant differentiations commute), 
the pull-back construction does indeed produce a closed three-form: 
\begin{equation}
    \nabla_{[\mu} n_{\nu \lambda \tau]} = \nabla_{[\mu} \left(N_{A B C} 
    \nabla_\nu X^A \nabla_\lambda X^B \nabla_{\tau]} X^C\right) \equiv 0 
    \ . 
\end{equation} 
In terms of the scalar fields $X^A$, we now have particle number density 
currents that are automatically conserved. \ Thus, another way of viewing 
the pull-back construction is that the fundamental fluid field variables 
are the $X^A$ (as evaluated on spacetime). \ In fact, the variations of 
$n_{\nu \lambda \tau}$ can now be taken with respect to variations of 
the $X^A$. 

Let us introduce the Lagrangian displacement on spacetime for the 
particles, to be denoted $\xi^\mu$. \ This is related to the variation 
$\delta X^A$ via a ``push-forward'' construction (which takes a 
variation in the matter space and pushes it forward to spacetime): 
\begin{equation} 
    \delta X^A = - \left(\nabla_\mu X^A\right) \xi^\mu \ . 
\end{equation} 
Using the fact that 
\begin{eqnarray} 
    \nabla_\nu \delta X^A &=& - \nabla_\nu \left(\left[\nabla_\mu 
                              X^A\right] \xi^\mu\right) \cr 
                           && \cr 
                          &=& - \left(\nabla_\mu X^A\right) \nabla_\nu 
                              \xi^\mu - \left(\nabla_\mu \nabla_\nu 
                              X^A\right) \xi^\mu 
\end{eqnarray}
means \cite{langlois98:_differ_rotat_superfl_ns} 
\begin{equation} 
    \delta n_{\nu \lambda \tau} = - \left(\xi^\sigma \nabla_\sigma 
    n_{\nu \lambda \tau} + n_{\sigma \lambda \tau} \nabla_\nu 
    \xi^\sigma + n_{\nu \sigma \tau} \nabla_\lambda \xi^\sigma + 
    n_{\nu \lambda \sigma} \nabla_\tau \xi^\sigma\right) = - 
    {\cal L}_{\xi} n_{\nu \lambda \tau} \ , \label{del3form} 
\end{equation} 
where ${\cal L}$ is the Lie derivative along $\xi^\mu$  
(cf.~Eq.~(\ref{liedev})). \ We can thus infer that 
\begin{equation} 
    \delta n^\mu = n^\sigma \nabla_\sigma \xi^\mu - \xi^\sigma 
                   \nabla_\sigma n^\mu - n^\mu \left(\nabla_\sigma 
                   \xi^\sigma + \frac{1}{2} g^{\sigma \rho} \delta 
                   g_{\sigma \rho}\right) \ . \label{delnvec} 
\end{equation} 
By introducing the decomposition 
\begin{equation} 
     n^\mu = n u^\mu \quad , \quad u_\mu u^\mu = - 1 \ , 
\end{equation}
we can furthermore show that 
\begin{equation}
   \delta n = - \nabla_\sigma\left(n \xi^\sigma\right) - n \left(
              u_\nu u^\sigma \nabla_\sigma \xi^\nu + \frac{1}{2} 
              \left[g^{\sigma \rho} + u^\sigma u^\rho\right] \delta 
              g_{\sigma \rho}\right) \ , \label{dens_perb} 
\end{equation} 
and 
\begin{equation} 
   \delta u^\mu = \left(\delta^\mu_\rho + u^\mu u_\rho\right) \left(
                  u^\sigma \nabla_\sigma \xi^\rho - \xi^\sigma 
                  \nabla_\sigma u^\rho\right) + \frac{1}{2} u^\mu 
                  u^\sigma u^\rho \delta g_{\sigma \rho} \ . 
                  \label{vel_perbs} 
\end{equation}

The Lagrangian variation resulting from the Lagrangian displacement is 
given by 
\begin{equation} 
    \Delta \equiv \delta + {\cal L}_\xi \ , 
\end{equation} 
from which it follows that 
\begin{equation} 
    \Delta n_{\mu \lambda \tau} = 0 \ , 
\end{equation} 
which is entirely consistent with the pull-back construction. \ We also 
find that
\begin{equation} 
    \Delta u^\mu = \frac{1}{2} u^\mu u^\sigma u^\rho \Delta 
                   g_{\sigma \rho} \ , \label{lagradelu} 
\end{equation} 
\begin{equation} 
    \Delta \epsilon_{\nu \lambda \tau \sigma} = \frac{1}{2} 
    \epsilon_{\nu \lambda \tau \sigma} g^{\mu \rho} \Delta g_{\mu \rho} 
    \ , \label{lagradeleps} 
\end{equation} 
and 
\begin{equation} 
    \Delta n = - \frac{n}{2} \left(g^{\sigma \rho} + u^\sigma u^\rho 
                \right) \Delta g_{\sigma \rho} \ . \label{lagradeln} 
\end{equation} 
These formulae (and their Newtonian analogues) have been adroitly used by 
Friedman and Schutz in their establishment of the so-called 
Chandrasekhar-Friedman-Schutz (CFS) instability 
\cite{chandrasekhar70:_grav_instab,friedman78:_lagran,%
friedman78:_secul_instab}, see Sec.~\ref{sec:cfs}. 

With a general variation of the conserved four-current in hand, we can 
now use an action principle to derive the equations of motion and the 
stress-energy-momentum tensor. \ The central quantity in the analysis 
is the so-called ``master'' function $\Lambda$, which is a function of the 
scalar $n^2 = - n_{\mu} n^{\mu}$. \ For this single fluid system, it is such 
that $- \Lambda$ corresponds to the local thermodynamic energy density. \ In 
the action principle, the master function is the Lagrangian density for the 
fluid: 
\begin{equation} 
    I_{fluid} = \int_{\cal M} \sqrt{- g} \Lambda {\rm d}^4 x \ . 
\end{equation} 

An unconstrained variation of $\Lambda(n^2)$ is with respect to $n^\mu$  
and the metric $g_{\mu \nu}$, and allows the four components of 
$n^\mu$ to be varied independently. \ It takes the form 
\begin{equation} 
     \delta \Lambda = \mu_{\mu} \delta n^{\mu} + \frac{1}{2} n^{\mu} 
                      \mu^{\nu} \delta g_{\mu \nu} \ , 
\end{equation} 
where 
\begin{equation} 
     \mu_{\mu} = \B n_\mu \quad , \quad 
     \B \equiv - 2 \frac{\partial \Lambda}{\partial n^2} \ . \label{mudef} 
\end{equation}
The use of the letter $\B$ is to remind us that this is a bulk fluid 
effect, which is present regardless of the number of fluids and 
constituents. \ The momentum covector $\mu_{\mu}$ is what we encountered 
earlier. \ It is dynamically, and thermodynamically, conjugate to 
$n^{\mu}$, and its magnitude can be seen to be the chemical potential of 
the particles by recalling that $\Lambda = - \rho$. \ Correctly 
identifying the momentum is essential for a concise and transparent 
formulation of fluids, e.g.~for constructing proofs about vorticity, and 
also for coupling in dissipative terms in the equations of motion (cf.~the 
discussion of Sec.~\ref{ppart}). \ For later convenience, we also introduce 
the momentum form defined as 
\begin{equation} 
    \mu^{\mu \nu \lambda} = \epsilon^{\mu \nu \lambda \rho} \mu_\rho \ . 
\end{equation} 

If the variation of the four-current was left unconstrained, the 
equations of motion for the fluid deduced from the above variation of 
$\Lambda$ would require, incorrectly, that the momentum covector $\mu_\mu$ 
should vanish in all cases.  \ This reflects the fact that the variation 
of the conserved four-current must be constrained, meaning that not all 
components of $n^\mu$ can be treated as independent. \ In terms of the 
constrained Lagrangian displacement of Eq.~(\ref{delnvec}), a first-order 
variation of the master function $\Lambda$  yields 
\begin{eqnarray}
    \delta \left(\sqrt{- g} \Lambda\right) &=& \frac{1}{2} \sqrt{- g} 
    \left(\Psi \delta^\mu{}_\lambda + n^\mu \mu_\lambda\right) g^{\lambda 
    \nu} \delta g_{\mu \nu} - \sqrt{- g} f_\nu \xi^{\nu} \cr 
    && + \nabla_\nu \left(\frac{1}{2}\sqrt{-g} \mu^{\nu \lambda \tau} 
       n_{\lambda \tau \mu} \xi^\mu\right) \ , 
\end{eqnarray} 
where the total divergence does not contribute to the field equations or 
stress-energy-momentum tensor (the divergence theorem implies that it 
becomes a boundary term in the action), the force density $f_\nu$ is as 
defined in Eq.~(\ref{fdens}), and the generalized pressure $\Psi$ is defined 
to be 
\begin{equation} 
     \Psi = \Lambda - n^{\mu} \mu_{\mu} \ . 
\end{equation} 
We have utilized the well-known formula \cite{mtw73} 
\begin{equation} 
     \delta \sqrt{- g} = \frac{1}{2} \sqrt{- g} g^{\mu \nu} \delta 
                         g_{\mu \nu} \ . \label{wellknown} 
\end{equation} 

At this point we can return to the view that $n^{\mu}$ is the fundamental 
field for the fluid. \ For vanishing metric variations, the principle of 
least action implies that the coefficient of $\xi^\nu$, as well as the 
total divergence (which yields the boundary term), must vanish. \ Thus, 
the equations of motion consist of the original conservation condition 
from Eq.~(\ref{consv}) and the Euler equation 
\begin{equation} 
     f_\nu = 2 n^{\mu} \omega_{\mu \nu} = 0 \ . \label{eueqn} 
\end{equation} 
We see that $\mu^{\nu \lambda \tau} n_{\lambda \tau \mu} \xi^\mu$, as 
evaluated on the boundary, is fixed during the variation. 

%

When $\delta g_{\mu \nu} \neq 0$, it is well-established in the 
relativity literature that the stress-energy-momentum tensor follows 
from a variation of the Lagrangian with respect to the metric, that is 
\begin{equation} 
     T^{\mu \nu} \delta g_{\mu \nu} \equiv \frac{2}{\sqrt{- g}} \delta 
     \left(\sqrt{- g} {\cal L}\right) = \left(\Psi \delta^\mu{}_\lambda + 
     n^\mu \mu_\lambda\right) g^{\lambda \nu} \delta g_{\mu \nu} 
     \ . \label{seten} 
\end{equation} 
When the complete set of field equations is satisfied then we see from 
Eq.~(\ref{euler1f}), which applies here, that it is automatically true 
that $\nabla_\mu T^\mu{}_\nu = 0$. 

Let us now recall the discussion of the point particle. \ There we 
pointed out that only the fully conservative form of Newton's Second Law 
follows from an action, i.e.~external or dissipative forces are excluded. 
\ However, we argued that a well-established form of Newton's Second Law 
is known that allows for external and/or dissipative forces, 
cf.~Eq.~(\ref{ppmodel}). \ There is thus much purpose in using the 
particular symbol $f_\nu$ in Eq.~(\ref{eueqn}). \ We may take the $f_\nu$ 
to be the relativistic analogue of the left-hand-side of Eq.~(\ref{ppmodel}) 
in every sense. \ In particular, when dissipation and/or external 
``forces'' act in a general relativistic setting, they are incorporated 
via the right-hand-side of Eq.~(\ref{eueqn}).  

\newpage

\section{The Two-Constituent, Single Fluid} \label{pbonemc} 

Generally speaking, the total energy density $\rho$ can be a function of 
independent parameters other than the particle number density $n_\n$, 
such as the entropy density $n_\s$, assuming that the system scales in the 
manner discussed in Sec.~\ref{thermo} so that only densities need enter 
the equation of state. \ (For later convenience we will introduce the 
constituent indices $\X,~\Y$, etc., which range over the two constituents 
$\{\n,\s\}$, and do not satisfy any kind of summation convention.) \ If 
there is no heat conduction, then this is still a single fluid problem, 
meaning that there is still just one unit flow velocity $u^\mu$ 
\cite{comer93:_hamil_multi_con}. \ This is what we mean by a two-constituent, 
single fluid. \ We assume that the particle number and entropy are both 
conserved along the flow, in the same sense as in Eq.~(\ref{comove}). \ 
Associated which each parameter there is thus a conserved current density 
four-vector, i.e.~$n^\mu_\n = n_\n u^\mu$ for the particle number density 
and $n_\s^\mu = n_\s u^\mu$ for the entropy density. \ Note that the ratio 
$x_\s = n_\s/n_\n$ is comoving in the sense that 
\begin{equation} 
     u^\mu \nabla_\mu \left(x_\s\right) = 0 \ . \label{comoving} 
\end{equation}

In terms of constituent indices $\X,~\Y$, the associated first law can be 
written in the form 
\begin{equation} 
     {\rm d} \rho = \sum_{\X = \{n,s\}} \mu^\X {\rm d} n_\X 
             =  - \sum_{\X = \{n,s\}} \mu^\X_\mu {\rm d} n^\mu_\X \ , 
\end{equation} 
since $\rho = \rho(n^2_{\n},n^2_{\s})$, where 
\begin{equation} 
     n^\mu_\X = n_\X u^\mu \quad , \quad 
     n^2_{\X} = - g_{\mu \nu} n^\mu_\X n^\nu_\X \ , 
\end{equation} 
and 
\begin{equation} 
     \mu^\X_\mu = g_{\mu \nu} \B^{\X} n^\nu_\X \quad , \quad 
     \B^{\X} \equiv 2 \frac{\partial \rho}{\partial n^2_{\X}} \ . 
\end{equation} 
Because of Eq.~(\ref{1stlaw}) $\mu_\s = T$ where $T$ is the temperature. 

Given that we only have one four-velocity, the system will still just have 
one fluid element per spacetime point. \ But unlike before, there will be 
an additional conserved number, $N_\s$, that can be attached to each 
worldline, like the particle number $N_\n$ of Fig.~\ref{pullback}. \ In 
order to describe the worldlines we can use the same three scalars 
$X^A(x^\mu)$ as before. \ But how do we get a construction that allows 
for the additional conserved number? \ Recall that the intersections of 
the worldlines with some hypersurface, say $t = 0$, is uniquely specified 
by the three $X^A(0,x^i)$ scalars. \ Each worldline will have also the 
conserved numbers $N_\n$ and $N_\s$ assigned to them. \ Thus, the values 
of these numbers can be expressed as functions of the $X^A(0,x^i)$. \ But 
most importantly, the fact that each $N_\X$ is conserved, means that this 
function specification must hold for all of spacetime, so that in 
particular the ratio $x_\s$ is of the form $x_\s(x^\mu) = x_\s(X^A(x^\mu))$. 
\ Consequently, we now have a construction whereby this ratio identically 
satisfies Eq.~(\ref{comoving}), and the action principle remains a 
variational problem just in terms of the three $X^A$ scalars. 

The variation of the action follows like before, except now a constituent 
index $\X$ must be attached to the particle number density current 
and three-form: 
\begin{equation} 
     n^\X_{\nu \lambda \tau} = \epsilon_{\nu \lambda \tau \mu} n^\mu_\X 
                               \ . 
\end{equation} 
Once again it is convenient to introduce the momentum form, now defined as 
\begin{equation} 
     \mu^{\mu \nu \lambda}_\X = \epsilon^{\mu \nu \lambda \rho} 
                                \mu^\X_\rho \ . \label{momformx} 
\end{equation} 
Since the $X^A$ are the same for each $n^\X_{\nu \lambda \tau}$, the 
above discussion indicates that the pull-back construction now is to be 
based on 
\begin{equation} 
     n^\X_{\nu \lambda \tau} = N^\X_{A B C} \left(\nabla_\nu X^A\right) 
     \left(\nabla_\lambda X^B\right) \nabla_\tau X^C \ , 
\end{equation} 
where $N^\X_{A B C}$ is completely antisymmetric and a function of the 
$X^A$. \ After some due deliberation, the reader should be convinced that 
the only thing required  here in addition to the previous subsection 
is to attach a $\X$ index to $n_{\nu \lambda \tau}$, $n^\mu$, and 
$n$ in Eqs.~(\ref{del3form}), (\ref{delnvec}), and (\ref{dens_perb}), 
respectively. \ If we now define the master function as 
\begin{equation} 
    \Lambda = - \rho 
\end{equation}
and the generalized pressure $\Psi$ to be 
\begin{equation} 
     \Psi = \Lambda - \sum_{\X = \{\n,\s\}} \mu^\X_\mu n^\mu_\X 
          = \Lambda + \sum_{\X = \{\n,\s\}} \mu^\X n_\X \ , 
\end{equation}
then the first-order variation for $\Lambda$ is 
\begin{eqnarray} 
     \delta \left(\sqrt{- g} \Lambda\right) &=& \frac{1}{2} \sqrt{- g} 
     \left(\Psi g^{\mu \nu} + \left[\Psi - \Lambda\right] u^\mu u^\nu 
     \right) \delta g_{\mu \nu} - \sqrt{- g} \left(\sum_{\X = \{\n,\s\}} 
     f^\X_\nu\right) \xi^\nu + \cr 
     &&  \nabla_\nu \left(\frac{1}{2} \sqrt{-g} \sum_{\X = \{\n,\s\}} 
     \mu^{\nu \lambda \tau}_\X n^\X_{\lambda \tau \mu} \xi^\mu\right) 
     \ , 
\end{eqnarray}  
where 
\begin{equation} 
     f^\X_\nu = 2 n^\mu_\X \omega^\X_{\mu \nu} \ , \label{forcex} 
\end{equation} 
and 
\begin{equation} 
     \omega^\X_{\mu \nu} = \nabla_{[\mu} \mu^\X_{\nu]} \ . 
\end{equation} 
The final equations of motion are 
\begin{equation} 
   \sum_{\X = \{\n,\s\}} f^\X_\nu = 0 \ , \label{eueqn1} 
\end{equation} 
and 
\begin{equation} 
   \nabla_\nu n^\nu_\X = 0 \ , 
\end{equation} 
whereas the stress-energy-momentum tensor takes the familiar form  
\begin{equation} 
    T^{\mu \nu} = \Psi g^{\mu \nu} + (\Psi - \Lambda) u^\mu u^\nu \ . 
\end{equation} 


\newpage

\section{The ``Pull-Back'' Formalism for Two Fluids} \label{2fluids} 
\label{sec:twofluids} 

Having discussed the single fluid model, and how one accounts for 
stratification, it is time to move on to the problem of modelling 
multi-fluid systems. \ We will experience for the first time novel 
effects due to the presence of a relative flow between two, 
interpenetrating fluids, and the fact that there is no longer a 
single, preferred rest-frame. \ This kind of formalism is necessary, for 
example, for the simplest model of a neutron star, since it is generally 
accepted that the inner crust is permeated by an independent neutron 
superfluid, and the outer core is thought to contain superfluid 
neutrons, superconducting protons, and a highly degenerate gas of 
electrons. \ Still unknown is the number of independent fluids that 
would be required for neutron stars that have colour-flavour-locked 
quark matter in the deep core \cite{alford00:_cfl}. \ The model can also 
be used to describe superfluid Helium and heat-conducting fluids, which is 
of importance for incorporation of dissipation (see 
Sec.~\ref{sec:viscosity}). \ We will focus on this example here. \ It 
should be noted that, even though the particular system we concentrate on 
consists of only two fluids it illustrates all new features of a general 
multi-fluid system. \ Conceptually, the greatest step is to go from one to 
two fluids. \ A generalization to a system with more degrees of freedoms 
is straightforward. 

In keeping with the previous section, we will rely on use of the constituent 
index, which for all remaining formulas of this section will range over 
$\X,\Y,\Z = \n,\s$. \ In the example that we consider the two fluids 
represent the particles ($\n$) and the entropy ($\s$). \ Once again, the 
number density four-currents, to be denoted $n^{\mu}_\X$, are taken to be 
separately conserved, meaning 
\begin{equation} 
    \nabla_{\mu} n^{\mu}_\X = 0 \ . \label{consv2} 
\end{equation} 
As before, we use the dual formulation, i.e. introduce the three-forms 
\begin{equation} 
    n^\X_{\nu \lambda \tau} = \epsilon_{\nu \lambda \tau \mu} 
                              n^{\mu}_\X \quad , \quad 
    n^{\mu}_\X = \frac{1}{3!} \epsilon^{\mu \nu \lambda \tau} 
                  n^\X_{\nu \lambda \tau} \ . 
\end{equation} 
Also like before, the conservation rules are equivalent to the individual 
three-forms being closed, i.e. 
\begin{equation} 
    \nabla_{[\mu} n^\X_{\nu \lambda \tau]} = 0 \ . \label{multiclosed} 
\end{equation} 
However, we need a formulation whereby such conservation obtains 
automatically, at least in principle. 

\begin{figure}[h] 
\centering 
\includegraphics[height=8cm,clip]{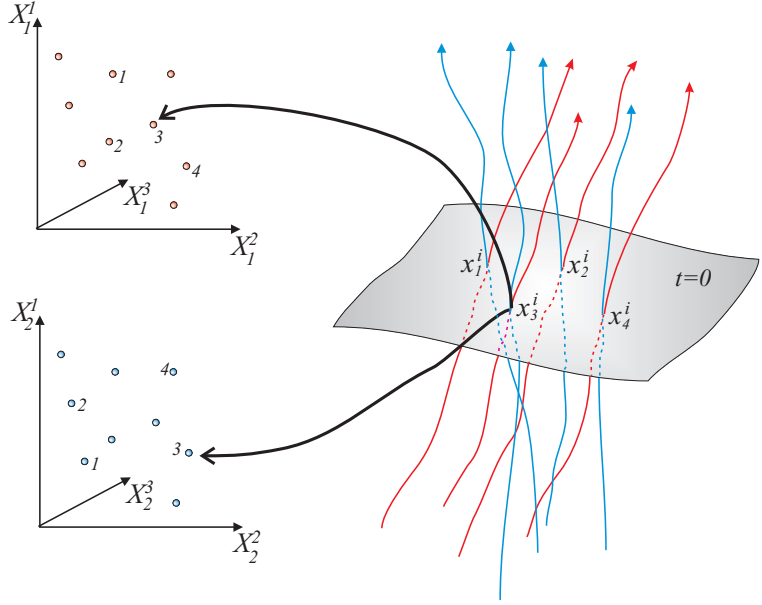} 
\caption{The pull-back from an $\X^{\rm th}$-constituent 
``fluid-particle'' worldline in spacetime, on the right, to 
``fluid-particle'' points in that $\X^{\rm th}$-constituent's 
three-dimensional matter space labelled by the coordinates 
$\{X^1_\X,X^2_\X,X^3_\X\}$, where the constituent index $\X = \n,\s$. \ 
There exist as many matter spaces as there are dynamically independent 
fluids, which for this case means two.} 
\label{pullback2} 
\end{figure}

We make this happen by introducing the three-dimensional matter space, the 
difference being that we now need two such spaces. \ These will be labelled 
by coordinates $X^A_\X$, and we recall that $A,B,C~{\rm etc} = 1,2,3$. \ 
This is depicted in Fig.~\ref{pullback2}, which indicates the important facts 
that (i) a given spatial point can be intersected by each fluid's worldline 
and (ii) the individual worldlines are not necessarily parallel at the 
intersection, i.e.~the independent fluids are interpenetrating {\em and} can 
exhibit a relative flow with respect to each other. \ Although we have not 
indicated this in Fig.~\ref{pullback2} (in order to keep the figure as 
uncluttered as possible) attached to each worldline of a given constituent 
will be a fixed number of particles $N^\X_1$, $N^\X_2$, 
etc.~(cf.~Fig.~\ref{pullback}). \ For the same reason, we have also not 
labelled (as in Fig.~\ref{pullback}) the ``pull-backs'' (represented by the 
arrows) from spacetime to the matter spaces. 

By ``pulling-back'' each constituent's three-form onto its respective 
matter space we can once again construct three-forms that are 
automatically closed on spacetime, i.e.~let 
\begin{equation} 
    n^\X_{\nu \lambda \tau} = N^\X_{A B C} \nabla_\nu X^A_\X 
    \nabla_\lambda X^B_\X \nabla_\tau X^C_\X \ , 
\end{equation} 
where $N^\X_{A B C}$ is completely antisymmetric in its indices and is a 
function of the $X^A_\X$. \ Using the same reasoning as in the single fluid 
case, the pull-back construction produces three-forms that are automatically 
closed, i.e.~they satisfy Eq.~(\ref{multiclosed}) identically. \ If we let 
the scalar fields $X^A_\X$ (as functions on spacetime) be the fundamental 
variables, they yield a representation for each particle number density 
current that is automatically conserved. \ The variations of the three-forms 
can now be derived by varying them with respect to the $X^A_\X$. 

Lagrangian displacements on spacetime for each fluid, to be denoted 
$\xi^\mu_\X$, will now be introduced. \ They are related to the 
variations $\delta X^A_\X$ via the ``push-forwards'' 
\begin{equation} 
    \delta X^A_\X = - \left(\nabla_\mu X^A_\X\right) \xi^\mu_\X \ . 
\end{equation} 
It should be clear that the analogues of Eqs.~(\ref{del3form}), 
(\ref{delnvec}), (\ref{dens_perb}), and (\ref{vel_perbs}) for this 
two-fluid case are given by the same formulas except that each 
displacement and four-current will now be associated with a constituent index, 
using the decomposition 
\begin{equation}
    n^\mu_\X = n_\X u^\mu_\X \quad , \quad u^\X_\mu u^\mu_\X = - 1 \ , 
\end{equation} 
where $u^\X_\mu = g_{\mu \nu} u^\nu_\X$. 

Associated with each constituent's Lagrangian displacement is its own 
Lagrangian variation. \ These are naturally defined to be
\begin{equation} 
    \Delta_\X \equiv \delta + {\cal L}_{\xi_\X} \ , 
\end{equation} 
so that it follows that 
\begin{equation} 
    \Delta_\X n^\X_{\mu \lambda \tau} = 0 \ , 
\end{equation} 
as expected for the pull-back construction. \ Likewise, two-fluid analogues 
of Eqs.~(\ref{lagradelu})--(\ref{lagradeln}) exist which take the same form 
except that the constituent index is attached. \ However, in contrast to 
the ordinary fluid case, there are many more options to consider. \ For 
instance, we could also look at the Lagrangian variation of the first 
constituent with respect to the second constituent's flow, 
i.e.~$\Delta_\s n_\n $, or the other way around, i.e.~$\Delta_\n n_\s$. \ 
The Newtonian analogues of these Lagrangian displacements were essential 
to an analysis of instabilities in rotating superfluid neutron stars 
\cite{andersson04:_canon_energy}. 

We are now in a position to construct an action principle that will yield 
the equations of motion and the stress-energy tensor. \ Again, the central 
quantity is the ``master'' function $\Lambda$, which is now a function of 
all the different scalars that can be formed from the $n^{\mu}_\X$, 
i.e.~the scalars $n_\X$ together with 
\begin{equation} 
    n^2_{\X \Y} = n^2_{\Y \X} = - g_{\mu \nu} n^\mu_\X n^\nu_\Y \ . 
\end{equation}
In the limit where all the currents are parallel, i.e.~the fluids are 
comoving, $- \Lambda$ corresponds again to the local thermodynamic 
energy density. \ In the action principle, $\Lambda$ is the Lagrangian 
density for the fluids. \ It should be noted that our choice to use only 
the fluid currents to form scalars implies that the system is ``locally 
isotropic'' in the sense that there are no a priori preferred directions, 
i.e.~the fluids are equally free to move in any direction. \ Structures 
like the crust believed to exist near the surface of  a neutron star 
generally would be locally anisotropic, eg.~sound waves in the lattice 
move in the preferred directions associated with the lattice. 

An unconstrained variation of $\Lambda$ with respect to the independent 
vectors $n^\mu_\X$ and the metric $g_{\mu \nu}$ takes the form 
\begin{equation} 
    \delta \Lambda = \sum_{\X = \{\n,\s\}} \mu^\X_\mu \delta n^{\mu}_\X + 
                     \frac{1}{2} g^{\lambda \nu} \left(\sum_{\X = 
                     \{\n,\s\}} n^\mu_\X \mu_\lambda^\X\right) \delta 
                     g_{\mu \nu} \ , 
\end{equation} 
where 
\begin{equation} 
    \mu^\X_{\mu} = g_{\mu \nu} \left(\B^{\X } n^\nu_\X + \A^{\X \Y} 
                   n^\nu_\Y\right) \ , \label{mudef2} 
\end{equation} 
\begin{equation} 
    \A^{\X \Y} = \A^{\Y \X} = - \frac{\partial \Lambda}{\partial 
                 n^2_{\X \Y}} \quad , \quad \X \neq \Y \ . \label{coef12} 
\end{equation} 
The momentum covectors $\mu^\X_\mu$ are each dynamically, and 
thermodynamically, conjugate to their respective number density currents 
$n^\mu_\X$, and their magnitudes are the chemical potentials. \ We see that 
$\A^{\X \Y}$ represents the fact that each fluid momentum  $\mu^\X_{\mu}$ 
may, in general, be given by a linear combination of the individual 
currents $n^{\mu}_\X$. \ That is, the current and momentum for a particular 
fluid do not have to be parallel. \ This is known as the entrainment effect. 
\ We have chosen to represent it by the letter $\A$ for historical reasons. 
\ When Carter first developed his formalism he opted for this notation, 
referring to the ``anomaly'' of having misaligned currents and momenta. \ 
It has since been realized that the entrainment is a key feature of most 
multi-fluid systems and it would, in fact, be anomalous to leave it out! 

In the general case, the momentum of one constituent carries along some 
mass current of the other constituents. \ The entrainment only vanishes 
in the special case where $\Lambda$ is independent of $n^2_{\X \Y}$ 
($\X \neq \Y$) because then we obviously have $\A^{\X \Y} = 0$. \ 
Entrainment is an observable effect in laboratory superfluids 
\cite{putterman74:_sfhydro,tilley90:_super} (eg.~via flow modifications 
in superfluid ${}^4{\rm He}$ and mixtures of superfluid ${}^3{\rm He}$ 
and ${}^4{\rm He}$). \ In the case of neutron stars, entrainment is an 
essential ingredient of the current best explanation for the so-called 
glitches \cite{radha69:_glitches,reichly69:_glitches}. \ Carter 
\cite{carter89:_covar_theor_conduc} has also argued that these ``anomalous'' 
terms are necessary for causally well-behaved heat conduction in 
relativistic fluids, and by extension necessary for building well-behaved 
relativistic equations that incorporate dissipation. 

In terms of the constrained Lagrangian displacements, a variation of 
$\Lambda$ now yields 
\begin{eqnarray} 
    \delta \left(\sqrt{- g} \Lambda\right) &=& \frac{1}{2} \sqrt{- g} 
    \left(\Psi \delta^\mu{}_\lambda + \sum_{\X = \{\n,\s\}} n^{\mu}_\X 
    \mu_\lambda^\X\right) g^{\lambda \nu} \delta g_{\mu \nu} -  
    \sqrt{- g} \sum_{\X = \{\n,\s\}} f^\X_{\nu} \xi^{\nu}_\X  \cr 
    && + \nabla_\nu \left(\frac{1}{2} \sqrt{-g} \sum_{\X = \{\n,\s\}} 
    \mu^{\nu \lambda \tau}_\X n^\X_{\lambda \tau \mu} \xi^\mu_\X\right) 
    \ , 
\end{eqnarray} 
where $f^\X_\nu$ is as defined in Eq.~(\ref{forcex}) except that the 
individual velocities are no longer parallel. \ The generalized pressure 
$\Psi$ is now 
\begin{equation}
     \Psi = \Lambda - \sum_{\X = \{\n,\s\}} n^{\mu}_\X \mu^\X_{\mu} \ . 
\end{equation} 
At this point we will return to the view that $n^{\mu}_\n$ and 
$n^{\mu}_\s$ are the fundamental variables for the fluids. \ Because 
the $\xi^\mu_\X$ are independent variations, the equations of motion 
consist of the two original conservation conditions of Eq.~(\ref{consv2}) 
plus two Euler type equations 
\begin{equation} 
     f^\X_{\nu} = 0 \ . \label{eueqn2} 
\end{equation} 
We also find that the stress-energy tensor is 
\begin{equation} 
     T^{\mu}{}_{\nu} = \Psi \delta^\mu{}_\nu + \sum_{\X = \{\n,\s\}} 
                       n^\mu_\X \mu^\X_\nu \ . \label{seten2} 
\end{equation} 
When the complete set of field equations is satisfied then it is 
automatically true that $\nabla_{\mu} T^{\mu}{}_{\nu} = 0$. \ One can 
also verify that $T_{\mu \nu}$ is symmetric. \ The momentum form 
$\mu^{\mu \nu \lambda}_\X$ entering the boundary term is the natural 
extension of Eq.~(\ref{momformx}) to this two-fluid case. 

It must be noted that Eq.~(\ref{eueqn2}) is significantly different from 
the multi-constituent version of Eq.~(\ref{eueqn1}). \ This is true even 
if one is solving for a static and spherically symmetric configuration, 
where the fluid four-velocities would all necessarily be parallel. \ 
Simply put, Eq.~(\ref{eueqn2}) still represents two independent 
equations. \ If one takes entropy as an independent fluid, then the 
static and spherically symmetric solutions will exhibit thermal 
equilibrium \cite{comer99:_quasimodes_sf}. \ This explains, for instance, 
why one must specify an extra condition (eg.~convective stability 
\cite{weinberg72:_book}) to solve for a double-constituent star with only 
one four-velocity. 
 

\newpage

\section{Speeds of Sound}
\label{sec:soundspeed}

Crucial to the understanding of black holes is the effect of spacetime 
curvature on the light-cone structures, that is, the totality of null 
vectors that emanate from each spacetime point. \ Crucial to the 
propagation of massless fields is the light-cone structure. \ In the case of 
fluids, it is both the speed of light and the speed (and/or speeds) of 
sound that dictate how waves propagate through the matter, and thus how 
the matter itself propagates. \ We give here a simple analysis that uses 
plane-wave propagation to derive the speed(s) of sound in the single-fluid, 
two-constituent single-fluid, and two-fluid cases, assuming that the metric 
variations vanish (see \cite{carter89:_covar_theor_conduc} for a more 
rigorous derivation). \ The analysis is local, assuming that the speed of 
sound is a locally defined quantity, and performed using local, Minkowski 
coordinates $x^\mu$. \ The purpose of the analysis is to illuminate how 
the presence of various constituents and multi-fluids impact the local 
dynamics. 

In a small region we will assume that the configuration of the matter with 
no waves present is locally isotropic, homogeneous, and static. \ Thus for 
the background $n^\mu_\X = \{n_\X,0,0,0\}$ and the vorticity 
$\omega^\X_{\mu \nu}$ vanishes. \ For all cases the general form of the 
variation of the force density $f^\X_\nu$ for each constituent is 
\begin{equation} 
    \delta f_\nu^\X = 2 n^\mu_\X \partial_{[\mu} \delta \mu^\X_{\nu]} \ . 
\end{equation} 
Similarly, the conservation of the $n^\mu_\X$ gives 
\begin{equation} 
    \partial_\mu \delta n^\mu_\X = 0 \ . \label{consvar} 
\end{equation} 
We are here using the point of view that the $n^\mu_\X$ are the fundamental 
fluid fields, and thus plane wave propagation means 
\begin{equation} 
    \delta n^\mu_\X = {A}^\mu_\X e^{i k_\nu x^\nu} \ , \label{pwave} 
\end{equation} 
where the amplitudes ${A}^\mu_\X$ and the wave vector $k_\mu$ are constant. \ 
Eqs.~(\ref{consvar}) and (\ref{pwave}) imply simply
\begin{equation} 
    k_\mu \delta n^\mu_\X = 0 \ ; \label{transverse} 
\end{equation} 
i.e.~the waves are ``transverse'' in the spacetime sense. \ We will give 
general forms for $\delta \mu^\X_\nu$ in what follows, and only afterwards 
assume static, homogeneous, and isotropic backgrounds. 
 
\subsection{Single Fluid Case}

Suppose that there is only one constituent, with index $\X = \n$. \ The 
master function $\Lambda$ then depends only on $n^2_{\n}$. \ The variation 
in the chemical potential due to a small disturbance $\delta n^\mu_\n$ is 
\begin{equation} 
     \delta \mu^\n_\mu = \B^{\n}_{\mu \nu} \delta n^\nu_\n \ , 
\end{equation} 
where 
\begin{equation} 
    \B^{\n}_{\mu \nu} = \B^{\n} g_{\mu \nu} - 2 
    \frac{\partial \B^{\n}}{\partial n^2_{\n}} g_{\mu \sigma} 
    g_{\nu \rho} n^\sigma_\n n^\rho_\n \ . \label{knn} 
\end{equation} 
The equation of motion is $\delta f^\n_\mu = 0$. \ It is not difficult to 
show, by using the condition of transverse wave propagation  
Eq.~(\ref{transverse}) and contracting with the spatial part of the wave 
vector $k^i = g^{i j} k_j$, that the equation of motion reduces to 
\begin{equation} 
    \left(\B^{\n } + \B^{\n}_{0 0} \frac{k_j k^j}{k^2_0}\right) k_i 
    \delta n^i_\n = 0 \ . 
\end{equation} 
The speed of sound is thus 
\begin{equation}
    {C}^2_S \equiv \frac{k^2_0}{k^i k_i} = 1 + \frac{\partial \ln 
    \B^{\n }}{\partial \ln n_\n} \ . 
\end{equation}
Given that a well-constructed fluid model should have ${C}^2_S \leq 1$, we 
see that the second term must be negative. \ This would ensure that the 
model is causal. 

\subsection{Two-constituent, Single Fluid Case}

Now consider the case when there are the two constituents with densities 
$n_\n$ and $n_\s$, two conserved density currents $n^\mu_\n$ and $n^\mu_\s$, 
two chemical potential covectors $\mu^\n_\mu$ and $\mu^\s_\mu$, but still 
only one four-velocity $u^\mu$. \ The master function $\Lambda$ depends on 
both $n^2_{\n}$ and $n^2_{\s}$ meaning that 
\begin{equation} 
    \delta \mu^\X_\mu = \B^{\X}_{\mu \nu} \delta n^\nu_\X 
                        + {\cal X}^{\X \Y}_{\mu \nu} \delta n^\nu_\Y \ , 
\end{equation} 
where 
\begin{equation} 
    {\cal X}^{\X \Y}_{\mu \nu} = - 2 \frac{\partial \B^{\X}}{\partial 
    n^2_{\Y}} g_{\mu \sigma} g_{\nu \rho} n^\sigma_\X n^\rho_\Y 
\end{equation} 
and $\B^{\s}_{\mu \nu}$ is as in Eq.~(\ref{knn}) except that each $\n$ 
is replaced with $\s$. \ We have chosen to use the letter ${\cal X}$ to 
represent what is a true multi-constituent effect, which arises due to 
composition gradients in the system. \ An alternative would have been to 
use $\C$ since the effect is due to the presence of different constituents. 
\ However, in his papers Carter tends to use $\B^{\s} = {\cal C}$, referring 
to the bulk entropy contribution as ``caloric''. \ Our chosen notation is 
intended to avoid confusion. \ It is also the case that the presence of the 
composition term ${\cal X}^{\X \Y}_{\mu \nu}$ has not been emphasized in 
previous work. \ This may be unfortunate since a composition variation is 
known to affect the dynamics of a system, eg.~by giving rise to the g-modes 
in a neutron star \cite{reisenegger92}. 

The fact that $n^\mu_\s$ is parallel to $n^\mu_\n$ implies that it is 
only the magnitude of the entropy density current that is independent. \ 
One can show that the condition of transverse propagation, as applied to 
both currents, implies 
\begin{equation}
    \delta n^\mu_\s = x_\s \delta n^\mu_\n \ . 
\end{equation} 
Now, we proceed as in the previous example, noting that the equation of 
motion is $\delta f^\n_\mu + \delta f^\s_\mu = 0$, which reduces to 
\begin{equation} 
    \left(\left[\B^{\n} + x^2_\s \B^{\s}\right] 
    \frac{k^2_0}{k^j k_j} - \left[\B^{\n} c^2_\n + x^2_\s \B^{\s} 
    c^2_\s - 2 x_{\s} {\cal X}^{\n \s}_{0 0} \right]\right) k_i 
    \delta n^i_\n = 0 \ , 
\end{equation} 
where 
\begin{equation} 
    c^2_\X \equiv 1 + \frac{\partial \ln \B^{\X}}{\partial \ln n_\X} 
                  \ . 
\end{equation} 
The speed of sound is thus 
\begin{equation} 
    {C}^2_S = \frac{\B^{\n} c^2_\n + x^2_\s \B^{\s} c^2_\s - 2 x_\s  
              {\cal X}^{\n \s}_{0 0}}{\B^{\n} + x^2_\s \B^{\s}} \ . 
\end{equation} 

\subsection{Two Fluid Case} 

The two-fluid problem is qualitatively different from the previous two cases, 
since there are now two independent density currents. \ This fact impacts 
the analysis in two crucial ways: (i) the master function $\Lambda$ depends 
on $n^2_{\n}$, $n^2_{\s}$, {\em and} $n^2_{\n \s} = n^2_{\s \n}$ 
(i.e.~entrainment is present), and (ii) the equations of motion, after 
taking into account the transverse flow condition, are doubled to 
$\delta f^\n_\mu = 0 = \delta f^\s_\mu$. \ As we will see, the key point is 
that there are now {\em two} sound speeds that must be determined. 

A variation of the chemical potential covectors that leaves the metric 
fixed takes the form 
\begin{equation} 
    \delta \mu^\X_\mu = \left(\B^{\X}_{\mu \nu} + \A^{\X \X}_{\mu \nu} 
    \right) \delta n^\nu_\X + \left({\cal X}^{\X \Y}_{\mu \nu} + 
    \A^{\X \Y}_{\mu \nu}\right) \delta n^\nu_\Y \ , 
\end{equation} 
where the complicated terms 
\begin{eqnarray} 
    \A^{\X \X}_{\mu \nu} &=& - g_{\mu \sigma} g_{\nu \rho} \left(
    \frac{\partial \B^{\X}}{\partial n^2_{\X \Y}} \left[n^\rho_\X 
    n^\sigma_\Y + n^\sigma_\X n^\rho_\Y\right] + \frac{\partial 
    \A^{\X \Y}}{\partial n^2_{\X \Y}} n^\sigma_\Y n^\rho_\Y\right) \ , \\ 
                                && \cr
    \A^{\X \Y}_{\mu \nu} &=& \A^{\X \Y} g_{\mu \nu} - g_{\mu \sigma} 
    g_{\nu \rho} \left(\frac{\partial \B^{\X}}{\partial n^2_{\X \Y}} 
    n^\sigma_\X n^\rho_\X + \frac{\partial \B^{\Y}}{\partial 
    n^2_{\X \Y}} n^\sigma_\Y n^\rho_\Y + \frac{\partial \A^{\X \Y}} 
    {\partial n^2_{\X \Y}} n^\sigma_\Y n^\rho_\X\right) \ , 
\end{eqnarray} 
exist solely because of entrainment. \ The same procedure as in the 
previous two examples leads to the dispersion relation 
\begin{equation} 
    \left(\B^{\n} - \left[\B^{\n}_{0 0} + \A^{\n \n}_{0 0}\right] 
    \frac{k^i k_i}{k^2_0}\right) \left(\B^{\s} - \left[
    \B^{\s}_{0 0} + \A^{\s \s}_{0 0}\right] \frac{k^j k_j}{k^2_0} 
    \right) - \left(\A^{\n \s} - {\cal X}^{\n \s}_{0 0} - \A^{\n \s}_{0 0} 
    \right)^2 = 0 \ . 
\end{equation} 
This is quadratic in $k^i k_i/k^2_0$ meaning there are two sound speeds. 
\ This is a natural result of the doubling of fluid degrees of freedom. 

The sound speed analysis is local, but its results are seen globally in the 
analysis of modes of oscillation of a fluid body. \ For a neutron star, the 
full spectrum of modes is quite impressive (see McDermott et al 
\cite{mcdermott88:_modes}): polar (or spheroidal) f-, p-, and g-modes and the 
axial (or toroidal) r-modes. \ Epstein \cite{epstein88:_acoust_proper_ns} was 
the first to suggest that there should be even more modes in superfluid 
neutron stars because the superfluidity allows the neutrons to move 
independently of the protons. \ Mendell \cite{mendell91:_superflnodiss} 
developed this idea further by using an analogy with coupled pendulums. \ He 
argued that the new modes should feature a counter-motion between the neutrons 
and protons, i.e.~as the neutrons move out radially, say, the protons will 
move in. \ This is in contrast to ordinary fluid motion that would have the 
neutrons and protons move in more or less ``lock-step''. \ Analytical and 
numerical studies \cite{lee95:_nonrad_osc_superfl_ns,%
lindblom95:_does_gravit_radiat_limit_angul,comer99:_quasimodes_sf,%
andersson01:_dyn_superfl_ns} have confirmed this basic picture and the new 
modes of oscillation are known as superfluid modes. 

\newpage

\section{The Newtonian limit and the Euler equations}
\label{sec:newton}

One reason relativistic fluids are needed is that they can be used to 
model neutron stars. \ However, even though neutron stars are clearly general 
relativistic objects, one often starts with good old Newtonian physics 
when considering new applications. \ The main reason is that effects (such as 
acoustic modes of oscillation) which are primarily due to the fluids that 
make up the star can often be understood qualitatively from just Newtonian 
calculations. \ There are also certain regimes in a neutron star where 
the Newtonian limit is sufficient quantitatively (such as the outer 
layers). 

There has been much progress recently in the analysis of Newtonian 
multiple fluid systems. \ Prix \cite{prix04:_multi_fluid} has developed 
an action-based formalism, analogous to what has been used here. \ Carter 
and Chamel \cite{Carter03:_newtI,Carter03:_newtII,Carter04:_newtIII} have 
done the same except that they use a fully spacetime covariant formalism. 
\ We will be somewhat less ambitious (for example, as in 
\cite{andersson01:_dyn_superfl_ns}) by extracting the Newtonian equations 
as the non-relativistic limit of the fully relativistic equations. \ Given 
the results of Prix and Carter and Chamel we can think of this exercise 
as a consistency check of our equations. 

The Newtonian limit consists of writing the general relativistic field 
equations to order $c^0$ where $c$ is the speed of light. \ The Newtonian 
equations are obtained in the limit that $c \to \infty$. \ The 
gravitational potential, denoted $\Phi$, is assumed to be small in the 
sense that 
\begin{equation}
     - 1 << \frac{\Phi}{c^2} \leq 0 \ . 
\end{equation}
To order $c^0$ the metric becomes 
\begin{equation} 
    {\rm d} s^2 = - c^2 \left(1 + \frac{2 \Phi}{c^2}\right)  {\rm d} t^2 + 
           g_{i j} {\rm d} x^i  {\rm d} x^j \ , 
\end{equation}
where the $x^i$ are Cartesian coordinates and $g_{i j}$ is the metric of 
Eq.~(\ref{flatmet}). 

If $\tau$ is the proper time as measured along a fluid element worldline, 
then the curve it traces out can be written 
\begin{equation} 
    x^{\mu}(\tau) = \{t(\tau),x^i(\tau)\} \ .
\end{equation} 
Recall that its four-velocity is given by (cf.~Eq.~(\ref{uvec})) 
\begin{equation} 
    u^{\mu} = \frac{{\rm d} x^\mu}{{\rm d} \tau} \ , 
\end{equation} 
but because the speed of light has been restored, $u_{\mu} u^{\mu} = - c^2$. \ 
The proper time is now given by 
\begin{equation} 
  - {\rm d} s^2 = c^2 {\rm d} \tau^2 = c^2 \left(1 + \frac{2 \Phi}{c^2} 
                  - \frac{g_{ij} v^i v^j}{c^2}\right) {\rm d} t^2 \ , 
\end{equation} 
where $v^i = {\rm d}x^i/{\rm d}t$ is the Newtonian three-velocity of the 
fluid. \ It is considered to be small in the sense that 
\begin{equation} 
     {\left|v^i\right| \over c} << 1 \ . 
\end{equation} 
Hence, to the correct order the four-velocity components are 
\begin{equation} 
     u^t = 1 - \frac{\Phi}{c^2} + \frac{v^2}{2 c^2} 
        \ , \ 
     u^i = v^i \ , 
\end{equation} 
where $v^2 = g_{i j} v^i v^j$. \ Note that the particle number current 
becomes 
\begin{equation} 
    n^{\mu} = n \left(u^{\mu}/c\right) \ . 
\end{equation}

In order to write the single fluid Euler equations, keeping terms to the 
required order, it is necessary to explicitly break up the ``master'' function 
into its mass part and internal energy part ${\cal E}$, i.e., to write 
$\Lambda$ as 
\begin{equation} 
    \Lambda = - m n c^2 - {\cal E}(n) \ , 
\end{equation} 
where $m$ is the particle mass. \ The internal energy is small compared to the 
mass, meaning 
\begin{equation} 
     0 \leq \frac{{\cal E}}{m n c^2} << 1 \ . 
\end{equation}
With this choice, a variation of $\Lambda$ that leaves the metric fixed 
yields: 
\begin{equation}
     {\rm d} \Lambda = - \left(m c^2 + \mu\right) {\rm d} n \ , 
\end{equation} 
where 
\begin{equation} 
     \mu = \frac{\partial {\cal E}}{\partial n} \ . 
\end{equation} 
The $\B^{\n}$ coefficient reduces to
\begin{equation} 
    \B^{\n} = m c^2 + \frac{\mu}{n} \ . 
\end{equation} 
In terms of these variables, the pressure $p$ is seen to be 
\begin{equation} 
     p = - \E + \mu n \ . 
\end{equation} 
The Newtonian limit of the general relativistic fluid equations, in a 
general coordinate basis, is then 
\begin{equation} 
    0 = \frac{\partial n}{\partial t} + \nabla_i \left(n v^i\right) \ ,  
\end{equation} 
and

\begin{equation} 
    0 = \frac{\partial}{\partial t} v^i + v^j \nabla_j v^i + g^{i j}  
        \nabla_j \left(\Phi + \frac{\mu}{m}\right) \ , 
\label{euleq1}\end{equation} 
where $\nabla_i$ is the Euclidean space covariant derivative. \ Of 
course, the gravitational potential $\Phi$ is determined by a Poisson 
equation with $m n$ as source. 

A Newtonian two-fluid system can be obtained in a similar fashion. \ As 
discussed in Sec.~\ref{sec:twofluids}, the main difference is that we 
obviously need two sets of worldlines, describable, say, by curves 
$x^\mu_\X(\tau_\X)$ where $\tau_\X$ is the proper time along a constituent's 
worldline. \ Of course, entrainment also comes into play. \ Its presence 
implies that the relative flow of the fluids is required to specify the local 
thermodynamic state of the system, and that the momentum of a given fluid is 
not simply proportional to that fluid's flux. \ This is the situation for 
superfluid He$^4$ \cite{putterman74:_sfhydro,tilley90:_super}, where the 
entropy can flow independently of the superfluid Helium atoms. \ Superfluid 
He$^3$ can also be included in the mixture, in which case there will be a 
relative flow of the He$^3$ isotope with respect to He$^4$, and relative 
flows of each with respect to the entropy \cite{voll_book}. 

Let us consider a two-fluid model like a mixture of He$^4$ and He$^3$, or 
neutrons and protons in a neutron star. \ We will denote the two fluids as 
fluids ${\rm a}$ and ${\rm b}$. \ The magnitude squared of the difference of 
three-velocities 
\begin{equation} 
    w_i^{\Y \X} =  {v}_{i}^{\Y} - {v}_{i}^{\X}  
\end{equation} 
is denoted $w^2$ so that the equation of state takes the form ${\cal E} = 
{\cal E}(n_{\rm a},n_{\rm b},w^2)$. \ Hence, 
\begin{equation} 
    {\rm d} {\cal E} = \mu^{\rm a} {\rm d} n_{\rm a} + \mu^{\rm b} {\rm d} 
                       n_{\rm b} + \alpha {\rm d} w^2 \ , 
\end{equation}
where 
\begin{equation}
     \mu^{\rm a} = \left.\frac{\partial \mathcal{E}}{\partial n_{\rm a}} 
                  \right|_{n_{\rm b},w^2} \ , \  
     \mu^{\rm b} = \left.\frac{\partial \mathcal{E}}{\partial n_{\rm b}} 
                   \right|_{n_{\rm a},w^2} \ , \ 
     \alpha = \left.\frac{\partial \mathcal{E}}{\partial w^2} 
              \right|_{n_{\rm a},n_{\rm b}} \ . 
\end{equation} 
The $\alpha$ coefficient reflects the effect of entrainment on the equation of 
state. \ Similarly, entrainment causes the fluid momenta to be modified to 
\begin{equation} 
        {p^\X_i \over m^\X} =  v^i_\X + 2 \frac{\alpha}{\rho_\X} 
                   w^i_{\Y \X}\ . 
\end{equation}

The number density of each fluid obeys a continuity equation: 
\begin{equation}
    \pd{n_{\X}}{t} + \nabla_{j} (n_{\X} v_{\X}^{j}) = 0 \ . \label{eq:Cont} 
\end{equation} 
Each fluid is also seen to satisfy an Euler-type equation, which ensures the 
conservation of total momentum. \ This equation can be written 
\begin{eqnarray} 
    \label{eq:MomEqu} 
    \left(\pd{}{t} + {v}^{j}_{\X}\nabla_{j} \right)   
    \left[{v}_{i}^{\X} + \varepsilon_{\X} w_i^{\Y\X} \right] + \nabla_{i} 
    (\Phi + \tilde{\mu}_{\X}) + \varepsilon_{\X} w_j^{\Y \X} \nabla_{i} 
    v^{j}_{\X} = 0 \ ,  
\end{eqnarray} 
where, 
\begin{equation} 
    \tilde{\mu}_\X = {\mu^\X \over m^\X} \ , 
\end{equation}
i.e.~$\tilde{\mu}_\X$ is the relevant chemical potential per unit mass, and 
the entrainment is included via the coefficients 
\begin{equation} 
   \varepsilon_\X = 2 \rho_\X \alpha\ . 
\end{equation} 
For a detailed discussion of these equations, see  
\cite{prix04:_multi_fluid,andersson05:_flux_con}. 

\newpage

\section{The CFS instability} 
\label{sec:cfs} 

Investigations into the stability properties of rotating self-gravitating 
bodies are of obvious relevance to astrophysics. \ By improving our 
understanding of the relevant issues we can hope to shed light on 
the nature of the various dynamical and secular instabilities that may 
govern the spin-evolution of rotating stars. \ The relevance of such 
knowledge for neutron star astrophysics may be highly significant, especially 
since instabilities may lead to detectable gravitational-wave signals. \ 
In this section we will review the Lagrangian perturbation framework developed 
by Friedman and Schutz \cite{friedman78:_lagran,friedman78:_secul_instab} for 
rotating non-relativistic stars. \ This will lead to criteria that can be 
used to decide when the oscillations of a rotating neutron star are unstable. 
\ We provide an explicit example proving the instability of the so-called 
r-modes at all rotation rates in a perfect fluid star. 

\subsection{Lagrangian perturbation theory}

Following \cite{friedman78:_lagran,friedman78:_secul_instab}, we work with 
Lagrangian variations. \ We have already seen that the Lagrangian perturbation 
$\Delta Q$ of a quantity $Q$ is related to the Eulerian variation $\delta Q$ 
by 
\begin{equation} 
    \Delta Q = \delta Q + \mathcal{L}_\xi Q 
\end{equation} 
where (as before) $\mathcal{L}_\xi$ is the Lie derivative that was introduced 
in Sec.~\ref{sec:gr}. \ The Lagrangian change in the fluid velocity follows 
from the Newtonian limit of Eq.~(\ref{vel_perbs}): 
\begin{equation} 
    \Delta v^i = \partial_t \xi^i 
\end{equation} 
where $\xi^i$ is the Lagrangian displacement. \ Given this, and 
\begin{equation} 
    \Delta g_{ij} = \nabla_i \xi_j + \nabla_j \xi_i 
\end{equation} 
we have 
\begin{equation}
    \Delta v_i = \partial_t \xi_i + v^j\nabla_i \xi_j + v^j \nabla_j \xi_i 
    \ . 
\end{equation} 

Let us consider the simplest case, namely a barotropic ordinary fluid for 
which $\mathcal{E}=\mathcal{E}(n)$. \ Then we want to perturb the continuity 
and Euler equations from the previous section. \ The conservation of mass for 
the perturbations follows immediately from the Newtonian limits of 
Eqs.~(\ref{dens_perb}) and (\ref{lagradelu}) (which recall automatically 
satisfy the continuity equation): 
\begin{equation} 
    \Delta n = - n \nabla_i \xi^i \quad , \quad  
    \delta n = - \nabla_i (n \xi^i) \ . 
\end{equation} 
Consequently, the perturbed gravitational potential follows from 
\begin{equation} 
    \nabla^2 \delta \Phi = 4 \pi G m \delta n = - 4\pi G m \nabla_i(n \xi^i) 
                           \ . 
\end{equation} 
In order to perturb the Euler equations we first rewrite Eq.~(\ref{euleq1}) as 
\begin{equation} 
    (\partial_t +\mathcal{L}_v) v_i +  \nabla_i \left(\tilde{\mu} +  \Phi 
    - \frac{1}{2}  v^2\right) = 0 \ , \label{euleq2} 
\end{equation} 
where $\tilde{\mu}= \mu/m$. \ This form is particularly useful since the 
Lagrangian variation commutes with the operator $\partial_t + \mathcal{L}_v$. 
\ Perturbing Eq.~(\ref{euleq2}) we thus have 
\begin{equation} 
    (\partial_t +\mathcal{L}_v) \Delta v_i + \nabla_i \left(\Delta \tilde{\mu} 
    + \Delta \Phi - \frac{1}{2} \Delta( v^2) \right) = 0 \ . \label{peuls} 
\end{equation} 

We want to rewrite this equation in terms of the displacement vector $\xi$. 
\ After some algebra one finds 
\begin{eqnarray} 
    \partial_t^2 \xi_i + 2 v^j \nabla_j \partial_t \xi_i +  (v^j \nabla_j)^2 
    \xi_i + \nabla_i \delta \Phi + \xi^j \nabla_i \nabla_j \Phi \nonumber \\ 
    - (\nabla_i \xi^j) \nabla_j\tilde{\mu} + \nabla_i \Delta \tilde{\mu} 
    = 0 \ . \label{peul2} 
\end{eqnarray} 
Finally, we need 
\begin{eqnarray} 
    \Delta \tilde{\mu} = \delta \tilde{\mu} + \xi^i\nabla_i \tilde{\mu} = 
    \left(\frac{\partial \tilde{\mu}}{\partial n} \right) \delta n 
    + \xi^i\nabla_i \tilde{\mu} \nonumber \\ 
    = \quad - \left(\frac{\partial \tilde{\mu}}{\partial n}\right) \nabla_i 
      (n \xi^i) + \xi^i\nabla_i \tilde{\mu} \ . 
\end{eqnarray} 
Given this, we have arrived at the following form for the perturbed 
Euler equation
\begin{eqnarray}
    \partial_t^2 \xi_i + 2 v^j \nabla_j \partial_t \xi_i + (v^j \nabla_j)^2 
    \xi_i + \nabla_i \delta \Phi +  \xi^j \nabla_i \nabla_j \left(\Phi + 
    \tilde{\mu}\right) \nonumber \\ 
    - \nabla_i \left[\left(\frac{\partial \tilde{\mu}}{\partial n}\right) 
    \nabla_j (n \xi^j)\right] = 0 \ . \label{peul3} 
\end{eqnarray}
This equation should be compared to Eq.~(15) of \cite{friedman78:_lagran}. 

Having derived the perturbed Euler equations, we are interested in 
constructing conserved quantities that can be used to assess the stability 
of the system. \ To do this, we first multiply Eq.~(\ref{peul3}) by the number 
density $n$, and then write the result (schematically) as 
\begin{equation} 
    A \partial_t^2 \xi + B \partial_t \xi + C \xi  = 0 \ , 
\end{equation} 
omitting the indices since there is little risk of confusion. \ Defining the 
inner product 
\begin{equation} 
    \left< \eta^i,\xi_i \right> = \int \eta^{i*} \xi_i {\rm d} V 
\end{equation} 
where the asterisk denotes complex conjugation, one can now show that 
\begin{equation} 
    \left< \eta, A\xi \right>=\left< \xi,A\eta \right>^* \qquad \mbox{ and } 
    \qquad 
    \left< \eta,B\xi \right> = - \left< \xi,B\eta \right>^* \ . 
\end{equation} 
The latter requires the background relation $\nabla_i (n v^i) = 0$, and holds 
provided that $n \to 0$ at the surface of the star. \ A slightly more 
involved calculation leads to 
\begin{equation}
    \left< \eta, C\xi \right> = \left< \xi, C\eta \right>^* \ . 
\end{equation} 
Inspired by the fact that the momentum conjugate to $\xi^i$ is  
$\rho(\partial_t + v^j \nabla_j)\xi^i$, we now consider the symplectic 
structure 
\begin{equation} 
    W(\eta,\xi) = \left<\eta, A\partial_t \xi + \frac{1}{2} B \xi\right> 
    - \left< A\partial_t \eta + \frac{1}{2} B \eta, \xi\right> \label{Wdef} 
\end{equation} 
where $\eta$ and $\xi$ both solve the perturbed Euler equation. \ Given this, 
it is straightforward to show that $W(\eta,\xi)$ is conserved, 
i.e.~$\partial_t W = 0$. \ This leads us to define the \emph{canonical energy} 
of the system as 
\begin{equation} 
    E_c = \frac{m}{2} W (\partial_t \xi,\xi) = \frac{m}{2} 
    \left\{ \left< \partial_t \xi , A \partial_t \xi \right> 
    + \left< \xi, C \xi \right> \right\} \ . 
\end{equation} 
After some manipulations, we arrive at the following explicit expression 
\begin{eqnarray} 
    E_c = \frac{1}{2} \int \left\{ \rho |\partial_t \xi|^2 
    - \rho | v^j \nabla_j \xi_i|^2 + \rho \xi^i \xi^{j*}\nabla_i \nabla_j 
    (\tilde{\mu} + \Phi) \right. \nonumber \\ 
    + \left. \left( \frac{\partial \mu}{\partial n} \right) | \delta n |^2 
    - \frac{1}{4 \pi G} |\nabla_i \delta \Phi|^2 \right\} {\rm d} V 
\end{eqnarray} 
which can be compared to Eq.~(45) of \cite{friedman78:_lagran}. \ In the 
case of an axisymmetric system, eg. a rotating star, we can also define a 
\emph{canonical angular momentum} as 
\begin{equation}
    J_c = - \frac{m}{2} W (\partial_\varphi \xi, \xi) = - \mbox{ Re }  
          \left< \partial_\varphi \xi, A\partial_t \xi + \frac{1}{2} B\xi 
          \right> \ . 
\end{equation} 
The proof that this quantity is conserved relies on the fact that (i) 
$W(\eta, \xi)$ is conserved for any two solutions to the perturbed Euler 
equations, and (ii) $\partial_\varphi$ commutes with $\rho v^j \nabla_j$ 
in axisymmetry, which means that if $\xi$ solves the Euler equations then 
so does $\partial_\varphi \xi$. 

As discussed by \cite{friedman78:_lagran,friedman78:_secul_instab}, the 
stability analysis is complicated by the presence of so-called ``trivial'' 
displacements. \ These trivials can be thought of as representing a 
relabeling of the physical fluid elements. \ A trivial displacement $\zeta^i$ 
leaves the physical quantities unchanged, i.e.~is such that $\delta n = 
\delta v^i = 0$. \ This means that we must have
\begin{eqnarray}
     \nabla_i (\rho \zeta^i) &=& 0 \ , \\ 
     \left( \partial_t + \mathcal{L}_v \right) \zeta^i &=& 0 \ . 
\end{eqnarray}
The solution to the first of these equations can be written 
\begin{equation} 
    \rho \zeta^i = \epsilon^{ijk} \nabla_j \chi_k 
\end{equation}
where, in order to satisfy the second equations, the vector $\chi_k$ must 
have time-dependence such that
\begin{equation}
    ( \partial_t + \mathcal{L}_v) \chi_k = 0 \ . 
\end{equation} 
This means that the trivial displacement will remain constant along the 
background fluid trajectories. \ Or, as Friedman and Schutz 
\cite{friedman78:_lagran} put it, the ``initial relabeling is carried along 
with the unperturbed motion.''

The trivials have the potential to cause trouble because they affect the 
canonical energy. \ Before one can use the canonical energy to assess the 
stability of a rotating configuration one must deal with this ``gauge 
problem''. \ To do this one should ensure that the displacement vector 
$\xi$ is orthogonal to all trivials. \ A prescription for this is provided 
by Friedman and Schutz \cite{friedman78:_lagran}. \ In particular, they 
show that the required canonical perturbations preserve the vorticity of 
the individual fluid elements. \ Most importantly, one can also prove that 
a normal mode solution is orthogonal to the trivials. \ Thus, normal mode 
solutions can serve as canonical initial data, and be used to assess 
stability. 

\subsection{Instabilities of rotating perfect fluid stars} 
\label{pfinstab}

The importance of the canonical energy stems from the fact that 
it can be used to test the stability of the system. \ In particular, 

\begin{itemize}

\item dynamical instabilities are only possible for motions such that 
$E_c=0$. \ This makes intuitive sense since the amplitude of a mode for 
which $E_c$ vanishes can grow without bounds and still obey the 
conservation laws. 

\item if the system is coupled to radiation (eg. gravitational waves) 
which carries positive energy away from the system (which should be 
taken to mean that $\partial_t E_c < 0$) then any initial data 
for which $E_c<0$ will lead to an unstable evolution. 

\end{itemize}

Consider a real frequency normal-mode solution to the perturbation equations, 
a solution of form $\xi = \hat{\xi} e^{i(\omega t+m\varphi)}$. \ One can 
readily show that the associated canonical energy becomes 
\begin{equation} 
    E_c = \omega \left[  \omega \left<{\xi} , A {\xi}\right> - \frac{i}{2} 
          \left<{\xi} , B{\xi}\right> \right] \label{Ec} 
\end{equation} 
where the expression in the bracket is real. \ For the canonical angular 
momentum we get 
\begin{equation} 
    J_c = -m \left[  \omega \left<{\xi} , A {\xi} \right> - \frac{i}{2} 
          \left< {\xi} , B{\xi} \right> \right] \ . \label{Jc} 
\end{equation} 
Combining Eq.~(\ref{Ec}) and (\ref{Jc}) we see that, for real frequency 
modes, we will have 
\begin{equation} 
    E_c = - \frac{\omega}{m} J_c = \sigma_p J_c \label{EJrel} 
\end{equation} 
where $\sigma_p$ is the pattern speed of the mode. 

Now notice that Eq.~(\ref{Jc}) can be rewritten as 
\begin{equation} 
    \frac{J_c}{\left< \hat{\xi}, \rho\hat{\xi} \right>} = - m\omega + 
    m \frac{\left<{\xi}, i \rho v^j \nabla_j {\xi} \right>}{\left< 
    \hat{\xi}, \rho\hat{\xi} \right>} \ . 
\end{equation}
Using cylindrical coordinates, and $v^j = \Omega \varphi^j $, one can 
show that 
\begin{equation} 
    - i \rho {{\xi}}_i^* v^j \nabla_j {\xi}^i = \rho \Omega [ m |\hat{\xi}|^2 
    + i ({\hat{\xi}}^* \times \hat{\xi})_z] \ . 
\end{equation} 
But 
\begin{equation}
  | ({\hat{\xi}}^* \times \hat{\xi})_z | \le | \hat{\xi} |^2 
\end{equation} 
and we must have (for uniform rotation) 
\begin{equation} 
    \sigma_p - \Omega \left( 1 + \frac{1}{m} \right) \le \frac{J_c/m^2}{
    \left<\hat{\xi}, \rho\hat{\xi}\right>} \le \sigma_p - \Omega \left(1 - 
    \frac{1}{m} \right) \ . \label{ineq1} 
\end{equation} 

Eq.~(\ref{ineq1}) forms a key part of the proof that rotating perfect fluid 
stars are generically unstable in the presence of radiation 
\cite{friedman78:_secul_instab}. \ The argument goes as follows: Consider 
modes with finite frequency in the $\Omega \to 0$ limit. \ Then 
Eq.~(\ref{ineq1}) implies that co-rotating modes (with $\sigma_p>0$) must 
have $J_c>0$, while counter-rotating modes (for which $\sigma_p < 0$) will 
have $J_c<0$. \ In both cases $E_c>0$, which means that both classes of 
modes are stable. \ Now consider a small region near a point where 
$\sigma_p=0$ (at a finite rotation rate). \ Typically, this corresponds to 
a point where the initially counter-rotating mode becomes co-rotating. \ In 
this region $J_c<0$. \ However, $E_c$ will change sign at the point where 
$\sigma_p$ (or, equivalently, the frequency $\omega$) vanishes. \ Since the 
mode was stable in the non-rotating limit this change of sign indicates the 
onset of instability at a critical rate of rotation. 

\subsection{The r-mode instability}

In order to further demonstrate the usefulness of the canonical energy, 
let us prove the instability of the inertial r-modes. \ For a general inertial 
mode we have (cf.~\cite{lockitch99:_r-modes} who provide a discussion of the 
single fluid problem using notation which closely resembles the one we adopt 
here) 
\begin{equation}
     v^i \sim \delta v^i \sim \dot{\xi}^i \sim \Omega 
     \qquad \mbox{and} \qquad 
     \delta \Phi \sim \delta n \sim \Omega^2 \ . 
\end{equation} 
If we also assume axial-led modes, like the r-modes, then we have $\delta v_r 
\sim \Omega^2$ and the continuity equation leads to 
\begin{equation} 
    \nabla_i \delta v^i \sim \Omega^3 \rightarrow 
    \nabla_i \xi^i \sim \Omega^2 \ . 
\end{equation} 

Under these assumptions we find that $E_c$ becomes (to order $\Omega^2$) 
\begin{equation} 
    E_c \approx \frac{1}{2} \int \rho \left[  |\partial_t {\xi}|^2 
    - |v^i \nabla_i{\xi}|^2  +  \xi^{i*} \xi^{j} \nabla_i \nabla_j \left(
    \Phi + \tilde{\mu}\right)\right] {\rm d} V \ . \label{ec1} 
\end{equation} 
We can rewrite the last term using the equation governing the axisymmetric 
equilibrium. \ Keeping only terms of order $\Omega^2$ we have
\begin{equation}
    \xi^{i*} \xi^{j} \nabla_i\nabla_j \left(\Phi + \tilde{\mu}\right) 
    \approx \frac{1}{2} \Omega^2  \xi^{i*} \xi^{j} \nabla_i \nabla_j 
    (r^2 \sin^2 \theta) \ . 
\end{equation} 
A bit more work then leads to 
\begin{equation} 
    \frac{1}{2} \Omega^2  \xi^{i*} \xi^{j} \nabla_i \nabla_j 
    (r^2 \sin^2 \theta) = \Omega^2 r^2 \left[ \cos^2 \theta |\xi^\theta|^2 + 
    \sin^2\theta | \xi^\varphi|^2  \right] 
\end{equation} 
and 
\begin{eqnarray} 
    | v^i  \nabla_i \xi_j |^2 = \Omega^2 \left\{ m^2 | \xi |^2  - 2imr^2 
    \sin \theta \cos \theta \left[ \xi^\theta \xi^{\varphi *} - \xi^\varphi 
    \xi^{\theta *} \right] \right. \nonumber \\ 
    + \left. r^2 \left[ \cos^2 \theta |\xi^\theta|^2 + \sin^2\theta | 
    \xi^\varphi|^2 \right] \right\} 
\end{eqnarray} 
which means that the canonical energy can be written in the form 
\begin{eqnarray} 
    E_c \approx - \frac{1}{2} \int \rho \left\{ (m \Omega - \omega)(m \Omega 
    + \omega) | \xi |^2 \right. \qquad \qquad \nonumber \\ 
    \left. -  2 i m \Omega^2 r^2 \sin \theta \cos \theta \left[
    \xi^\theta \xi^{\varphi *} - \xi^\varphi \xi^{\theta *} \right]
    \right\} {\rm d} V 
\end{eqnarray}
for an axial-led mode. 

Introducing the axial stream function $U$ we have 
\begin{eqnarray} 
    \xi^\theta &=& - \frac{iU}{r^2 \sin \theta} \partial_\varphi Y_l^m 
    e^{i \omega t} \ , \\ 
    \xi^\varphi &=& \frac{iU}{r^2 \sin \theta} \partial_\theta Y_l^m 
    e^{i\omega t} \ , 
\end{eqnarray} 
where $Y_l^m=Y_l^m(\theta,\varphi)$ are the standard spherical harmonics. 
\ This leads to 
\begin{equation} 
    |\xi|^2 = \frac{|U|^2}{r^2} \left[  \frac{1}{\sin^2 \theta} 
    | \partial_\varphi Y_l^m |^2 + |\partial_\theta Y_l^m|^2 \right] 
\end{equation} 
and 
\begin{eqnarray} 
     ir^2 \sin \theta \cos \theta \left[ \xi^\theta \xi^{\varphi *} - 
     \xi^\varphi \xi^{\theta *} \right] \qquad \qquad \qquad \qquad 
     \nonumber \\ = \frac{1}{r^2} \frac{ \cos \theta}{\sin \theta} m |U|^2 
     \left[ Y_l^m \partial_\theta Y_l^{m*} +  Y_l^{m *} \partial_\theta 
     Y_l^{m}\right] \ . 
\end{eqnarray} 

After performing the angular integrals, we find that 
\begin{equation} 
    E_c = - \frac{ l(l+1) }{2} \left\{ (m \Omega - \omega)(m \Omega + \omega) 
          - \frac{2 m^2 \Omega^2}{l(l+1)} \right\} \int \rho |U|^2 {\rm d} r \ . 
\end{equation}
Combining this with the r-mode frequency \cite{lockitch99:_r-modes} 
\begin{equation} 
    \omega = m \Omega \left[ 1 - \frac{2}{l(l+1)} \right] 
\end{equation}
we see that $E_c < 0$ for all $l>1$ r-modes, i.e.~they are all unstable. \ 
The $l=m=1$ r-mode is a special case, leading to $E_c=0$. 

\subsection{The relativistic problem} 

The theoretical framework for studying stellar stability in general relativity 
was mainly developed during the 1970s, with key contributions from 
Chandrasekhar and Friedman \cite{cf72a,cf72b} and Schutz \cite{bfs72a,bfs72b}. 
\ Their work extends the Newtonian analysis discussed above. \ There are 
basically two reasons why a relativistic analysis is more complicated than the 
Newtonian one. \ Firstly, the problem is algebraically more complex because 
one must solve the Einstein field equations in addition to the fluid equations 
of motion. \ Secondly, one must account for the fact that a general 
perturbation will generate gravitational waves. \ The work culminated in a 
series of papers \cite{fs0,friedman78:_lagran,friedman78:_secul_instab,jf} 
in which the role that gravitational radiation plays in these problems was 
explained, and a foundation for subsequent research in this area was 
established. \ The main result was that gravitational radiation acts in the 
same way in the full theory as in a post-Newtonian analysis of the problem. \ 
If we consider a sequence of equilibrium models, then a mode becomes secularly 
unstable at the point where its frequency vanishes (in the inertial frame). 
\ Most importantly, the proof does not require the completeness of the modes 
of the system.  

\newpage

\section{Modelling dissipation}
\label{sec:viscosity}

Although the inviscid model provides a natural starting point for any 
investigation of the dynamics of a fluid system, the effects of various 
dissipative mechanisms are often key to the construction of a realistic 
model. \ Consider, for example, the case of neutron star oscillations and 
possible instabilities. \ While it is interesting from the conceptual point 
of view to establish that an instability (such as the gravitational-wave 
driven instability of the fundamental f-mode or the inertial r-mode 
discussed above) may be present in an ideal fluid, it is crucial to 
establish that the instability actually has opportunity to grow on a 
reasonably short timescale. \ To establish this, one must consider the most 
important damping mechanisms and work out whether they will suppress the 
instability or not. \ A recent discussion of these issues in the context of 
the r-mode instability can be found in \cite{narev}. 

From the point of view of relativistic fluid dynamics, it is clear already 
from the outset that we are facing a difficult problem. \ After all, the 
Fourier theory of heat conduction leads to instantaneous propagation of 
thermal signals. \ The fact that this non-causality is built into the 
description is unattractive already in the context of the Navier-Stokes 
equations. \ After all, one would expect heat to propagate at roughly the 
mean molecular speed in the system. \ For a relativistic description 
non-causal behaviour would be truly unacceptable. \ As work by Lindblom and 
Hiscock \cite{hislin83} has established, there is a deep connection between 
causality, stability, and hyperbolicity of a dissipative model. \ One would 
expect an acceptable model to be hyperbolic, not allowing signals to 
propagate superluminally. 

Our aim in this section is to discuss the three main models that exist in 
the literature. \ We first consider the classic work of Eckart 
\cite{eckart40:_rel_diss_fluid} and Landau and Lifshitz 
\cite{landau59:_fluid_mech}, which is based on a seemingly natural extension 
of the inviscid equations. \ However, a detailed analysis of Lindblom and 
Hiscock \cite{hiscock85:_rel_diss_fluids,hislin87} has demonstrated that 
these descriptions have serious flaws and must be considered unsuitable for 
practical use. \ However, having discussed these models it is relatively 
easy to extend them in the way proposed by Israel and Stewart 
\cite{stew77,Israel79:_kintheo2,Israel79:_kintheo1}. \ Their description, the 
derivation of which was inspired by early work of Grad \cite{grad} and 
M\"uller \cite{muller} and results from relativistic kinetic theory, 
provides a framework that is generally accepted as meeting the key criteria 
for a relativistic model \cite{hislin83}. \ Finally, we describe Carter's 
more recent approach to the problem. \ This model is elegant because it 
makes maximal use of a variational principle argument. \ The construction is 
also more general than that of Israel and Stewart. \ In particular, it shows 
how one would account for several dynamically independent interpenetrating 
fluid species. \ This extension is important for, for example, the 
consideration of relativistic superfluid systems.  

\subsection{The ``standard'' relativistic models}

It is natural to begin by recalling the equations that describe the evolution 
of a perfect fluid, see Sec.~\ref{sec:perfect_fluid}. \ For a single 
particle species, we have the number current $n^\mu$ which satisfies 
\begin{equation} 
    \nabla_\mu n^\mu = 0 \ , \qquad \mbox{ where } \qquad n^\mu = n u^\mu \ .  
\end{equation} 
The stress energy tensor 
\begin{equation} 
    T^\mu{}_\nu = p \perp^\mu_\nu + \rho u^\mu u_\nu \ , 
\end{equation} 
where $p$ and $\rho$ represent the pressure and the energy density, 
respectively, satisfies 
\begin{equation} 
     \nabla_\mu T^\mu{}_\nu = 0 \ . \label{divT} 
\end{equation} 
In order to account for dissipation we need to introduce additional fields. 
\ First introduce a vector $\nu^\mu$ representing particle diffusion. \ That 
is, let 
\begin{equation} 
    n^\mu \longrightarrow n u^\mu + \nu^\mu \ , 
\end{equation} 
and assume that the diffusion satisfies the constraint $u_\mu \nu^\mu = 0$. 
\ This simply means that it is purely spatial according to an observer moving 
with the particles in the inviscid limit, exactly what one would expect from 
a diffusive process. \ Next we introduce the heat flux $q^\mu$ and the 
viscous stress tensor, decomposed into a trace-part $\tau$ (not to be confused 
with the proper time) and a trace-free bit $\tau^{\mu \nu}$, such that 
\begin{equation} 
    T^{\mu\nu} \longrightarrow (p + \tau) \perp^{\mu\nu} + \rho u^\mu u^\nu 
    + q^{(\mu} u^{\nu)} + \tau^{\mu \nu}  \ , \label{Tdiss1} 
\end{equation} 
subject to the constraints 
\begin{equation} 
    u^\mu q_\mu = u^\mu \tau_{\mu\nu} = \tau^\mu_{\ \mu} = \tau_{\mu\nu} - 
    \tau_{\nu\mu} = 0 \ . 
\end{equation} 
That is, both the heat flux and the trace-less part of the viscous stress 
tensor are purely spatial in the matter frame, and $\tau^{\mu\nu}$ is also 
symmetric. \ So far, the description is quite general. \ The constraints 
have simply been imposed to ensure that the problem has the anticipated 
number of degrees of freedom. 

The next step is to deduce the permissible form for the additional fields 
from the second law of thermodynamics. \ The requirement that the total 
entropy must not decrease leads to the entropy flux $s^\mu$ having to be 
such that 
\begin{equation} 
    \nabla_\mu s^\mu \ge 0 \ . 
\end{equation}
Assuming that the entropy flux is a combination of all the available vectors, 
we have 
\begin{equation} 
    s^\mu = s u^\mu + \beta q^\mu  - \lambda \nu^\mu \ , 
\end{equation}
where $\beta$ and $\lambda$ are yet to be specified. \ It is easy to work out 
the divergence of this. \ Then using the component of Eq.~(\ref{divT}) along 
$u^\mu$, i.e. 
\begin{equation} 
    u_\mu \nabla_\nu T^{\mu\nu} = 0 \ , 
\end{equation} 
and the thermodynamic relation 
\begin{equation} 
    n \nabla_\mu x_\s = \frac{1}{T} \nabla_\mu \rho - \frac{p + \rho}{n T} 
    \nabla_\mu n \ , 
\end{equation} 
which follows from assuming the equation of state $s = s(\rho,n)$, and we 
recall that $x_\s=s/n$, one can show that 
\begin{eqnarray} 
     \nabla_\mu s^\mu &=& q^\mu \left( \nabla_\mu \beta - \frac{1}{T} u^\nu 
     \nabla_\nu u_\mu \right) +  \left( \beta - \frac{1}{T} \right) 
     \nabla_\mu q^\mu \nonumber \\ 
     &-& \left( x_\s + \lambda - \frac{p + \rho}{n T}\right) \nabla_\mu 
     \nu^\mu - \nu^\mu\nabla_\mu \lambda - \frac{\tau}{T} \nabla_\mu u^\mu 
     - \frac{\tau^{\mu \nu}}{T} \nabla_\mu u_\nu \ . \label{2ndlaw} 
\end{eqnarray} 
We want to ensure that the right-hand side of this equation is positive 
definite (or indefinite). \ An easy way to achieve this is to make the 
following identifications 
\begin{equation}
    \beta = 1/T \ , 
\end{equation} 
and 
\begin{equation} 
    \lambda = - x_\s + \frac{p + \rho}{n T} \ . 
\end{equation} 
Here we note that $\lambda = g/nT$ where $g$ is the Gibbs free energy 
density. \ We also identify 
\begin{equation} 
    \nu^\mu = - \sigma T^2 \perp^{\mu\nu} \nabla_\nu \lambda 
\end{equation}
where the ``diffusion coefficient'' $\sigma \ge 0$, and the projection is 
needed in order for the constraint $u_\mu \nu^\mu = 0$ to be satisfied. \ 
Furthermore, we can use 
\begin{equation} 
    \tau = -\zeta \nabla_\mu u^\mu 
\end{equation} 
where $\zeta \ge 0$ is the coefficient of bulk viscosity, and 
\begin{equation} 
    q^\mu = - \kappa T \perp^{\mu\nu} \left(\frac{1}{T} \nabla_\nu T + 
            u^\alpha \nabla_\alpha u_\nu \right) 
\end{equation} 
with $\kappa \ge 0$ the heat conductivity coefficient. \ To complete the 
description, we need to rewrite the final term in Eq.~(\ref{2ndlaw}). \ To 
do this it is useful to note that the gradient of the four-velocity can 
generally be written 
\begin{equation}
    \nabla_\mu u_\nu = \sigma_{\mu\nu} + \frac{1}{3} \perp_{\mu\nu} \theta + 
     \omega_{\mu\nu} - a_\nu u_\mu  
\end{equation}
where the acceleration is defined as 
\begin{equation} 
    a_\mu = u^\beta \nabla_\beta u_\mu \ , 
\end{equation} 
the expansion is $\theta=\nabla_\mu u^\mu$ and the shear is given by 
\begin{equation} 
    \sigma_{\mu\nu} = \frac{1}{2} \left( \perp^\alpha_\nu \nabla_\alpha 
    u_\mu + \perp^\alpha_\mu \nabla_\alpha u_\nu \right) - \frac{1}{3} 
    \perp_{\mu\nu} \theta  \ . 
\end{equation} 
Finally, the ``vorticity'' follows from\epubtkFootnote{It is important to note 
the difference between the vorticity formed from the momentum and the 
``vorticity'' in terms of the velocity. \ They differ because of the 
entrainment, and one can show that while the former is conserved along the 
flow, the latter is not.} 
\begin{equation}
    \varpi_{\mu\nu} = \frac{1}{2} \left( \perp^\alpha_\nu \nabla_\alpha u_\mu 
    - \perp^\alpha_\mu \nabla_\alpha u_\nu  \right)  \ . 
\end{equation} 
Since we want $\tau^{\mu\nu}$ to be symmetric, trace-free, and purely spatial 
according to an observer moving along $u^\mu$, it is useful to introduce the 
notation 
\begin{equation} 
    \left< A_{\mu\nu} \right> = \frac{1}{2} \perp^\rho_\mu \perp^\sigma_\nu 
    \left( A_{\rho\sigma} + A_{\sigma\rho} - \frac{2}{3} \perp_{\rho \sigma} 
    \perp^{\gamma\delta} A_{\gamma\delta}\right) \ , 
\end{equation} 
where $A_{\mu\nu}$ is a general two-form. \ In the case of the gradient of 
the four-velocity, it is easy to show that this leads to 
\begin{equation} 
    \left< \nabla_\mu u_\nu \right> = \sigma_{\mu\nu} 
\end{equation} 
and therefore it is natural to use 
\begin{equation} 
    \tau^{\mu\nu} = - \eta \sigma^{\mu\nu} \ , 
\end{equation} 
where $\eta \ge 0$ is the shear viscosity coefficient. \ Given these 
relations, we can write 
\begin{equation} 
    T \nabla_\mu s^\mu = \frac{q^\mu q_\mu}{\kappa T} + \frac{\tau}{\zeta} 
    + \frac{\nu^\mu \nu_\mu}{\sigma T^2 } + \frac{\tau^{\mu\nu} 
    \tau_{\mu\nu}}{2 \eta} \ge 0 \ . 
\end{equation} 
By construction, the second law of thermodynamics is satisfied. 

The model we have written down is quite general. \ In particular, it is 
worth noticing that we did not yet specify the four-velocity $u^\mu$. \ By 
doing this we can obtain from the above equations both the formulation due to 
Eckart \cite{eckart40:_rel_diss_fluid} and that of Landau and Lifshitz 
\cite{landau59:_fluid_mech}. \ To arrive at the Eckart description, we 
associate $u^\mu$ with the flow of particles. \ Thus we take $\nu^\mu=0$ 
(or equivalently $\sigma =0$). \ This prescription has the advantage of 
being easy to implement. \ The Landau and Lifshitz model follows if we 
choose the four velocity to be a timelike eigenvector of the stress-energy 
tensor. \ From Eq.~(\ref{Tdiss1}) it is easy to see that, by setting 
$q^\mu = 0$ we get 
\begin{equation} 
    u_\mu T^{\mu\nu} = - \rho u^\nu \ . 
\end{equation} 
This is equivalent to setting $\kappa = 0$. \ Unfortunately, these models, 
which have been used in most applications of relativistic dissipation to 
date, are not at all satisfactory. \ While they pass the key test set by 
the second law of thermodynamics, they fail several other requirements of a 
relativistic description. \ A detailed analysis of perturbations away from 
an equilibrium state \cite{hiscock85:_rel_diss_fluids} demonstrates serious 
pathologies. \ The dynamics of small perturbations tends to be dominated 
by rapidly growing instabilities. \ This suggests that these formulations 
are likely to be practically useless. \ From the mathematical point of view 
they are also not acceptable since, being non-hyperbolic, they do not admit 
a well-posed initial-value problem. 

\subsection{The Israel-Stewart approach}

From the above discussion we know that the most obvious strategy for 
extending relativistic hydrodynamics to include dissipative processes leads 
to unsatisfactory results. \ At first sight, this may seem a little bit 
puzzling because the approach we took is fairly general. \ Yet, the 
formulation suffers from pathologies. \ Most importantly, we have not managed 
to enforce causality. \ Let us now explain how this problem can be solved. 

Our previous analysis was based on the assumption that the entropy current 
$s^\mu$ can be described as a linear combination of the various fluxes in 
the system, the four-velocity $u^\mu$, the heat-flux $q^\mu$ and the diffusion 
$\nu^\mu$. \ In a series of papers, Israel and Stewart  
\cite{stew77,Israel79:_kintheo2,Israel79:_kintheo1} contrasted this 
``first-order'' theory with relativistic kinetic theory. \ Following early 
work by M\"uller \cite{muller} and connecting with Grad's 14-moment kinetic 
theory description \cite{grad}, they concluded that a satisfactory model 
ought to be ``second order'' in the various fields. \ If we, for simplicity, 
work in the Eckart frame (cf.~Lindblom and Hiscock \cite{hislin83}) this 
means that we would use the Ansatz 
\begin{equation}
    s^\mu = s u^\mu + \frac{1}{T} q^\mu - \frac{1}{2 T} \left(\beta_0 \tau^2 
    + \beta_1 q_\alpha q^\alpha + \beta_2 \tau_{\alpha\beta} 
    \tau^{\alpha \beta} \right) u^\mu + \frac{\alpha_0 \tau q^\mu}{T} + 
    \frac{\alpha_1 \tau^{\mu}_{\ \nu} q^\nu}{T} \ . \label{2ndorder} 
\end{equation} 
This expression is arrived at by asking what the most general form of a 
vector constructed from all the various fields in the problem may be. \ Of 
course, we now have a number of new (so far unknown) parameters. \ The three 
coefficients $\beta_0$, $\beta_1$, and $\beta_2$ have a thermodynamical 
origin, while the two coefficients $\alpha_0$ and $\alpha_1$ represent the 
coupling between viscosity and heat flow. \ From the above expression, we see 
that in the frame moving with $u^\mu$  the effective entropy density is 
given by 
\begin{equation} 
    - u_\mu s^\mu =  s - \frac{1}{2 T} \left( \beta_0 \tau^2 + \beta_1 
    q_\alpha q^\alpha + \beta_2 \tau_{\alpha\beta} \tau^{\alpha \beta}\right) 
    \ . 
\end{equation} 
Since we want the entropy to be maximized in equilibrium, when the extra 
fields vanish, we should have $[\beta_0, \beta_1, \beta_2]\ge0$. \ We also 
see that the entropy flux, 
\begin{equation} 
     \perp^\mu_\nu s^\nu = \frac{1}{T} \left[ (1 +  \alpha_0 \tau) q^\mu + 
     \alpha_1 \tau^{\mu\nu} q_\nu \right] \ , 
\end{equation}
is affected only by the parameters $\alpha_0$ and $\alpha_1$. 

Having made the assumption Eq.~(\ref{2ndorder}) the rest of the calculation 
proceeds as in the previous case. \ Working out the divergence of the entropy 
current, and making use of the equations of motion, we arrive at 
\begin{eqnarray} 
    \nabla_\mu s^\mu &=& - { 1 \over T} \tau \left[ \nabla_\mu u^\mu + 
    \beta_0 u^\mu \nabla_\mu \tau - \alpha_0 \nabla_\mu q^\mu - \gamma_0 T 
    q^\mu \nabla_\mu \left( \frac{\alpha_0}{T} \right) + \frac{\tau T}{2} 
    \nabla_\mu \left( {\beta_0 u^\mu \over T} \right) \right] \nonumber \\ 
    &-& \frac{1}{T} q^\mu \left[ \frac{1}{T} \nabla_\mu T + u^\nu 
    \nabla_\nu u_\mu + \beta_1 u^\nu \nabla_\nu q_\mu - \alpha_0 \nabla_\mu 
    \tau - \alpha_1 \nabla_\nu \tau^\nu_{\ \mu} \right. \nonumber \\ 
    &+& \left. \frac{T}{2} q_\mu \nabla_\nu \left(\frac{\beta_1 u^\nu}{T} 
    \right) - (1 - \gamma_0) \tau T \nabla_\mu \left( \frac{\alpha_0}{T} 
    \right) - (1 - \gamma_1) T \tau^\nu_{\ \mu} \nabla_\nu \left( 
    \frac{\alpha_1}{T}\right)\right] \nonumber \\ 
    &-&  \frac{1}{T} \tau^{\mu\nu} \left[ \nabla_\mu u_\nu + \beta_2 u^\alpha 
    \nabla_\alpha \tau_{\mu\nu} - \alpha_1 \nabla_\mu q_\nu + \frac{T}{2}
    \tau_{\mu\nu} \nabla_\alpha \left( \frac{\beta_2 u^\alpha}{T} \right)
    \right. \nonumber \\ 
    &-& \left. \gamma_1 T q_\mu \nabla_\nu \left( \frac{\alpha_1}{T} \right) 
    \right] \ . 
\end{eqnarray} 
In this expression it should be noted that we have introduced (following 
Lindblom and Hiscock) two further parameters, $\gamma_0$ and $\gamma_1$. \ 
They are needed because, without additional assumptions it is not clear how 
the ``mixed'' quadratic term should be distributed. \ A natural way to fix 
these parameters is to appeal to the Onsager symmetry principle  
\cite{Israel79:_kintheo1}, which leads to the mixed terms being distributed 
``equally'' and hence $\gamma_0 = \gamma_1 = 1/2$. 

Denoting the comoving derivative by a dot, i.e.~using $u^\mu \nabla_\mu \tau 
= \dot{\tau}$ etcetera, we see that the second law for thermodynamics is 
satisfied if we choose 
\begin{equation} 
    \tau = - \zeta \left[ \nabla_\mu u^\mu + \beta_0 \dot{\tau} - \alpha_0 
    \nabla_\mu q^\mu - \gamma_0 T q^\mu \nabla_\mu \left( \frac{\alpha_0}{T} 
    \right) + \frac{\tau T}{2} \nabla_\mu \left( \frac{\beta_0 u^\mu}{T}  
    \right) \right] \ , 
\end{equation} 
\begin{eqnarray} 
    q^\mu &=& -\kappa T \perp^{\mu\nu} \left[ \frac{1}{T} \nabla_\nu T +  
    \dot{u}_\nu + \beta_1 \dot{q}_\nu - \alpha_0 \nabla_\nu \tau 
    - \alpha_1 \nabla_\alpha \tau^\alpha_{\ \nu} +  \frac{T}{2} q_\nu 
    \nabla_\alpha \left(\frac{\beta_1 u^\alpha}{T} \right)  \right. \nonumber 
    \\ 
    &-& \left. (1 - \gamma_0) \tau T \nabla_\nu \left( \frac{\alpha_0}{T} 
    \right) - (1 - \gamma_1) T \tau^\alpha_{\ \nu} \nabla_\alpha \left( 
    \frac{\alpha_1}{T}\right) + \gamma_2 \nabla_{[\nu}u_{\alpha]} q^\alpha
    \right] \ , 
\end{eqnarray} 
and 
\begin{eqnarray} 
    \tau_{\mu\nu} &=& - 2 \eta \left[  \beta_2 \dot{\tau}_{\mu\nu} + 
    \frac{T}{2} \tau_{\mu\nu} \nabla_\alpha \left( \frac{\beta_2 u^\alpha}{T} 
    \right) \right. \nonumber \\ 
    &+& \left. \left< \nabla_\mu u_\nu - \alpha_1 \nabla_\mu q_\nu - \gamma_1 
    T q_\mu \nabla_\nu \left( \frac{\alpha_1}{T} \right) + \gamma_3 
    \nabla_{[\mu} u_{\alpha]} \tau_\nu^{\ \alpha}  \right> \right] 
\end{eqnarray} 
where the angular brackets denote symmetrization as before. \ In these 
expression we have added yet another two terms, representing the coupling to 
vorticity. \ These bring further ``free'' parameters $\gamma_2$ and 
$\gamma_3$. \ It is easy to see that we are allowed to add these terms since 
they do not affect the entropy production. \ In fact, a large number of 
similar terms may, in principle, be considered (see note added in proof in 
\cite{hislin83}). \ The presence of coupling terms of the particular form 
that we have introduced is suggested by kinetic theory  
\cite{Israel79:_kintheo1}. 


What is clear from these complicated expressions is that we now have 
evolution equations for the dissipative fields. \ Introducing characteristic 
``relaxation'' times
\begin{equation}
    t_0 = \zeta \beta_0 \ , \quad t_1 = \kappa \beta_1 \ , \quad \mbox{and} 
    \quad t_2 = 2 \eta \beta_2
\end{equation}
the above equations can be written
\begin{eqnarray} 
    t_0 \dot{\tau} + \tau &=& -\zeta [ ...] \ , \\ 
    t_1 \perp^{\mu\nu} \dot{q}^\nu + q^\mu &=& -\kappa T \perp^{\mu\nu} [ ...] 
    \ , \\
    t_2 \dot{\tau}_{\mu\nu} + \tau_{\mu\nu} &=& - 2\eta [...] \ .
\end{eqnarray}

A detailed stability analysis by Hiscock and Lindblom \cite{hislin83} shows 
that the Israel-Stewart theory is causal for stable fluids. \ Then the 
characteristic velocities are subluminal and the equations form a hyperbolic 
system. \ An interesting aspect of the analysis concerns the potentially 
stabilizing role of the extra parameters ($\beta_0,...,\alpha_0,...$). \ 
Relevant discussions of the implications for the nuclear equation of state 
and the maximum mass of neutron stars have been provided by Olson and 
Hiscock \cite{olshis89,ols01}. \ A more detailed mathematical stability 
analysis can be found in the work of Kreiss et al \cite{kreiss}. 

Although the Israel-Stewart model resolves the problems of the first-order 
descriptions for near equilibrium situations, difficult issues remain to be 
understood for nonlinear problems. \ This is highlighted in work by Hiscock 
and Lindblom \cite{hislin88} and Olson and Hiscock \cite{olshis89b}. \ They 
consider nonlinear heat conduction problems and show that the Israel-Stewart 
formulation becomes non-causal and unstable for sufficiently large deviations 
from equilibrium. \ The problem appears to be more severe in the Eckart frame 
\cite{hislin88} than in the frame advocated by Landau and Lifshitz 
\cite{olshis89b}. \ The fact that the formulation breaks down in nonlinear 
problems is not too surprising. \ After all, the basic foundation is a 
``Taylor expansion'' in the various fields. \ However, it raises important 
questions. \ There are many physical situations where a reliable nonlinear 
model would be crucial, eg.~heavy-ion collisions and supernova core 
collapse. \ This problem requires further thought. 

\subsection{Carter's canonical framework} 

The most recent attempt to construct a relativistic formalism for dissipative 
fluids is due to Carter \cite{carter91}. \ His approach is less 
``phenomenological'' than the ones we have considered so far in that it is 
based on making maximal use of variational principle arguments. \ The 
construction is also extremely general. \ On the one hand this makes it more 
complex. \ On the other hand this generality could prove useful in more 
complicated cases, eg.~for investigations of multi-fluid dynamics and/or 
elastic media. \ Given the potential that this formalism has for future 
applications, it is worth discussing it in detail. 

The overall aim is to generalize the variational formulation described in 
Sec.~\ref{sec:pullback} in such a way that viscous ``stresses'' are accounted 
for. \ Because the variational foundations are the same, the number currents 
$n_\X^\mu$ play a central role ($\X$ is a constituent index as before). \ In 
addition we can introduce a number of viscosity tensors 
$\tau^{\mu\nu}_\Sigma$, which we assume to be symmetric (even though it is 
clear that such an assumption is not generally correct 
\cite{andersson05:_flux_con}). \ The index $\Sigma$ is ``analogous'' to the 
constituent index, and represents different viscosity contributions. \ It is 
introduced in recognition of the fact that it may be advantageous to consider 
different kinds of viscosity, eg.~bulk- and shear viscosity, separately. \ As 
in the case of the constituent index, a repeated index $\Sigma$ does not 
imply summation. 

The key quantity in the variational framework is the master function 
$\Lambda$. \ As it is a function of all the available fields, we will now have 
$\Lambda(n_\X^\mu, \tau_\Sigma^{\mu\nu}, g_{\mu\nu})$. \ Then a general 
variation leads to
\begin{equation}
    \delta \Lambda = \sum_\X \mu_\rho^\X \delta n_\X^\rho + \frac{1}{2} 
    \sum_\Sigma \pi^\Sigma_{\mu \nu} \delta \tau_\Sigma^{\mu\nu} 
    + \frac{\partial \Lambda}{\partial g^{\mu\nu}} \delta g^{\mu\nu} \ . 
\end{equation}
Since the metric piece is treated in the same way as in the non-dissipative 
problem we will concentrate on the case $\delta g^{\mu\nu}=0$ here. \ This is, 
obviously, no real restriction since we already know how the variational 
principle couples in the Einstein equations in the general case. \ In the 
above formula we recognize the momenta $\mu^\X_\rho$ that are conjugate to 
the fluxes. \ We also have an analogous set of ``strain'' variables which are 
defined by 
\begin{equation} 
    \pi^\Sigma_{\mu\nu} = \pi^\Sigma_{(\mu\nu)} = 2 \frac{\partial 
    \Lambda}{\partial \tau^{\mu\nu}_\Sigma} \ . 
\end{equation} 

As in the non-dissipative case, the variational framework suggests that the 
equations of motion can be written as a force-balance equation; 
\begin{equation} 
    \nabla_\mu T^\mu_{\nu} = \sum_\X f_\nu^\X + \sum_\Sigma f^\Sigma_\nu = 0 
\end{equation} 
where the generalized forces work out to be 
\begin{equation} 
    f_\rho^\X = \mu_\rho^\X \nabla_\sigma n_\X^\sigma + n_\X^\sigma 
    \nabla_{[\sigma}\mu_{\rho]}^\X 
\end{equation}  
and 
\begin{equation} 
    f_\rho^\Sigma = \pi_{\rho\nu}^\Sigma \nabla_\sigma 
    \tau_\Sigma^{\sigma\nu} + \tau_\Sigma^{\sigma\nu} 
    \left( \nabla_\sigma \pi^\Sigma_{\rho\nu} - \frac{1}{2} \nabla_\rho 
    \pi^\Sigma_{\sigma\nu} \right) \ . 
\end{equation} 
Finally, the stress-energy tensor becomes 
\begin{equation} 
    T^\mu{}_\nu = \Psi g^\mu{}_\nu + \sum_\X \mu_\nu n_\X^\mu + 
    \sum_\Sigma \tau_\Sigma^{\mu\sigma} \pi^\Sigma_{\sigma \nu} 
\end{equation} 
with the generalized pressure given by 
\begin{equation} 
    \Psi =  \Lambda - \sum_\X \mu_\rho^\X n^\rho_\X - \frac{1}{2} 
    \sum_\Sigma \tau_\Sigma^{\rho\sigma} \pi^\Sigma_{\rho\sigma} \ .
\end{equation}

For reasons that will become clear shortly, it is useful to introduce a set 
of ``convection vectors''. \ In the case of the currents, these are taken as 
proportional to the fluxes. \ This means that we introduce $\beta_\X^\mu$ 
according to 
\begin{equation} 
    h_\X \beta_\X^\mu = n_\X^\mu \ , \quad \mu^\X_\nu \beta_\X^\nu = -1 
    \quad \longrightarrow \quad  h_\X = - \mu_\nu^\X n_\X^\nu \ . 
\end{equation} 
With this definition we can introduce a projection operator 
\begin{equation} 
    \perp_\X^{\mu\nu} = g^{\mu\nu} + \mu_\X^\mu \beta_\X^\nu \quad 
    \longrightarrow \quad \perp_{\X\nu}^\mu \beta_\X^\nu = 
    \perp_\X^{\mu\nu} \mu^\X_\nu = 0 \ . 
\end{equation} 
From the definition of the force density $f_\X^\mu$ we can now show that 
\begin{equation} 
    \nabla_\mu n_\X^\mu = - \beta_\X^\mu f^\X_\mu 
\end{equation} 
and 
\begin{equation}  
    h_\X {\cal L}_\X  \mu_\nu^\X = \perp^\X_{\nu\sigma} f_\X^\sigma 
\end{equation} 
where the Lie-derivative along $\beta_\X^\mu$ is ${\cal L}_\X = 
{\cal L}_{\beta_\X^\mu}$. \ We see that the component of the force which is 
parallel to the convection vector $\beta_\X^\mu$ is associated with particle 
conservation. \ Meanwhile, the orthogonal component represents the ``change 
in momentum'' along $\beta_\X^\mu$. 

In order to facilitate a similar decomposition for the viscous stresses, it 
is natural to choose the conduction vector as a unit null eigenvector 
associated with $\pi_\Sigma^{\mu\nu}$. \ That is, we take 
\begin{equation} 
    \pi^\Sigma_{\mu\nu} \beta_\Sigma^\nu = 0 
\end{equation} 
together with
\begin{equation}
    u_\mu^\Sigma = g_{\mu\nu} \beta_\Sigma^\nu \quad \mbox{ and } \quad 
    u_\mu^\Sigma \beta_\Sigma^\mu = - 1 \ . 
\end{equation}
Introducing the projection associated with this conduction vector, 
\begin{equation}
    \perp^\Sigma_{\mu\nu} = g_{\mu\nu} + u^\Sigma_\mu u^\Sigma_\nu  
    \quad \longrightarrow \quad \perp^\Sigma_{\mu\nu} \beta_\Sigma^\nu = 0 
    \ . 
\end{equation} 
Once we have defined $\beta_\Sigma^\mu$, we can use it to reduce the degrees 
of freedom of the viscosity tensors. \ So far, we have only required them to 
be symmetric. \ However, in the standard case one would expect a viscous 
tensor to have only six degrees of freedom. \ To ensure that this is the case 
in the present framework we introduce the following degeneracy condition 
\begin{equation} 
    u_\mu^\Sigma \tau_\Sigma^{\mu\nu} = 0 \ . 
\end{equation}
That is, we require the viscous tensor $\tau_\Sigma^{\mu\nu}$ to be purely 
spatial according to an observer moving along $u^\mu_\Sigma$. \ With these
definitions one can show that 
\begin{equation}
    \beta_\Sigma^\mu {\cal L}_\Sigma \pi^\Sigma_{\mu\nu} = 0 
\end{equation}
where ${\cal L}_\Sigma= {\cal L}_{\beta_\Sigma^\mu}$ is the Lie-derivative 
along $\beta_\Sigma^\mu$, and 
\begin{equation}
    \tau_\Sigma^{\mu\nu} {\cal L}_\Sigma \pi^\Sigma_{\mu\nu} = - 2 
    \beta_\Sigma^\mu f^\Sigma_\mu \ . 
\end{equation} 

Finally, let us suppose that we choose to work in a given frame, moving with 
four-velocity $u^\mu$ (and that the associated projection operator 
$\perp^\mu_\nu$). \ Then we can use the following decompositions for the 
conduction vectors 
\begin{equation} 
    \beta_\X^\mu = \beta_\X(u^\mu + v_\X^\mu) \quad \mbox{and} \quad 
    \beta_\Sigma^\mu = \beta_\Sigma(u^\mu + v_\Sigma^\mu) \ . 
\end{equation}
We see that $\mu^\X = 1/\beta_\X$ represents a chemical type potential for 
species $\X$ with respect to the chosen frame. \ At the same time, it is 
easy to see that $\mu^\Sigma= 1/\beta_\Sigma$ is a Lorentz contraction 
factor. \ Using the norm of $\beta^\mu_\Sigma$ we have 
\begin{equation} 
    \beta^\mu_\Sigma \beta^\Sigma_\mu = - \beta^2_\Sigma \left(1 - 
    v_\Sigma^2\right) = - 1 
\end{equation} 
where $v_\Sigma^2 = v_\Sigma^\mu v^\Sigma_\mu$. \ Thus 
\begin{equation}
    \mu^\Sigma = 1/\beta_\Sigma = \sqrt{ 1 - v_\Sigma^2} 
\end{equation}
clearly analogous to the standard Lorentz contraction formula. 

So far the construction is quite formal, but we are now set to make contact 
with the physics. \ First we note that the above results allow us to
demonstrate that 
\begin{equation} 
    u^\sigma \nabla_\rho T^\rho_{\ \sigma} + \sum_\X \left( \mu^\X 
    \nabla_\sigma n_\X^\sigma + v_\X^\sigma f^\X_\sigma\right) 
    + \sum_\Sigma \left( v_\Sigma^\sigma f^\Sigma_\sigma + \frac{1}{2} 
    \mu^\Sigma \tau_\Sigma^{\rho\sigma} {\cal L}_\Sigma 
    \pi^\Sigma_{\rho\sigma} \right) = 0 \ . 
\end{equation} 
Recall that similar results were central to expressing the second law of 
thermodynamics in the previous two sections. \ To see how things work out 
in the present formalism, let us single out the entropy fluid (with index 
$\s$) by defining $s^\mu = n_\s^\mu$ and $T = \mu_\s$. \ To simplify the 
final expressions it is also useful to assume that the remaining species 
are governed by conservation laws of the form 
\begin{equation} 
    \nabla_\mu n_\X^\mu = \Gamma_\X 
\end{equation} 
subject to the constraint of total baryon conservation 
\begin{equation} 
    \sum_{\X\neq \s} \Gamma_\X = 0 \ . 
\end{equation} 
Given this, and the fact that the divergence of the stress-energy tensor 
should vanish, we have 
\begin{equation}
    T \nabla_\mu s^\mu = - \sum_{\X \neq\s} \mu^\X \Gamma_\X - \sum_\X 
    v_\X^\mu f^\X_\mu - \sum_\Sigma \left(v_\Sigma^\mu f^\Sigma_\mu + 
    \frac{1}{2} \mu^\Sigma \tau^{\mu\nu}_\Sigma {\cal L}_\Sigma 
    \pi^\Sigma_{\mu\nu} \right) \ . 
\end{equation}
Finally, we can bring the remaining two force contributions together by 
introducing the linear combinations 
\begin{equation}
    \sum_\X \zeta^\X v_\X^\mu = 0 \quad \mbox{and} \quad \sum_\X 
    \zeta^\X_\Sigma v_\X^\mu = v_\Sigma^\mu 
\end{equation} 
constrained by 
\begin{equation} 
    \sum_\X \zeta^\X = \sum_\X \zeta^\X_\Sigma = 1 \ . 
\end{equation}
Then defining 
\begin{equation} 
    \tilde{f}^\X_\nu = f^\X_\nu + \sum_\Sigma \zeta^\X_\Sigma f^\Sigma_\nu 
\end{equation} 
we have
\begin{equation} 
    T \nabla_\mu s^\mu = - \sum_{\X \neq \s} \mu^\X \Gamma_\X - 
    \sum_\X v_\X^\mu \tilde{f}^\X_\mu - \frac{1}{2} \sum_\Sigma 
    \mu^\Sigma \tau^{\mu\nu}_\Sigma {\cal L}_\Sigma \pi^\Sigma_{\mu\nu} \ge 0
    \label{2nd_final} \ . 
\end{equation} 
The three terms in this expression represent, respectively, the entropy 
increase due to i) chemical reactions, ii) conductivity, and iii) viscosity. 
\ The simplest way to ensure that the second law of thermodynamics is 
satisfied is to make each of the three terms positive definite. 

At this point, the general formalism must be completed by some suitably 
simple model for the various terms. \ A reasonable starting point would be 
to assume that each term is linear. \ For the chemical reactions this would 
mean that we expand each $\Gamma_\X$ according to 
\begin{equation}
    \Gamma_\X = - \sum_{\Y\neq s} {\cal C}_{\X\Y} \mu^\Y 
\end{equation} 
where ${\cal C}_{\X\Y}$ is a positive definite (or indefinite) matrix 
composed of the various reaction rates. \ Similarly, for the conductivity 
term it is natural to consider ``standard'' resistivity such that 
\begin{equation} 
    \tilde{f}^\X_\rho = - \sum_{\Y} {\cal R}^{\X\Y}_{\rho\sigma} v_\Y^\sigma 
    \ . 
\end{equation} 
Finally, for the viscosity we can postulate a law of form 
\begin{equation}
    \tau_\Sigma^{\mu\nu} = - \eta^{\mu\nu\rho\sigma}_\Sigma {\cal L}_\Sigma 
    \pi^\Sigma_{\rho\sigma}  \ , 
\end{equation}
where we would have, for an isotropic model, 
\begin{equation}
    \eta^{\mu\nu\rho\sigma}_\Sigma = \eta_\Sigma \perp_\Sigma^{\mu(\rho} 
    \perp_\Sigma^{\sigma)\nu} + \frac{1}{3} (\eta_\Sigma - \zeta_\Sigma) 
    \perp_\Sigma^{\mu\nu}\perp_\Sigma^{\rho\sigma} 
\end{equation} 
where the coefficients $\eta_\Sigma$ and $\zeta_\Sigma$ can be recognized as 
representing shear- and bulk viscosity, respectively. 

A detailed comparison between Carter's formalism and the Israel-Stewart 
framework has been carried out by Priou \cite{priou91}. \ He concludes that 
the two models, which are both members of a large family of dissipative 
models, have essentially the same degree of generality and that they are 
equivalent in the limit of linear perturbations away from a thermal 
equilibrium state. \ Providing explicit relations between the main parameters 
in the two descriptions, he also emphasizes the key point that analogous 
parameters may not have the same physical interpretation. 

In developing his theoretical framework, Carter argued in favour of an 
``off the peg'' model for heat conducting models \cite{carter_heat}. \ This 
model is similar to that introduced in Sec.~\ref{sec:twofluids}, and was 
intended as a simple, easier to use alternative to the Israel-Stewart 
construction. \ In the particular example discussed by Carter, he chooses to 
set the entrainment between particles and entropy to zero. \ This was done in 
order to simplify the discussion. \ But, as a discussion by Olson and Hiscock 
\cite{olshis90} shows, it has disastrous consequences. \ The resulting model 
violates causality in two simple model systems. \ As discussed by Priou 
\cite{priou91} and Carter and Khalatnikov \cite{carter92:_momen_vortic_helic}, 
this breakdown emphasizes the importance of the entrainment effect in these 
problems. 

\subsection{Remaining issues}

We have discussed some of the models that have been constructed in order to 
incorporate dissipative effects in a relativistic fluid description. \ We 
have seen that the most obvious ways of doing this, the ``text-book'' 
approach of Eckart \cite{eckart40:_rel_diss_fluid} and Landau and Lifshitz 
\cite{landau59:_fluid_mech}, fail completely. \ They do not respect causality 
and have serious stability problems. \ We have also described how this 
problem can be fixed by introducing additional dynamical fields. \ We 
discussed the formulations of Israel and Stewart 
\cite{stew77,Israel79:_kintheo2,Israel79:_kintheo1} and Carter 
\cite{carter91} in detail. \ From our discussion it should be clear that 
these models are  examples of an extremely large family of possible theories 
for dissipative relativistic fluids. \ Given this wealth of possibilities, 
can we hope to find the ``right'' model? \ To some extent, the answer to this 
question relies on the extra parameters one has introduced in the theory. \ 
Can they be constrained by observations? \ This question has been discussed 
by Geroch \cite{ger95} and Lindblom \cite{lind96}. \ The answer seems to be 
no, we should not expect to be able to use observations to single out a 
preferred theoretical description. \ The reason for this is that the different 
models relax to the Navier-Stokes form on very short timescales. \ Hence, one 
will likely only be able to constrain the standard shear- and bulk viscosity 
coefficients etc. \ Related questions concern the practicality of the 
different proposed schemes. \ To a certain extent, this is probably a matter 
of taste. \ Of course, it may well be that the additional parameters 
required in a particular model are easier to extract from microphysics 
arguments. \ This could make this description easier to use in practice, 
which would be strong motivation for preferring it. \ Of course, there is no 
guarantee that the same formulation will be ideal for all circumstances. 
\ Clearly, there is scope for a lot more research in this problem area. 

\newpage

\section{Heavy ion collisions} 
\label{heavyion} 

Relativistic fluid dynamics has regularly been used as a tool to model heavy 
ion collisions. \ The idea of using hydrodynamics to study the process of 
multiparticle production in high-energy hadron collisions can be traced back 
to work by, in particular, Landau in the early 1950s, see \cite{bel_lan}. \ 
In the early days these phenomena were observed in cosmic rays. \ The idea to 
use hydrodynamics was resurrected as collider data became available 
\cite{carrut} and early simulations were carried out at Los Alamos 
\cite{amsden1,amsden2}. \ More recently, modelling has primarily been focussed 
on reproducing data from, for example, CERN. \ A useful review of this active 
area of research can be found in \cite{Clare_Strottman}. 

In a hydrodynamics based model, a high-energy nuclear collision is viewed in 
the following way: In the centre of mass frame two Lorentz contracted nuclei 
collide and, after a complex microscopic process, a hot dense plasma is 
formed. \ In the simplest description this matter is assumed to be in local 
thermal equilibrium. \ The initial thermalization phase is, of course, out of 
reach for hydrodynamics. \ In the model, the state of matter is simply 
specified by the initial conditions, eg.~in terms of distributions of fluid 
velocities and thermodynamical quantities. \ Then follows a hydrodynamical 
expansion, which is described by the standard conservation equations for 
energy/momentum, baryon number and other conserved quantities, such as 
strangeness, isotopic spin, etc.~(see \cite{elze} for a variational 
principle derivation of these equations). \ As the expansion proceeds, the 
fluid cools and becomes increasingly rarefied. \ This eventually leads to the 
decoupling of the constituent particles, which then do not interact until 
they reach the detector. 

Fluid dynamics provides a well defined framework for studying the stages 
during which matter becomes highly excited and compressed and, later, expands 
and cools down. \ In the final stage when the nuclear matter is so dilute 
that nucleon-nucleon collisions are infrequent, hydrodynamics ceases to be 
valid. \ At this point additional assumptions are necessary to predict the 
number of particles, and their energies, which may be formed (to be compared 
to  data obtained from the detector). \ These are often referred to as the 
``freeze-out'' conditions. \ The problem is complicated by the fact that 
the ``freeze-out'' typically occurs at a different time for each fluid cell. 

Even though the application of hydrodynamics in this area has led to 
useful results, it is clear that the theoretical foundation for this 
description is not a trivial matter. \ Basically, the criteria required for 
the equations of hydrodynamics to be valid are 1) many degrees of freedom in 
the system, 2) a short mean free path, 3) a short mean stopping length, 4) a 
sufficient reaction time for thermal equilibration, and 5) a short de Broglie 
wavelength (so that quantum mechanics can be ignored). \ An interesting 
aspect of the hydrodynamic models is that they make use of concepts largely 
outside traditional nuclear physics: e.g., thermodynamics, statistical 
mechanics, fluid dynamics, and of course elementary particle physics. \ This 
is natural since the very hot, highly excited matter has a large number of 
degrees of freedom. \ But it is also a reflection of the basic lack of 
knowledge. \ As the key dynamics is uncertain, it is comforting to resort to 
standard principles of physics like the conservation of momentum and energy. 

Another key reason why hydrodynamic models are favoured is the simplicity 
of the input. \ Apart from the initial conditions which specify the masses 
and velocities, one needs only an equation of state and an Ansatz for the 
thermal degrees of freedom. \ If one includes dissipation one must in 
addition specify the form and magnitude of the viscosity and heat 
conduction. \ The fundamental conservation laws are incorporated into the 
Euler equations. In return for this relatively modest amount of input, one 
obtains the differential cross sections of all the final particles, the 
composition of clusters, etc. \ Of course, before one can confront the 
experimental data, one must make additional assumptions about the freeze-out, 
chemistry, etc. \ A clear disadvantage of the hydrodynamics model is that 
much of the microscopic dynamics is lost. 

Let us discuss some specific aspects of the hydrodynamics that has been used 
in this area. \ As we will recognize, the issues that need to be addressed 
for heavy-ion collisions are very similar to those faced in studies of 
relativistic dissipation theory and multi-fluid modelling. \ The one key 
difference is that the problem only requires special relativity, there is no 
need to worry about the spacetime geometry. \ Of course, it is still 
convenient to use a fully covariant description since one is then not tied 
down to the use of a particular set of coordinates. 

In many studies of heavy ions a particular frame of reference is chosen. \ As 
we know from our discussion of dissipation and causality, see 
Sec.~\ref{sec:viscosity}, this is an issue that must be approached with some 
care. \ In the context of heavy-ion collisions it is common to choose $u^\mu$ 
as the velocity of either energy transport (the Landau-Lifshitz frame) or 
particle transport (the Eckart frame). \ It is recognized that the Eckart 
formulation is somewhat easier to use and that one can let $u^\mu$ be either 
the velocity of nucleon or baryon number transport. \ On the other hand, 
there are cases where the Landau-Lifshitz picture has been viewed as more 
appropriate. \ For instance, when ultrarelativistic nuclei collide they 
virtually pass through one another leaving the vacuum between them in a 
highly excited state causing the creation of numerous particle-antiparticle 
pairs. \ Since the net baryon number in this region vanishes, the Eckart 
definition of the four-velocity cannot easily  be employed. \ This discussion 
is a clear reminder of the situation for viscosity in relativity, and the 
resolution is likely the same. \ A true frame-independent description will 
need to include several distinct fluid components. 

Multi-fluid models have, in fact, often been considered for heavy-ion 
collisions. \ One can, for example, treat the target and projectile nuclei as 
separate fluids to admit interpenetration, thus arriving at a two-fluid 
model. \ One could also use a relativistic multi-fluid model to allow for 
different species, e.g.~nucleons, deltas, hyperons, pions, kaons, etc. \ Such 
a model could account for the varying dynamics of the different species and 
their mutual diffusion and chemical reactions. \ The derivation of such a 
model would follow closely our discussion in Sec.~\ref{sec:twofluids}. \ 
In the heavy-ion community, it has been common to confuse the issue somewhat 
by insisting on choosing a particular local rest frame at each space-time 
point. \ This is, of course, complicated since the different fluids move at 
different speeds relative to any given frame. \ For the purpose of studying 
heavy ion collisions in the baryon-rich regions of space, the standard option 
seems to be to define the ``baryonic Lorentz frame''. \ This is the local 
Lorentz frame in which the motion of the center-of-baryon number (analogous 
to the center-of-mass) vanishes. 

The main problem with the one-fluid hydrodynamics model is the requirement 
of thermal equilibrium. \ In the hydrodynamic equations of motion it is 
implicitly assumed that local thermal equilibrium is ``imposed'' via an 
equation of state of the matter. \ This means that the relaxation timescale 
and the mean-free path should be much smaller than both the hydrodynamical 
timescale and the spatial size of the system. \ It seems reasonable to wonder 
if these conditions can be met for hadronic and nuclear collisions. \ On the 
other hand, from the kinematical point of view, apart from the use of the 
equation of state, the equations of hydrodynamics are nothing but 
conservation laws of energy and momentum, together with other conserved 
quantities such as charge. \ In this sense, for any process where the 
dynamics of flow is an important factor, a hydrodynamic framework is a 
natural first step. \ The effects of a finite relaxation time and mean-free 
path might be implemented at a later stage by using an effective equation 
of state, incorporating viscosity and heat conductivity, or some simplified 
transport equations. \ This does, of course, lead us back to the challenging 
problem of designing a causal relativistic theory for dissipation, see 
Sec.~\ref{sec:viscosity}. \ In the context of heavy-ion collisions no 
calculations have yet been performed using a fully three-dimensional, 
relativistic theory which includes dissipation. \ In fact, considering the 
obvious importance of entropy, it is surprising that so few calculations 
have been reported for either relativistic or nonrelativistic hydrodynamics 
(although see \cite{kapusta}). \ An interesting comparison of different 
dissipative formulations is provided in \cite{Muronga1,Muronga2}.  

\newpage

\section{Superfluids and broken symmetries} 
\label{sec:superfluid} 
 
In this section we discuss models that result when additional constraints are 
made on the properties of the fluids. \ We focus on the modelling of 
superfluid systems, using as our example the case of superfluid He$^4$. \ The 
equations describing more complex systems are readily obtained by generalizing 
our discussion. \ We contrast three different descriptions: (i) the model that 
follows from the variational framework that has been our prime focus so far 
if we impose that one constituent is irrotational, (ii) the potential 
formulation due to Khalatnikov and collaborators, and (iii) a ``hybrid'' 
formulation which has been used in studies of heavy-ion collisions with broken 
symmetries. 

\subsection{Superfluids} 

Neutron star physics provides ample motivation for the need to develop 
a relativistic description of superfluid systems. \ As the typical core 
temperatures (below $10^8$~K) are far below the Fermi temperature of the 
various constituents (of the order of $10^{12}$~K for baryons) neutron stars 
are extremely cold on the nuclear temperature scale. \ This means that, just 
like ordinary matter at near absolute zero temperature, the matter in the 
star will most likely freeze to a solid or become superfluid. \ While the 
outer parts of the star, the so-called crust, form an elastic lattice the 
inner parts of the star are expected to be superfluid. \ In practice, this 
means that we \underline{must} be able to model mixtures of superfluid 
neutrons and superconducting protons. \ It is also \underline{likely} that 
we need to understand superfluid hyperons and colour superconducting quarks. 
\ There are many hard physics questions that need to be considered if we are 
to make progress in this area. \ In particular, we need to make contact with 
microphysics calculations that determine the various parameters of such 
multi-fluid systems. \ However, we will ignore this aspect and focus on 
the various fluid models that have been used to describe relativistic 
superfluids. 

One of the key features of a pure superfluid is that it is irrotational. \ 
Bulk rotation is mimicked by the formation of vortices, slim ``tornadoes'' 
representing regions where the superfluid degeneracy is broken. \ In 
practice, this means that one would often, eg.~when modelling global neutron 
star oscillations, consider a macroscopic model where one ``averages'' over 
a large number of vortices. \ The resulting model would closely resemble the 
standard fluid model. \ Of course, it is important to remember that the 
vortices are present on the microscopic scale, and that they may affect the 
values of various parameters in the problem. \ There are also unique effects 
that are due to the vortices, eg.~the mutual friction that is thought to 
be the key agent that counteracts relative rotation between the neutrons and 
protons in a superfluid neutron star core \cite{mendII}. 

For the present discussion, let us focus on the simplest model problem of 
superfluid He$^4$. \ We then have two fluids, the superfluid Helium atoms 
with particle number density $n_\n$ and the entropy with particle number 
density $n_\s$. \ From the derivation in Sec.~\ref{sec:perfect_fluid} we 
know that the equations of motion can be written 
\begin{equation}
    \nabla_\mu n_\X^\mu = 0 
\end{equation} 
and
\begin{equation}
    n_\X^\mu \nabla_{[\mu} \mu^\X_{\nu]} = 0 \ . 
\end{equation} 
To make contact with other discussions of the superfluid problem 
\cite{carter92:_momen_vortic_helic,carter94:_canon_formul_newton_superfl,%
carter95:_equat_state,carter98:_relat_supercond_superfl},
we will use the notation $s^\mu= n_\s^\mu$ and $\Theta_\nu = \mu_\nu^\s$. \ Then 
the equations that govern the motion of the entropy obviously become 
\begin{equation}
    \nabla_\mu s^\mu = 0 \qquad \mbox{ and} \qquad s^\rho \nabla_{[\rho} 
    \Theta_{\nu]} = 0 \ . \label{ent_eom} 
\end{equation} 
Now, since the superfluid constituent is irrotational we will have 
\begin{equation} 
    \nabla_{[\sigma} \mu^\n_{\nu]} = 0 \ . 
\end{equation} 
Hence, the second equation of motion is automatically satisfied once we 
impose that the fluid is irrotational. \ The particle conservation law is, 
of course, unaffected. This example shows how easy it is to specify the 
equations that we derived earlier to the case when one (or several) 
components are irrotational. \ It is worth emphasizing that it is the 
momentum that is quantized, not the velocity. 

It is instructive to contrast this description with the potential formulation 
due to Khalatnikov and colleagues \cite{khal1,khal2}. \ We can obtain this 
alternative formulation in the following way 
\cite{carter94:_canon_formul_newton_superfl}. \ First of all, we know that 
the irrotationality condition implies that the particle momentum can be 
written as a gradient of a scalar potential $\alpha$ (say). \ That is, we 
have 
\begin{equation}
    v_\rho = - { \mu^\n_\rho \over m} = - \nabla_\rho \alpha \ . \label{kh_eq1} 
\end{equation}
Here $m$ is the mass of the Helium atom and $v_\rho$ is what is traditionally 
(and somewhat confusedly) referred to as the ``superfluid velocity''. \ We see 
that it is really a rescaled momentum. \ Next assume that the momentum of the 
remaining fluid (in this case, the entropy) is written 
\begin{equation}
    \mu^\s_\rho = \Theta_\rho = \kappa_\rho + \nabla_\rho \phi \ . 
\end{equation}
Here $\kappa_\rho$ is Lie transported along the flow provided that 
$s^\rho \kappa_\rho=0$ (assuming that the equation of motion 
Eq.~(\ref{ent_eom}) is satisfied). \ This leads to 
\begin{equation}
    \s^\rho \nabla_\rho \phi = s^\rho \Theta_\rho \ . 
\end{equation} 
There is now no loss of generality in introducing further scalar potentials 
$\beta$ and $\gamma$ such that $\kappa_\rho = \beta \nabla_\rho \gamma$, 
where the potentials are constant along the flow-lines as long as 
\begin{equation}
    s^\rho \nabla_\rho \beta = s^\rho \nabla_\rho \gamma = 0 \ . 
\end{equation} 
Given this, we have 
\begin{equation}
    \Theta_\rho = \nabla_\rho \phi + \beta \nabla_\rho \gamma \ . 
\end{equation}
Finally, comparing to Khalatnikov's formulation \cite{khal1,khal2} we 
define $\Theta_\rho = - \kappa w_\rho$ and let $\phi \to \kappa \zeta$ and 
$\beta \to \kappa \beta$. \ Then we arrive at the final equation of motion
\begin{equation}
    - \frac{\Theta_\rho}{\kappa} = w_\rho = - \nabla_\rho \zeta - \beta 
    \nabla_\rho \gamma \ . \label{kh_eq2}
\end{equation} 
Eqs.~(\ref{kh_eq1}) and (\ref{kh_eq2}), together with the standard particle 
conservation laws, are the key equations in the potential formulation. \ As 
we have seen, the content of this description is identical to that of the 
canonical variational picture that we have focussed on in this review. \ We 
have also seen how the various quantities can be related. \ Of course, one 
has to exercise some care in using this description. \ After all, referring 
to the rescaled momentum as the ``superfluid velocity'' is clearly misleading 
when the entrainment effect is in action.

\subsection{Broken symmetries}

In the context of heavy-ion collisions, models accounting for broken 
symmetries have sometimes been considered. \ At a very basic level, a model 
with a broken $U(1)$ symmetry should correspond to the superfluid model 
described above. \ However, at first sight our equations differ from those 
used in, for example, \cite{son,pujol,zhang}. \ Since we are keen to convince 
the reader that the variational framework we have discussed in this article 
is able to cover all cases of interest (in fact, we believe that it is more 
powerful than alternative formulations) a demonstration that we can 
reformulate our equations to get those written down for a system with a 
broken $U(1)$ symmetry has some merit. \ The exercise is also of interest 
since it connects with models that have been used to describe other 
superfluid systems.

Take as the starting point the general two-fluid system. \ From the 
discussion in Section~\ref{sec:twofluids}, we know that the momenta are in 
general related to the fluxes via
\begin{equation}
    \mu^\X_\rho = {\cal B}^\X n_\rho^\X + {\cal A}^{\X\Y} n_\rho^\Y \ .  
\end{equation}
Suppose that, instead of using the fluxes as our key variables, we want to 
consider a ``hybrid'' formulation based on a mixture of fluxes and momenta. 
\ In the case of the particle-entropy system discussed in the previous 
section, we can then use
\begin{equation}
    n_\rho^\n = \frac{1}{{\cal B}^\n} \mu_\rho^\n - 
    \frac{{\cal A}^{\n\s}}{{\cal B}^\n} n_\rho^\s \ . 
\end{equation}
Let us impose irrotationality on the fluid by representing the momentum as 
the gradient of a scalar potential $\varphi$. \ With $\mu_\rho^\n = 
\nabla_\rho \varphi$ we get
\begin{equation}
    n_\rho^\n = \frac{1}{{\cal B}^\n} \nabla_\rho \varphi - 
                \frac{{\cal A}^{\n\s}}{{\cal B}^\n} n_\rho^\s \ . 
\end{equation}
Now take the preferred frame to be that associated with the entropy flow, 
i.e.~introduce the unit four velocity $u^\mu$ such that $n_\s^\mu = n_\s 
u^\mu = su^\mu$. \ Then we have
\begin{equation}
    n_\rho^\n = n u_\rho - V^2 \nabla_\rho \varphi 
\end{equation}
where we have defined
\begin{equation}
    n \equiv - \frac{s {\cal A}^{\n\s}}{{\cal B}^\n} \qquad \mbox{and} 
    \qquad V^2 = - \frac{1}{{\cal B}^\n} \ . 
\end{equation}
With these definitions, the particle conservation law becomes
\begin{equation}
    \nabla_\rho n_\n^\rho = \nabla_\rho \left( n u^\rho - V^2 \nabla^\rho 
                            \varphi \right) = 0 \ . 
\end{equation}
The chemical potential in the entropy frame follows from
\begin{equation} 
    \mu = - u^\rho \mu^\n_\rho = - u^\rho \nabla_\rho \varphi \ . 
\end{equation}  
One can also show that the stress-energy tensor becomes
\begin{equation}
    T^\mu{}_\nu  = \Psi \delta^\mu{}_\nu + (\Psi+ \rho) u^\mu u_\nu - V^2 
                   \nabla^\mu \varphi \nabla_\nu \varphi
\end{equation}
where the generalized pressure is given by $\Psi$ as usual, and we have 
introduced
\begin{equation}
    \Psi + \rho = {\cal B}^\s s^2 + {\cal A}^{\s \n} s n \ .  
\end{equation}
The equations of motion can now be obtained from $\nabla_\mu T^\mu{}_\nu = 
0$. \ (Keep in mind that the equation of motion for $\X=\n$ is automatically 
satisfied once we impose irrotationality, as in the previous section.) \ This 
essentially completes the set of equations written down by, for example, Son 
\cite{son}. \ The argument in favour of this formulation is that it is close 
to the microphysics calculations, which means that the parameters may be 
relatively straightforward to obtain. \ Against the description is the fact 
that it is a, not very elegant, hybrid where the inherent symmetry amongst 
the different constituents is lost, and there is also a risk of confusion 
since one is treating a momentum as if it were a velocity.

\newpage

\section{Final remarks}

In writing this review, we have tried to discuss the different building 
blocks that are needed if one wants to construct a relativistic theory for 
fluids. \ Although there are numerous alternatives, we opted to base our 
discussion of the fluid equations of motion on the variational approach 
pioneered by Taub \cite{taub54:_gr_variat_princ} and in recent years 
developed considerably by Carter 
\cite{carter83:_in_random_walk,carter89:_covar_theor_conduc,carter92:_brane}. 
\ This is an appealing strategy because it leads to a natural formulation 
for multi-fluid problems. \ Having developed the variational framework, we 
discussed applications. \ Here we had to decide what to include and what to 
leave out. \ Our decisions were not based on any particular logic, we simply 
included topics that were either familiar to us, or interested us at the 
time. \ That may seem a little peculiar, but one should keep in mind that 
this is a ``living'' review. \ Our intention is to add further applications 
when the article is updated. \ On the formal side, we could consider how 
one accounts for elastic media and magnetic fields, as well as technical 
issues concerning relativistic vortices (and cosmic strings). \ On the 
application side, we may discuss many issues for astrophysical fluid flows 
(like supernova core collapse, jets, gamma-ray bursts, and cosmology). 

In updating this review we will obviously also correct the mistakes that 
are sure to be found by helpful colleagues. \ We look forward to receiving 
any comments on this review. \ After all, fluids describe physics at many 
different scales and we clearly have a lot of physics to learn. \ The only 
thing that is certain is that we will enjoy the learning process!

\newpage

\section{Acknowledgments} 

Several colleagues have helped us develop our understanding of relativistic 
fluid dynamics. \ We are particularly indebted to Brandon Carter, David 
Langlois, Reinhard Prix, and Bernard Schutz. 

NA acknowledges support from PPARC via grant no PPA/G/S/2002/00038 and 
Senior Research Fellowship no PP/C505791/1. \ GLC is supported by NSF grant 
no PHY-0457072.

\newpage

\bibliography{livrev_biblio}

\end{document}